\pdfoutput=1

\documentclass[11pt]{article}

\usepackage[preprint]{acl}

\usepackage{times}
\usepackage{latexsym}

\usepackage[T1]{fontenc}

\usepackage[utf8]{inputenc}

\usepackage{microtype}

\usepackage{inconsolata}

\usepackage{graphicx}
\usepackage{subcaption}
\usepackage{booktabs}
\usepackage{multirow}
\usepackage{array}    
\usepackage{amsmath}
\usepackage{tabu}
\usepackage{enumitem}
\usepackage{diagbox}
\usepackage{amssymb}

\definecolor{darkgreen}{RGB}{50,100,0}
\definecolor{darkred}{RGB}{200, 0, 0}
\newcommand{\cmark}{\textcolor{darkgreen}{\ding{51}}} 
\newcommand{\xmark}{\textcolor{darkred}{\ding{55}}} 
\newcommand{\xd}[1]{\textcolor{black}{#1}}

\usepackage{pifont}
\usepackage{tikz}

\newcommand{\cxmark}{%
    \tikz[baseline=(char.base)]{
        \node(char)[shape=rectangle, inner sep=0] {\textcolor{darkgreen}{\ding{51}}};
        \draw[darkred, thick, scale=0.16] (0.3,-0.2) -- (-0.2,0.3);
    }
}

\usepackage{color, xspace}
\usepackage[table]{xcolor}
\newcommand{\ie}{\emph{i.e.,}\xspace}
\newcommand{\eg}{\emph{e.g.,}\xspace}

\newcommand{\dname}{\textsc{MMDocIR}\xspace}

\definecolor{lightgray}{gray}{0.9}

\title{\dname: Benchmarking Multimodal Retrieval for Long Documents}

\author{Kuicai Dong$^{\dagger}$, Yujing Chang$^{\dagger}$, Derrick Goh Xin Deik$^{\dagger}$,   \\
        {\bf Dexun Li, Ruiming Tang, Yong Liu} \\
        Huawei Technologies Co., Ltd. \\ 
        correspond to $\{$dong.kuicai; liu.yong6$\}$@huawei.com}

\makeatletter
\AtBeginDocument{
\usepackage[hypcap=false]{caption}
\let\@oldmaketitle\@maketitle
\renewcommand{\@maketitle}{\@oldmaketitle
  \vspace{-35pt}
  \centering
  \includegraphics[width=\linewidth, trim={0.0em 0em 0.5em 0.0em}, clip]{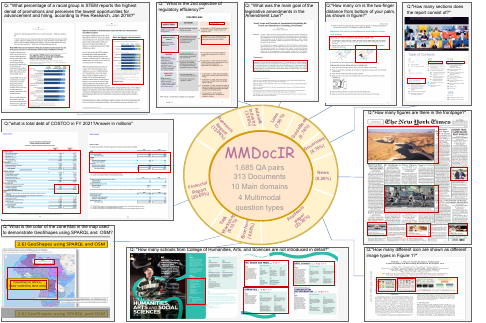}
  \vspace{-1em}
  \captionof{figure}{
    \dname evaluation set comprises 313 long documents and 1,658 queries across 10 domains. For each query, page-level labels are provided via selected screenshots. Red boundary boxes represent layout-level labels.
  }
  \label{fig:bench_overview}
  \vspace{17pt}
 }
}
\makeatother

\begin{document}
\maketitle
\def\thefootnote{$\dagger$}\footnotetext{These authors contributed equally}
\def\thefootnote{\arabic{footnote}}

\begin{abstract}

Multimodal document retrieval aims to identify and retrieve various forms of multimodal content, such as figures, tables, charts, and layout information from extensive documents. Despite its increasing popularity, there is a notable lack of a comprehensive and robust benchmark to effectively evaluate the performance of systems in such tasks. To address this gap, this work introduces a new benchmark, named MMDocIR, that encompasses two distinct tasks: page-level and layout-level retrieval. The former evaluates the performance of identifying the most relevant pages within a long document, while the later assesses the ability of detecting specific layouts, providing a more fine-grained measure than whole-page analysis. A layout refers to a variety of elements, including textual paragraphs, equations, figures, tables, or charts. The MMDocIR benchmark comprises a rich dataset featuring 1,685 questions annotated by experts and 173,843 questions with bootstrapped labels, making it a valuable resource in multimodal document retrieval for both training and evaluation. Through rigorous experiments, we demonstrate that (i) visual retrievers significantly outperform their text counterparts, (ii) MMDocIR training set effectively enhances the performance of multimodal document retrieval and (iii) text retrievers leveraging VLM-text significantly outperforms retrievers relying on OCR-text.
Our dataset is available at \url{https://mmdocrag.github.io/MMDocIR/}.

\end{abstract}

\begin{figure*}
    \centering
    \includegraphics[width=0.9\linewidth]{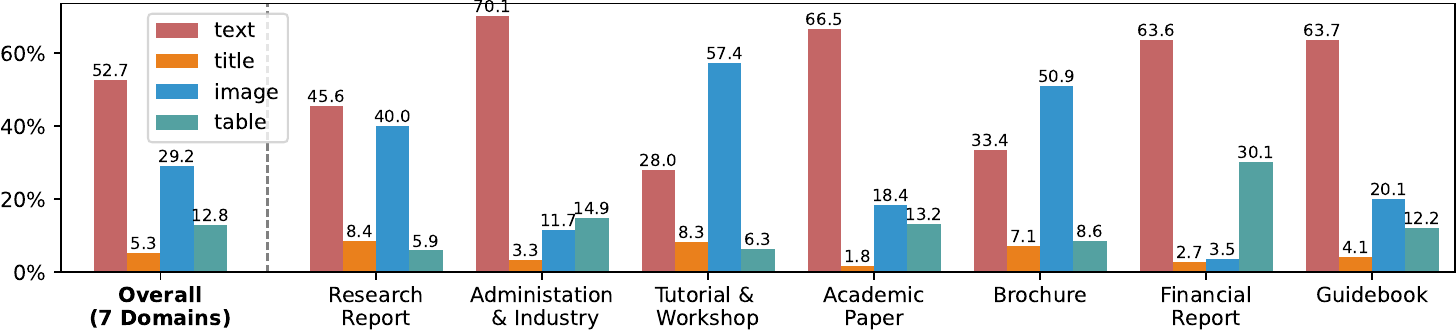}
    \vspace{-0.5em}
    \caption{Area ratio of different modalities (1) in overall and (2) by domains in MMLongBench-Doc benchmark. Note that the white spaces, headers, and footers are removed from the area counting.}
    \label{fig:mmlongbench_layout}
\end{figure*}

\section{Introduction}
\label{sec:introduction}

\begin{table*}[t]
\centering
\small
\renewcommand{\arraystretch}{0.9}
\resizebox{\linewidth}{!}{
\begin{tabular}{l@{\hskip 2pt}|c@{\hskip -4pt}c@{\hskip 2.5pt}c@{\hskip 6.5pt}c@{\hskip 7pt}c|c@{\hskip 2pt}c@{\hskip 2.5pt}c@{\hskip 1pt}|c@{\hskip 2pt}c@{\hskip 1pt}}
\toprule
\multirow{2}{*}{\textbf{Benchmarks}}  & \multicolumn{5}{c|}{\textbf{Question}} & \multicolumn{3}{c|}{\textbf{Document}} & \multicolumn{2}{c}{\textbf{Label}} \\
& Type & Expert? & IR? & \#Num & Evidence Type & Domain & \#Pages & Source & Page & Layout  \\
\midrule
DocCVQA & VQA question & \cmark & \cmark & 20 & TXT/L & Finance & 1.0 & \cmark & \cmark & \xmark\\
SciMMIR & Image caption & \xmark & \xmark & 530k & TAB/I & Science & 1.0 & \xmark & \xmark & \xmark \\
ViDoRe & VQA question & \cxmark & \xmark & 3,810 & TXT/C/TAB/I & Multiple & 1.0 & \xmark & \cmark & \xmark  \\
PDF-MVQA  & Search query & \xmark & \cmark & 260k & TXT/TAB/I & Biomedical & 9.6 & \cmark & \cmark & \cmark \\
MMLongBench-Doc & VQA question & \cmark & \xmark & 1,082 & TXT/C/TAB/I & Multiple & 47.5 & \cmark & \cmark & \xmark \\
Wiki-SS & Natural question  & \xmark & \cmark &  3,610 & TXT & Wikipedia & 1.0 & \xmark & \cmark & \xmark \\
DocMatix-IR & VQA question & \xmark & \xmark & 5.61m & TXT/C/TAB/I & Multiple & 4.2 & \cmark & \cmark & \xmark \\
\midrule
\textbf{\dname (eval)} & VQA question & \cmark & \cmark & 1,658 & TXT/C/TAB/I & Multiple & 65.1 & \cmark & \cmark & \cmark \\
\textbf{\dname (train)} & VQA question & \cxmark & \cmark & 73.8k & TXT/C/TAB/I & Multiple & 49.3 & \cmark & \cmark & \cxmark \\
\bottomrule
\end{tabular}
}
\vspace{-0.8em}
\caption{\dname versus existing document IR datasets. \textbf{TXT/C/TAB/I} refers to text/chart/table/image.}
\vspace{-0.8em}
\label{tab:dataset_comparison}
\end{table*}

Multimodal document retrieval~\cite{DBLP:conf/icdar/HassanCG13, lee2024unifiedmultimodalinterleaveddocument} aims to retrieve information from visually rich documents based on user queries. Unlike traditional document retrieval~\cite{zhang-etal-2022-multi,chen2023walking, dong-etal-2024-mc, wang-etal-2023-retrieval} and long-context~\cite{yang2025llmreason} QA which primarily deals with textual data, multimodal document retrieval imposes substantially greater demands on understanding multimodal elements such as images, tables, charts, and layout designs. Such elements often carry significant information that plain text fails to convey~\cite{cui2021documentai, DBLP:conf/wincom/SassiouiBOECO23}: tables reveal structured data patterns, charts visualize trends or correlations, images offer contextual and semantic cues, etc. Combining these visual elements enriches the quality of retrieved content. Our analysis of MMLongBench-Doc benchmark~\cite{ma2024mmlongbenchdoc} in Figure~\ref{fig:mmlongbench_layout} shows that: text occupies only 52.7\% of content area, while images and tables account for 29.2\% and 12.8\% respectively. This highlights the need for retrieval systems that effectively handle multimodal and cross modal~\cite{zhang2025cross, dong2025docresearcher} information.

However, as shown on Table~\ref{tab:dataset_comparison}, existing benchmarks exhibit several critical limitations that undermine comprehensive evaluation of multimodal retrieval systems. The \textbf{key limitations}  include: 
\textbf{1. Question Quality:} Many questions used in existing benchmarks are directly sourced from datasets for Visual Question Answering (VQA) tasks. Some questions often assume the input is already relevant, making it not suited for meaningful evaluation of retrieval capabilities.
\textbf{2. Document Completeness and Diversity}: Existing benchmarks often provide only partial documents, limiting the ability to evaluate within full document context. Additionally, the narrow range of document domains further restricts their applicability across diverse use-cases in real-world. 
\textbf{3. Retrieval Granularity:} Most benchmarks support only page-level retrieval. Such granularity is often insufficient, as user queries frequently target specific elements, such as figures or tables, rather than entire pages. 

To address these gaps, we introduce \textbf{\dname}, a multimodal document information retrieval benchmark. \dname is designed for \textbf{two key} tasks: \textbf{\textit{page-level} \xspace} and \textbf{\textit{layout-level}} retrieval.
\textbf{(1)} The page-level retrieval identifies the most relevant pages within a document to answer user query.
\textbf{(2)} The layout-level retrieval targets the most relevant layouts.
A layout is an element on the document page where the element could be a paragraph, a heading, an equation, a table, a figure, or a chart (see Appendix \ref{appendix:data_demo_layout} for more examples).
Such task supports more precise and context-aware retrieval that pinpoint specific elements to address user queries.
To support both tasks, we develop \textbf{\dname evaluation set} that comprises 313 documents, each averaging 65.1 pages, along with 1,658 modified queries derived from MMLongBench-Doc and DocBench~\cite{zou2024docbench}. 
The queries are annotated with 2,107 page-level and 2,638 layout-level labels.
The page labels are specific pages that contain the evidence needed to answer the query.\footnote{While MMLongBench-Doc provided initial page labels, our meticulous review lead to corrections in 21.3\% of them.}
The layout labels consist of precisely drawn bounding boxes around the key evidence within the identified pages.
In addition, we introduce the \textbf{\dname training set}, designed to support retriever training. It contains 73,843 questions sourced from 7 DocQA datasets.
To construct this set, we manually collect 6,878 documents and apply a semi-automatic pipeline to annotate the ground truth labels.

By leveraging \dname, we conduct a comprehensive evaluation on multimodal document retrieval across two retriever types: visual-driven and text-driven.
\textit{\textbf{Visual-driven retrievers}} \cite{ma2024dse,faysse2024colpali}, leverage vision language models (VLMs) 
to capture rich multimodal cues and generate embeddings for both queries and documents.
In contrast, \textit{\textbf{text-driven retrievers}}~\cite{karpukhin-etal-2020-dense, khattab-etal-2020-colbert, xiao2023-bge} rely on OCR or VLM to first convert the multimodal content into text, subsequently employing language models (LMs)
to generate embeddings for both queries and documents.
Our extensive experiments reveal that visual-driven retrievers consistently outperform their text-driven counterparts, often by a significant margin. 
In summary, our contributions are threefold:
\begin{itemize}[leftmargin=*, itemsep=-0.3em, topsep=0.1em]
    \item \textbf{Dual-task Retrieval Framework}: We propose a dual-task retrieval framework ($\S$~\ref{sec:definition}) that supports page-level and fine-grained layout-level multimodal document retrieval.
    
    \item \textbf{MMDocIR Benchmark}: We introduce Multimodal Document Information Retrieval benchmark. The evaluation set ($\S$~\ref{sec:eval_dataset}) consists of 313 documents with expert-annotated labels for 1,658 questions. The training set ($\S$~\ref{sec:train_dataset}) consists of 6,878 documents and labels for 73,843 questions.

    \item We conduct extensive experiments and comparisons of both text and visual retrievers ($\S$~\ref{sec:experiment}), demonstrating clear advantage of incorporating visual content in multimodal retrieval tasks.
    
\end{itemize}

\section{Dual-Task Retrieval Definition} \label{sec:definition}

Let $\mathcal{D}$ be a document corpora consisting of document pages: $\mathcal{P} = \{p_1, p_2, \dots, p_n\}$, and layouts: $\mathcal{L} = \{l_1, l_2, \dots, l_m\}$ extracted via layout detection.\footnote{A document page usually comprises about 5 to 15 layouts, depending on its complexity.}. 
The objective is to perform document retrieval at both page-level and layout-level.
Specifically, given query $Q$, the task is to retrieve the top $k$ pages and layouts most relevant to $Q$, where $k << n$ and $k << m$. The relevance of pages ($p$) and layouts ($l$) to $Q$ is measured by similarity scores, $\textrm{Sim}(Q,p)$ and $\textrm{Sim}(Q,l)$ respectively.
The retrieval system consists of two phases: (1) an offline indexing phase, where pages and layouts from $\mathcal{P}$ and $\mathcal{L}$ are encoded into vectors, and (2) an online querying phase, in which a query $Q$ is encoded into a vector, which is then compared against the offline-indexed vectors using similarity scores $\textrm{Sim}(Q,p)$ for pages and $\textrm{Sim}(Q,l)$ for layouts.

\section{\dname: Evaluation Set}
\label{sec:eval_dataset}

\subsection{Document Corpora Collection} \label{ssec:eval_doc_corpora}

After a comprehensive review of existing DocVQA datasets, we select MMLongBench-Doc \cite{ma2024mmlongbenchdoc} and DocBench \cite{zou2024docbench} to facilitate our benchmark construction (see Appendix \ref{appendix:eval_doc_corpora} for our selection criteria).
MMLongBench-Doc is a long-context, multimodal benchmark comprising 1,091 questions across 135 documents with 47.5 pages on average. 
DocBench emphasizes long document understanding, consisting of 1,102 questions across 229 documents, each with an average length of 77.5 pages. 
Both datasets offer corpora from diverse domains with expert-annotated questions that require evidence from various modalities. Consequently, we curate a set of 364 documents and 2,193 questions for our subsequent annotation.

\subsection{Annotation Process}
\label{ssec:dataset_label}

\paragraph{Question Filtering and Revision.}
To ensure that the questions in \dname are optimally suited for document retrieval tasks, we identify four specific types of questions (see Appendix \ref{appendix:dataset_filter}) that do not align well with the objectives of IR.
By filtering and refining these questions, we ensure the integrity and relevance of \dname, resulting in 1,658 questions for subsequent annotation.

\paragraph{Page-level Annotation.}
We annotate page labels that precisely identify the exact pages containing ground truth evidence. Given that documents in \dname contain 65.1 pages on average, pinpointing relevant pages is highly non-trivial,  akin to finding a needle in haystack, which demands careful inspection and document understanding. 
Our annotation process is described as follows:
\begin{itemize}[leftmargin=*, itemsep=-0.5em, topsep=0.0em]
    \item For \textbf{DocBench}: we manually annotate page labels for all 864 questions from scratch, by carefully reviewing each document and locating the pages containing answer evidence.  
    \item For \textbf{MMLongBench-Doc}: we rigorously review and validate the answers and page labels of 794 questions. This effort results in corrections to 10 answers and 169 page labels\footnote{Common errors in page labeling: annotators starting page indexing at 1 rather than 0, missing labels for questions spanning multiple pages, and incorrect or absent page labels.}.
\end{itemize}
Through these efforts, we successfully obtain page-level labels for a total of 1,658 questions in \dname.

\paragraph{Layout-level Annotation.}
To enhance the granularity of our benchmark, we extend our annotations to include layout-level labels, identifying specific layout elements as evidence. Compared to page annotation, layout-level labeling is significantly more complex and labor-intensive. 
Our method of annotations is as follows:
\begin{itemize}[leftmargin=*, itemsep=-0.5em, topsep=0.0em]
\item \textbf{Layout Detection}. We begin by utilizing MinerU \cite{wang2024mineru} to automatically parse all documents and detect all layouts (\eg layout type and bounding boxes).\footnote{These elements are categorized into five main types: text, image, table, title, and equation. The bounding boxes are represented by coordinates of top-right and bottom-left corners.}

\item \textbf{Evidence Identification}. We identify the layouts that contain necessary answer evidence. In case where MinerU fails to detect evidentiary element, we manually annotate the bounding boxes, accounting for 7\% of the total layout-level labels.
\end{itemize}
Ultimately, this meticulous process leads to the annotation of 2,638 layout labels in \dname.

\subsection{Quality Control}

\begin{table}[t]
\setlength{\tabcolsep}{4.5pt}
    \centering
    \small
    \resizebox{\linewidth}{!}{%
    \begin{tabular}{ c|ccc|ccc}
     \toprule
     \multirow{2}{*}{Consistency} & \multicolumn{3}{c|}{Page Labels} &  \multicolumn{3}{c}{Layout Labels} \\
      & Prec. & Recall & F1 & Prec. & Recall & F1 \\
     \midrule
     A$\leftarrow$B & 95.7 &96.1 & 95.9 & 88.1 & 86.8 & 87.4\\
     B$\leftarrow$A & 94.3 & 94.6 & 94.4 & 85.9 & 87.5 & 86.7\\
     \midrule
     Average & 95.0 & 95.4 & 95.2 & 87.0 & 87.2 & 87.1\\
     \bottomrule
    \end{tabular}}
    \vspace{-1em}
    \caption{Annotation consistency between group A \& B.}
    \vspace{-1em}
    \label{tab:IAA}
\end{table}

To ensure annotation quality and reliability in \dname, we have adopted a rigorous 3-stage quality control process. We split questions into two parts. Each group is responsible for annotating approximately 1,000 questions, with an overlap of 400 questions serving the need for cross-validation.
The process for quality control works as follows:
\begin{itemize}[leftmargin=*, itemsep=-0.5em, topsep=0.0em]

    \item \textbf{Overlap Scoring}: For the 400 overlapping questions, A$\leftarrow$B evaluates A's labels with B's labels as ground truth, and vice versa for B$\leftarrow$A.

    \item \textbf{Cross-Evaluation}: We cross-evaluate and achieve F$_1$ score of 95.2 and 87.1 for page and layout labels, as shown in Table~\ref{tab:IAA}. We then identify and fix the discrepancies.

    \item \textbf{Random Cross-Validation}: We randomly cross-validate 50\% of the remaining annotations. In the cases where we have different opinions, we discuss to achieve mutually-agreed annotations. 
\end{itemize}

\begin{figure}[t]
    \centering
    \begin{subfigure}{0.335\linewidth}
        \includegraphics[width=\linewidth, trim={0.1em 0 0.1em 0}, clip]{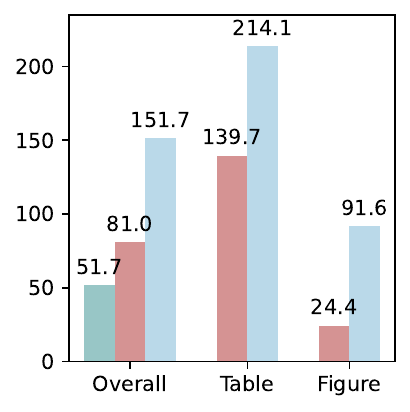}
        \vspace{-1.5em}
        \caption{\#words/layout}
        \label{fig:subimg1}
    \end{subfigure}
    \begin{subfigure}{0.645\linewidth}
        \includegraphics[width=\linewidth, trim={2em 0 0.8em 0}, clip]{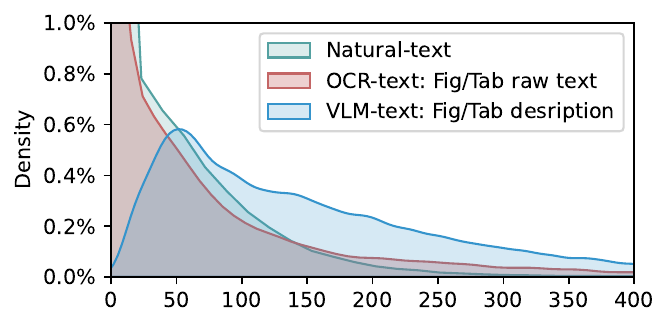}
        \vspace{-1.5em}
        \caption{Distribution density: \#words/layout}
        \label{fig:subimg2}
    \end{subfigure}
    \begin{subfigure}{0.98\linewidth}
        \includegraphics[width=\linewidth]{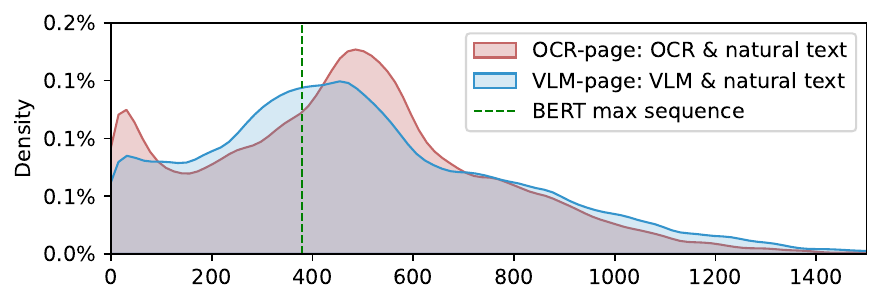}
        \vspace{-1.8em}
        \caption{Distribution density: \#words/page}
        \label{fig:subimg3}
    \end{subfigure}
    \vspace{-0.5em}
    \caption{Distribution of OCR/VLM-text by length.}
    \vspace{-1.0em}
    \label{fig:distribution_fig}
\end{figure}

\subsection{Multimodal content as OCR/VLM-text}
\label{ssec:mm-text}
To apply multimodal retrieval to text retrievers, we convert multimodal layouts (\eg tables or figures) into text.
Specifically, we extract text using OCR~\cite{2007TessOverview} (``OCR-text'') and generate detailed descriptions using VLMs~\cite{2024gpt4o, 2024qwenvl} (``VLM-text''). As a result, each image layout is represented in three formats: original image, OCR-text, and VLM-text.

\noindent
For \textbf{layouts}, the average word length and distribution of OCR-text and VLM-text of \dname are shown in Figure \ref{fig:subimg1} and \ref{fig:subimg2}. Notably, the length of VLM-text is 1.5 and 3.8 times of OCR-text for table (with more structured numbers) and figure (mostly with visual elaboration), respectively.\footnote{Tables typically contain structured numbers, which can be well recorded by text. Figures contain mostly visual elaboration, making it hard to extract raw text.}

\noindent
For \textbf{pages}, we construct two variants by combining the natural text with either OCR-text and VLM-text for each page, resulting in OCR-page and VLM-page representations. The average word length is 477 and 505 for OCR-page and VLM-page respectively, with their distribution shown in Figure \ref{fig:subimg3}.

\begin{table*}[t] 
\small
\renewcommand{\arraystretch}{0.9}
\setlength{\tabcolsep}{4.5pt}
    \centering
    \resizebox{\linewidth}{!}{%
  \begin{tabular}{l@{\hskip 2pt}|c@{\hskip 2pt}c@{\hskip 3pt}c@{\hskip 1pt}c@{\hskip 1pt}c@{\hskip 2.5pt}c@{\hskip 2.5pt}c|c@{\hskip 4.5pt}c@{\hskip 5.5pt}c@{\hskip 4.5pt}c@{\hskip 4.5pt}|c@{\hskip 4.5pt}c@{\hskip 3.5pt}c@{\hskip 2.5pt}c@{\hskip 2.5pt}}
    \toprule
    
    \multirow{3}{*}{\textbf{\dname}} &  \multicolumn{7}{c|}{Document Statistics} & \multicolumn{4}{c|}{Questions (\%)} & \multicolumn{4}{c}{Modality Distribution \%} \\
    & \multirow{2}{*}{\#Doc} & \multirow{2}{*}{\#QA} & \#Page & \#Lay & \#Page & \#Lay & \multirow{2}{*}{\%Lay} & \multirow{2}{*}{Text} & \multirow{2}{*}{Image} & \multirow{2}{*}{Table} & Lay/ & \multirow{2}{*}{Text} & \multirow{2}{*}{Image} & \multirow{2}{*}{Table} & \multirow{2}{*}{Title}\\
    & & & Label & Label & /Doc & /Page  & & & & & Meta & & & \\
    
    \midrule
    \textbf{Eval Domains}  & 313 & 1,658 & 2,107 & 2,638 & 65.1 & 8.3 & 41.8 &44.7 &21.7 &37.4 &11.5 & 60.4 & 18.8 & 16.7 & 4.1 \\
    \midrule
    - Research Report     & 34 & 200 & 318 & 400 & 39.4 & 6.0 & 39.1 & 45.0& 17.5 & 74.5 & 13.5 & 45.6 & 40.0 & 5.9 & 8.4 \\
    - Admin \& Industry   & 10 & 59 & 82 & 113 & 16.8 & 9.1 & 45.1 & 78.0 & 20.3 & 13.5 & 13.5 & 70.1 & 11.7 & 14.9 & 3.2 \\
    - Tut \& Workshop     & 17 & 102 & 165 & 225 & 57.5 & 4.1 & 43.8 & 37.2 & 61.7 & 24.5 & 9.8 & 28.0 & 57.3 & 6.3 & 8.3 \\
    - Academic Paper      & 75 & 386 & 473 & 571 & 19.5 & 10.1 & 48.4 & 28.8 & 25.7 & 50.0 & 10.4 & 74.6 & 12.8 & 11.1 & 1.5 \\
    - Brochure            & 15 & 76 & 121 & 178 & 30.3 & 9.7 & 41.1 & 60.5 & 52.6 & 18.4 & 36.8 & 33.3 & 50.8 & 8.5 & 7.0 \\
    - Financial Report    & 51 & 343 & 394 & 477 & 169.5 & 9.2 & 44.8 & 28.0 & 13.1 & 54.5 & 5.3 & 60.3 & 7.9 & 29.2 & 2.6 \\
    - Guidebook           & 22 & 112 & 168 & 223 & 78.4 & 10.0 & 33.6 & 51.8 & 54.4 & 26.8 & 17.8 & 63.7 & 20.0 & 12.1 & 4.1 \\
    - Government    & 44 & 111 & 116 & 132 & 68.9 & 6.9 & 45.4 & 69.37 & 2.7 & 0 & 7.6 & 88.2 & 3.7 & 5.7 & 2.4 \\
    - Laws          & 44 & 132 & 133 & 149 & 58.5  & 6.0 & 31.2 & 62.1 & 0 & 10.6 & 27.3 & 83.8 & 1.6 & 12.3 & 2.2 \\
    - News          & 1  & 137 & 137 & 170 & 50.0  & 73.6 & 72.3 & 70.1 & 1.5 & 0 & 28.5 & 48.5 & 39.8 & 0.0 & 11.6 \\

  \bottomrule
  \end{tabular}
  }
  \vspace{-1em}
\caption{Detailed statistics of \dname evaluation set. ``\#Lay/Page'' is averaging layouts per page, reflecting layout complexity. ``\%Lay'' is the area ratio of useful layouts (excluding white spaces, headers, and footers).}
\vspace{-0.5em}
\label{tab:eval_doc_stats}
\end{table*}

\begin{table*}[t] 
\renewcommand{\arraystretch}{0.9}
\small
    \centering
    \resizebox{\linewidth}{!}{%

    \begin{tabular}{l@{\hskip 2.5pt}|c@{\hskip 2.5pt}|c@{\hskip 7.5pt}c@{\hskip 6.5pt}c@{\hskip 4.5pt}c@{\hskip 4.5pt}c|c@{\hskip 7.5pt}c@{\hskip 5.5pt}c@{\hskip 7.5pt}c@{\hskip 7.5pt}|c@{\hskip 3.5pt}c@{\hskip 2.5pt}}
    \toprule

    \multirow{3}{*}{\textbf{\dname}} &  \multirow{3}{*}{Domain} & \multicolumn{5}{c|}{Document Statistics} & \multicolumn{4}{c|}{Evidence Modality (\%)} & \multicolumn{2}{c}{Labels} \\
    
    & & \multirow{2}{*}{\#Doc} & \multirow{2}{*}{\#QA} & \#Page & \#Lay & \multirow{2}{*}{\%Lay} & \multirow{2}{*}{Text} & \multirow{2}{*}{Image} & \multirow{2}{*}{Table} & \multirow{2}{*}{Title} & \multirow{2}{*}{Page} & \multirow{2}{*}{Lay} \\
    
    & & & & /Doc & /Page  & & & & & & \\
    
    \midrule
    \textbf{Train Subsets} & assorted docs & 6,878 & 73,843 & 32.6 & 6.32 & 42.6 & 49.3 & 34.3 & 10.8 & 4.9 & \cmark & \cxmark \\
    \tabucline[.4pt on 2pt off 1.5pt]{-} \\[-7.5pt] 
    - MP-DocVQA  & health/ind. docs & 875   & 15,266 & 46.8  & 6.9 & 38.8 & 57.3 & 18.0 & 22.7 & 1.9 & \cmark & \xmark \\
    - SlideVQA  & diverse slides & 2,011 & 11,066 & 49.3  & 4.4 &  42.3 & 30.1 & 56.2 & 4.7 & 8.8 & \cmark & \xmark \\
    - TAT-DQA   & annual reports & 163   & 15,814 & 147.3 & 9.2 &  42.2 & 66.4 & 4.4 & 26.5 & 2.7 & \cmark & \cmark \\
    - arXivQA & arXiv papers & 1,579 & 12,314 & 18.4  & 7.9 & 50.0 & 70.4 & 22.3 & 2.8 & 1.0 & \cmark & \cmark \\
    - SciQAG & science papers & 1,197 & 4,976 & 9.0   & 9.1 & 53.7 & 61.8 & 28.0 & 6.7 & 1.5 & \cmark & \cmark \\
    - DUDE & assorted docs & 779   & 3,173 & 15.6  & 7.4 & 42.5  & 57.1 & 24.7 & 15.2 & 2.9 & \cmark & \cmark \\
    - CUAD & legal contracts & 274   & 11,234 & 29.6  & 7.4 & 24.7 & 89.3 & 2.5 & 6.4 & 1.1 & \cmark & \xmark \\
  \bottomrule
  \end{tabular}
  }
  \vspace{-1em}
\caption{\dname training set statistics about our collected documents, questions, and constructed labels.}
\vspace{-1em}
\label{tab:train_doc_stats}
\end{table*}

\subsection{Statistics and Analysis}
\textbf{Document Analysis. \xspace}
As shown in Table~\ref{tab:eval_doc_stats}, \dname evaluation set includes 313 long documents, averaging 65.1 pages, categorized into 10 domains.
Different domains feature distinct multimodal distribution. 
For instance, reports, tutorials, workshops, and brochures predominantly contain images, whereas financial and industry documents are table-rich. In contrast, government and legal documents primarily comprise text. 
The overall modality distribution is as follows: text (60.4\%), image (18.8\%), table (16.7\%), and others (4.1\%), with fine-grained distribution shown in Figure \ref{subfig:mmdocir_eval}.

\noindent
\textbf{Question and Annotation Analysis. \xspace}
\dname includes 1,658 questions, and 2,107 page and 2,638 layout labels. The evidence spans 4 modalities: text (44.7\%), image (21.7\%), table (37.4\%), and layout/meta\footnote{The ``layout/meta'' refers to evidence related to layout information and meta-data statistics.} (11.5\%).  
Notably, \dname presents several challenges: 254 questions require cross-modal understanding, 313 questions require evidence across multiple pages, and 637 questions require reasoning over multiple layouts.

\section{\dname: Training Set}
\label{sec:train_dataset}

\subsection{Document Corpus Collection} \label{ssec:doc_corpora}

After screening related DocVQA datasets, we collect our training set corpora from 7 datasets, namely MP-DocVQA \cite{tito2023mp-docvqa}, SlideVQA \cite{tanaka2023slidevqa}, TAT-DQA \cite{zhu2022tatdqa}, SciQAG \cite{wan2024sciqag}, DUDE \cite{landeghem2023dude}, and CUAD \cite{2021HendrycksCUAD}.
Since most of these datasets do not provide original document, we invest significant efforts in tracing and recovering the original document, as detailed in Appendix \ref{appendix:train_doc}.

\subsection{Label Construction and statistics}
We use semi-automated construction pipeline to generate page-level and layout-level labels for datasets that lack them, referring to Appendix \ref{appendix:labels} and \ref{appendix:hard_neg} for more details of construction process. Notably, layout annotations are missing from most existing datasets, as we manage to obtain or construct layout-level labels for only 4 datasets.
The overall statistics (\eg document information, modality distribution, domain, etc) of \dname training set are summarized in Table~\ref{tab:train_doc_stats}.

\begin{table*}[t] 
\small
\setlength{\tabcolsep}{2.5pt}
\renewcommand{\arraystretch}{0.8}
    \centering
    \resizebox{\linewidth}{!}{%
  \begin{tabu}{lll|cccccccccc|cc}

    \toprule

    \multicolumn{3}{c}{\multirow{2}{*}{\diagbox{Method}{Domain}}} 
     & Resear. & Admin & Tutori.\& & Acade. & \xspace Broch- & Finance & Guide- & Govern- & \multirow{2}{*}{\xspace Laws \xspace} & \multirow{2}{*}{\xspace News\xspace}  & \multicolumn{2}{c}{Average} \\
     & & & Report & \&Indu. & Worksh. & Paper & ure & Report & book & ment & & & \xspace Macro & Micro \xspace \\
     
    \midrule
   
    \parbox[t]{2.5mm}{\multirow{11}{*}{\rotatebox[origin=c]{90}{Recall@$k=1$}}} &
    \parbox[t]{2.0mm}{\multirow{6}{*}{\rotatebox[origin=c]{90}{VLM-text}}}
     & DPR          & 32.3 & 25.5 & 27.0 & 31.0 & 28.4 & 18.8 & 23.5 & 31.2 & 38.3 & 16.1 & 27.2 & 26.9 \\
    && ColBERT      & 48.6 & 42.8 & 51.1 & 46.2 & 36.0 & 36.8 & 49.6 & \underline{60.9} & 59.5 & 26.3 & 45.8 & 44.9 \\
    && BGE          & 48.8 & 30.9 & 47.1 & 40.8 & 37.6 & 28.4 & 43.4 & 51.9 & 48.9 & 28.5 & 40.6 & 39.6 \\
    && E5           & 48.1 & 30.0 & 50.4 & 39.4 & 41.1 & 29.7 & 40.9 & 52.8 & 51.1 & 24.1 & 40.8 & 39.5 \\
    && Contriever   & 45.5 & 31.2 & 49.8 & 41.5 & 39.4 & 29.4 & 45.2 & 55.3 & 51.1 & 20.4 & 40.9 & 39.7 \\
    && GTE          & 46.5 & 26.3 & 48.7 & 38.9 & 35.9 & 27.0 & 46.2 & 50.1 & 45.8 & 24.1 & 38.9 & 37.9 \\
    \tabucline[.2pt on 1pt off 1.5pt]{2-15} \\[-7.5pt]  
    
    & \parbox[t]{1.5mm}{\multirow{5}{*}{\rotatebox[origin=c]{90}{Image}}}
    & DSE$_{\mathrm{wiki-ss}}$  & 53.0 & 50.0 & 54.0 & 48.7 & 45.1 & 43.0 & 51.5 & 46.9 & 54.2 & 33.6 & 48.0 & 47.5 \\
    && DSE$_{\mathrm{docmatix}}$  & 52.3 & 40.4 & 56.1 & 51.7 & 45.8 & 43.5 & 53.8 & 53.7 & 58.3 & \underline{46.7} & 50.2 & 50.1 \\
    && ColPali  & 56.0 & \textbf{51.8} & \textbf{58.6} & 55.9 & 52.0 & \underline{47.2} & 57.9 & 53.9 & \underline{64.0} & 32.8 & 53.0 & 52.7 \\
    && DPR-Phi3$_{\mathrm{ours}}$  & \textbf{58.9} & \underline{50.4} & \underline{57.4} & \underline{59.0} & \textbf{57.3} & 44.6 & \textbf{63.8} & 50.5 & \textbf{64.4} & 35.0 & \underline{54.1} & \underline{53.7} \\
    && Col-Phi3$_{\mathrm{ours}}$ & \underline{56.7} & \underline{50.4} & 56.9 & \textbf{61.3} &\underline{54.8} & \textbf{50.7} & \underline{60.8} & \textbf{61.3} & 63.6 & \textbf{54.0} & \textbf{57.0} & \textbf{57.1} \\
    \midrule

    \parbox[t]{1.2mm}{\multirow{11}{*}{\rotatebox[origin=c]{90}{Recall@$k=3$}}} &
    \parbox[t]{1.5mm}{\multirow{6}{*}{\rotatebox[origin=c]{90}{VLM-text}}}
     & DPR          & 52.2 & 44.2 & 43.5 & 54.6 & 52.0 & 35.1 & 44.4 & 53.9 & 57.2 & 25.5 & 46.3 & 46.2 \\
    && ColBERT      & 70.1 & 64.4 & 70.3 & 72.3 & 59.1 & 55.3 & 71.1 & \textbf{81.3} & 70.8 & 34.3 & 64.9 & 64.8 \\
    && BGE          & 71.5 & 48.2 & 68.8 & 65.7 & 56.2 & 46.5 & 66.1 & 69.9 & 72.0 & 32.1 & 59.7 & 59.6 \\
    && E5           & 68.4 & 45.7 & 68.1 & 63.7 & 60.1 & 44.0 & 69.3 & 72.3 & 78.8 & 32.8 & 60.3 & 59.3 \\
    && Contriever   & 69.4 & 55.3 & 68.3 & 64.9 & 56.9 & 46.2 & 69.9 & 71.1 & 72.0 & 32.1 & 60.6 & 59.7 \\
    && GTE          & 71.1 & 44.5 & 67.2 & 64.4 & 54.3 & 43.0 & 70.6 & 71.9 & 68.2 & 31.4 & 58.7 & 58.3 \\
    \tabucline[.2pt on 1pt off 1.5pt]{2-15} \\[-7.5pt] 
    
    & \parbox[t]{1.5mm}{\multirow{5}{*}{\rotatebox[origin=c]{90}{Image}}}
    & DSE$_{\mathrm{wiki-ss}}$  & 75.4 & 65.0 & 73.9 & 79.8 & 69.5 & 63.5 & 75.4 & 71.5 & 81.4 & 50.4 & 70.6 & 71.4 \\
    && DSE$_{\mathrm{docmatix}}$  & 75.4 & 67.5 & 73.3 & 80.0 & 66.3 & 61.6 & 72.8 & 76.4 & 82.6 & \underline{57.7} & 71.4 & 71.8 \\
    && ColPali  & 77.6 & \underline{71.8} & \textbf{79.4} & 83.4 & \textbf{72.6} & \underline{66.1} & \textbf{80.0} & \underline{80.4} & \textbf{86.4} & 49.6 & \underline{74.7} & \underline{75.0} \\
    && DPR-Phi3$_{\mathrm{ours}}$  & \textbf{80.3} & 66.5 & \underline{77.6} & \underline{83.9} & \underline{71.9} & 63.8 & \underline{79.8} & 71.4 & \underline{84.5} & 55.5 & 73.5 & 74.3 \\
    && Col-Phi3$_{\mathrm{ours}}$ & \underline{80.2} & \textbf{74.1} & 77.4 & \textbf{84.8} & 69.1 & \textbf{67.7} & 78.7 & 79.5 & 81.8 & \textbf{69.3} & \textbf{76.3} & \textbf{76.8} \\
    \midrule

    \parbox[t]{1.2mm}{\multirow{11}{*}{\rotatebox[origin=c]{90}{Recall@$k=5$}}} &
    \parbox[t]{1.5mm}{\multirow{6}{*}{\rotatebox[origin=c]{90}{VLM-text}}}
     & DPR          & 66.5 & 60.1 & 56.0 & 68.9 & 58.8 & 43.8 & 57.1 & 68.6 & 64.8 & 33.6 & 57.8 & 57.8 \\
    && ColBERT      & 78.8 & 74.0 & 78.7 & 82.3 & 66.1 & 60.8 & 77.0 & \textbf{88.5} & 78.0 & 38.7 & 72.3 & 72.3 \\
    && BGE          & 79.5 & 65.8 & 71.3 & 76.8 & 62.4 & 56.0 & 77.2 & 77.4 & 79.5 & 38.0 & 68.4 & 68.5 \\
    && E5           & 76.9 & 64.2 & 75.3 & 74.4 & 67.4 & 52.0 & 78.5 & 78.6 & 82.6 & 40.9 & 69.1 & 67.9 \\
    && Contriever   & 77.2 & 67.1 & 76.7 & 75.2 & 65.1 & 53.7 & 75.4 & 79.2 & 83.3 & 39.4 & 69.2 & 68.3 \\
    && GTE          & 77.4 & 62.6 & 74.7 & 75.8 & 62.0 & 51.8 & 77.8 & 80.0 & 75.0 & 39.4 & 67.6 & 67.2 \\
    \tabucline[.2pt on 1pt off 1.5pt]{2-15} \\[-7.5pt] 
    
    & \parbox[t]{1.5mm}{\multirow{5}{*}{\rotatebox[origin=c]{90}{Image}}}
    & DSE$_{\mathrm{wiki-ss}}$  & 84.0 & \textbf{80.2} & 78.7 & 87.0 & 75.7 & \underline{73.0} & 82.0 & 77.3 & 88.3 & 58.4 & 78.5 & 79.2 \\
    && DSE$_{\mathrm{docmatix}}$  & 82.1 & 77.2 & 79.6 & 87.8 & 73.9 & 72.4 & 81.7 & 83.1 & 89.4 & \underline{67.9} & 79.5 & 80.1 \\
    && ColPali  & 84.6 & \underline{79.3} & \underline{82.3} & 89.0 & \underline{79.8} & 72.1 & \underline{86.7} & \underline{84.9} & \textbf{92.4} & 56.9 & 80.8 & 81.0 \\
    && DPR-Phi3$_{\mathrm{ours}}$  & \textbf{86.9} & 76.2 & \textbf{85.3} & \underline{91.9} & \textbf{80.0} & 71.2 & \textbf{87.1} & 79.5 & \underline{92.0} & 61.3 & \underline{81.1} & \underline{81.8} \\
    && Col-Phi3$_{\mathrm{ours}}$ & \underline{86.3} & 78.8 & 81.2 & \textbf{92.4} & 79.0 & \textbf{73.8} & 85.3 & \textbf{85.1} & 87.1 & \textbf{73.0} & \textbf{82.2} & \textbf{83.0} \\

  \bottomrule
  \end{tabu}
  }
\vspace{-1em}
\caption{Main results for page-level retrieval, with the best results in \textbf{boldface} and second best results \underline{underlined}. For clarity, we omit results using VLM-text (Refer to Table \ref{tab:appendix_page_recall} for full results).}
\vspace{-1em}
\label{tab:main_page_recall}
\end{table*}

\section{Experiment}
\label{sec:experiment}

\begin{table*}[t] 
\renewcommand{\arraystretch}{0.8 }
\setlength{\tabcolsep}{2.5pt}
\small
    \centering
    \resizebox{\linewidth}{!}{%
  \begin{tabu}{lll|cccccccccc|cc}
    \toprule

    \multicolumn{3}{c}{\multirow{2}{*}{\diagbox{Method}{Domain}}} 
     & Resear. & Admin & Tutori.\& & Acade. & \xspace Broch- & Finance & Guide- & Govern- & \multirow{2}{*}{\xspace Laws \xspace} & \multirow{2}{*}{\xspace News\xspace}  & \multicolumn{2}{c}{Average} \\
     & & & Report & \&Indu. & Worksh. & Paper & ure & Report & book & ment & & & \xspace Macro & Micro \xspace \\
     
    \midrule
   
    \parbox[t]{2.5mm}{\multirow{11}{*}{\rotatebox[origin=c]{90}{Recall@$k=1$}}} &

    \parbox[t]{2.0mm}{\multirow{6}{*}{\rotatebox[origin=c]{90}{VLM-text}}}
    & DPR          & 11.6 & 9.5 & 19.2 & 19.2 & 14.9 & 15.9 & 15.8 & 25.6 & 34.7 & 27.0 & 19.3 & 19.2 \\
    && ColBERT      & 22.0 & 14.9 & 28.0 & 28.3 & 17.9 & 29.7 & 21.1 & \textbf{52.6} & \textbf{54.5} & \textbf{44.5} & \underline{31.3} & 31.4 \\
    && BGE          & 19.2 & 15.2 & 24.6 & 28.7 & 12.8 & 27.6 & 19.7 & 47.0 & 52.3 & 35.8 & 28.3 & 29.0 \\
    && E5           & 15.9 & 8.8 & 27.7 & 24.3 & 14.6 & 21.8 & 14.7 & 45.6 & \underline{53.0} & 40.5 & 26.7 & 26.4 \\
    && Contriever   & \textbf{23.4} & 7.5 & 28.2 & 26.8 & 17.1 & 25.7 & 16.1 & 43.6 & 51.5 & \underline{42.9} & 28.3 & 28.9 \\
    && GTE          & 17.5 & 10.5 & 23.0 & 27.2 & 14.5 & 26.3 & 14.4 & 39.8 & 49.2 & 38.3 & 26.1 & 27.1 \\
    \tabucline[.2pt on 1pt off 1.5pt]{2-15} \\[-7.5pt]

    & \parbox[t]{1.5mm}{\multirow{5}{*}{\rotatebox [origin=c]{90}{Image}}}
    & DSE$_{\mathrm{wiki-ss}}$  & 20.6 & 15.1 & 31.0 & 31.1 & 20.1 & 29.2 & 22.0 & 39.3 & 37.5 & 35.8 & 28.2 & 29.2 \\
    && DSE$_{\mathrm{docmatix}}$  & 19.9 & 11.4 & 31.5 & 30.1 & 17.8 & 30.0 & 20.8 & 46.5 & 39.4 & 31.4 & 27.9 & 29.1 \\
    && ColPali   & 22.5 & 21.3 & 36.6 & 30.9 & \underline{26.8} & \textbf{32.1} & 19.3 & \underline{52.5} & 51.8 & 33.6 & \textbf{32.7} & \textbf{32.5} \\
    && DPR-Phi3$_{\mathrm{ours}}$  & 21.1 & \textbf{22.1} & \underline{36.8} & \textbf{35.2} & 25.6 & 28.7 & \textbf{24.1} & 38.3 & 35.4 & 27.4 & 29.5 & 30.2  \\
    && Col-Phi3$_{\mathrm{ours}}$ & \underline{22.6} & \underline{22.0} & \textbf{37.5} & \underline{34.9} & \textbf{28.9} & \underline{30.3} & \underline{22.7} & 50.2 & 45.1 & 26.3 & 31.1 & \underline{31.6}\\
    
    \midrule

    \parbox[t]{2.5mm}{\multirow{11}{*}{\rotatebox[origin=c]{90}{Recall@$k=5$}}} &

    \parbox[t]{2.0mm}{\multirow{6}{*}{\rotatebox[origin=c]{90}{VLM-text}}}
    & DPR          & 31.0 & 25.7 & 36.7 & 44.9 & 33.0 & 34.1 & 34.9 & 49.9 & 56.3 & 51.1 & 39.8 & 40.4 \\
    && ColBERT      & 41.8 & 37.7 & 53.7 & \underline{61.8} & 35.1 & 52.4 & 46.1 & \textbf{83.2} & 70.1 & 62.5 & \textbf{54.4} & \textbf{56.0} \\
    && BGE          & 41.0 & 28.1 & 52.7 & 59.2 & 36.7 & 46.0 & \textbf{50.7} & 72.0 & 71.5 & 59.9 & 51.8 & 53.2 \\
    && E5           & 35.4 & 28.1 & 51.7 & 58.5 & 33.2 & 41.2 & 40.2 & \underline{79.7} & \textbf{77.9} & \textbf{64.6} & 51.1 & 51.8 \\
    && Contriever   & 40.2 & 29.2 & 54.1 & 57.9 & 36.8 & 47.1 & 44.6 & 68.8 & \underline{76.2} & 61.9 & 51.7 & 53.0 \\
    && GTE          & 36.7 & 25.6 & 51.2 & 56.6 & 39.7 & 46.6 & \underline{46.4} & 72.2 & 74.3 & \underline{63.2} & 51.3 & 52.3 \\
    \tabucline[.2pt on 1pt off 1.5pt]{2-16} \\[-7.5pt]  

    & \parbox[t]{1.5mm}{\multirow{5}{*}{\rotatebox [origin=c]{90}{Image}}}
    & DSE$_{\mathrm{wiki-ss}}$  & 42.4 & 32.9 & \underline{56.3} & 58.5 & 39.8 & 50.6 & 41.6 & 68.6 & 60.9 & 50.0 & 50.2 & 52.1 \\
    && DSE$_{\mathrm{docmatix}}$  & 39.6 & 36.2 & 53.9 & 57.5 & 33.7 & \underline{52.5} & 42.8 & 69.4 & 63.1 & 48.9 & 49.8 & 51.9 \\
    && ColPali   & 40.7 & \textbf{45.9} & 54.9 & 58.5 & \underline{42.6} & 51.2 & 45.7 & 76.8 & 74.5 & 48.9 & \underline{54.0} & 54.3 \\
    && DPR-Phi3$_{\mathrm{ours}}$  & \underline{45.5} & 37.7 & \textbf{57.0} & \textbf{62.9} & 41.4 & 51.1 & 45.5 & 65.1 & 60.8 & 49.3 & 51.6 & 53.9 \\
    && Col-Phi3$_{\mathrm{ours}}$ & \textbf{46.4} & \underline{38.2} & 53.1 & \underline{61.8} & \textbf{45.0} & \textbf{54.6} & 45.7 & 68.8 & 65.7 & 43.8 & 52.3 & \underline{54.5} \\

    \midrule

    \parbox[t]{2.5mm}{\multirow{11}{*}{\rotatebox[origin=c]{90}{Recall@$k=10$}}} &

    \parbox[t]{2.0mm}{\multirow{6}{*}{\rotatebox[origin=c]{90}{VLM-text}}}
     & DPR          & 42.2 & 33.1 & 52.1 & 56.2 & 39.9 & 43.5 & 44.0 & 62.8 & 61.7 & 59.7 & 49.5 & 50.5 \\
    && ColBERT      & 51.0 & 48.7 & 60.6 & 69.8 & 43.9 & \textbf{61.6} & 53.7 & \textbf{88.4} & 74.8 & 66.4 & \underline{61.9} & \textbf{63.7} \\
    && BGE          & 51.1 & 38.7 & 62.1 & 71.5 & 41.9 & 55.6 & \textbf{58.7} & 80.8 & 78.7 & 63.5 & 60.3 & 62.4 \\
    && E5           & 45.3 & 38.6 & 62.0 & 70.5 & 45.6 & 50.0 & 55.3 & \underline{87.1} & \underline{82.4} & \underline{66.8} & 60.4 & 61.2 \\
    && Contriever   & 49.9 & 41.3 & 62.0 & 70.5 & 44.8 & 56.5 & 54.5 & 81.3 & 78.0 & 64.9 & 60.4 & 62.2 \\
    && GTE          & 48.6 & 41.1 & 61.5 & 68.8 & 44.3 & 56.9 & 58.0 & 83.0 & 77.5 & \textbf{66.9} & 60.7 & 62.2 \\
    \tabucline[.2pt on 1pt off 1.5pt]{2-16} \\[-7.5pt]

    & \parbox[t]{1.5mm}{\multirow{5}{*}{\rotatebox [origin=c]{90}{Image}}}
    & DSE$_{\mathrm{wiki-ss}}$  & 55.9 & 41.3 & 61.5 & 68.1 & 47.8 & \underline{60.7} & 54.2 & 72.9 & 68.3 & 54.4 & 58.5 & 61.1 \\
    && DSE$_{\mathrm{docmatix}}$  & 53.7 & 43.3 & 59.6 & 66.5 & 44.7 & 59.1 & 50.3 & 75.4 & 69.2 & 53.7 & 57.5 & 59.9  \\
    && ColPali   & 53.6 & \textbf{54.1} & 64.4 & 69.5 & \underline{48.8} & \underline{60.7} & 54.0 & 81.9 & \textbf{82.5} & 50.4 & \textbf{62.0} & 63.2 \\
    && DPR-Phi3$_{\mathrm{ours}}$  &  \textbf{58.1} & 49.1 & \textbf{67.0} & \textbf{74.7} & 48.4 & 57.9 & \underline{57.8} & 68.7 & 66.2 & 54.4 & 60.2 & 62.8 \\
    && Col-Phi3$_{\mathrm{ours}}$ & \underline{57.7} & \underline{50.5} & \underline{66.6} & \underline{72.3} & \textbf{50.7} & 59.3 & 53.6 & 68.5 & 74.8 & 57.5 & 61.1 & \underline{63.3} \\

  \bottomrule
  \end{tabu}
  }
  \vspace{-1em}
\caption{Main results for layout-level retrieval (Refer to Table \ref{tab:appendix_layout_recall} for full results with VLM-text and Hybrid inputs).}
\vspace{-1em}
\label{tab:main_layout_recall}
\end{table*}

\subsection{Evaluation Metric}
The retriever scores each page or layout in the document based on its relevance to the question, and returns the top $k$ candidates with the highest scores.
Recall$@k$ is defined as the proportion of ground truth page/layout evidence successfully retrieved.
For \textbf{page matching}, the recall is straightforwardly computed with page indices.
For \textbf{layout matching}, we calculate recall based on the overlaps between the bounding boxes of retrieved layouts and gold-standard layouts.
Unlike page retrieval, where boundaries are unambiguous, layout detection tools can produce differing bounding boxes for the same content. Our ground-truth layouts include both MinerU outputs and manual annotations for cases where MinerU misses elements.
During our evaluation, retrievers are provided with MinerU predicted layouts (\eg bboxes and types), but some ground truth bboxes cannot be exactly matched to them. Therefore, simple binary classifications (matched or not matched) are insufficient. Overlap-based recall offers a nuanced and realistic evaluation, especially where perfect alignment is not guaranteed.

\subsection{Baseline Models and Setting} \label{ssec:baseline}
We evaluate 6 state-of-the-art text retrievers: namely DPR, ColBERT, BGE, E5, Contriever, and GTE (see Appendix~\ref{appendix:text_retriever}). 
Additionally, we evaluate 5 VLM-based retrievers: 3 off-the-shelf models, namely DSE$_{\mathrm{wiki-ss}}$, DSE$_{\mathrm{docmatix}}$, and ColPali (see Appendix~\ref{appendix:visual_retriever}), and 2 models trained using \dname training set (see Appendix~\ref{appendix:train}).
Among all retrievers, ColBERT, ColPali, and Col-Phi3$_{\mathrm{ours}}$ represent query/document as a list of token-level embeddings, while the other retrievers represent query/document as a single dense embedding.
All retrievers are adapted to a dual-task setting:
\begin{itemize}[leftmargin=*, itemsep=-0.5em, topsep=0.1em]
    \item \textbf{Page Retrieval}: For text retrievers, we use the text from OCR-page or VLM page as described in Section \ref{ssec:mm-text}.
    For visual retrievers, we directly utilize document page screenshots.
    
    \item \textbf{Layout Retrieval}: Text retrievers process multimodal layouts using OCR or VLM text (see Section \ref{ssec:mm-text}).
    Visual retrievers\footnote{Most visual retrievers are not explicitly trained on text query-doc pairs, this setup constitutes out-of-domain data.} process textual layouts using either \textbf{Image} input (cropped image of textual area) or \textbf{Hybrid} input (original text, as VLM can directly encode text).
\end{itemize}

\subsection{Main Results for Page-level Retrieval}
\label{ssec:main_page}

Table~\ref{tab:main_page_recall} presents the main results for page-level retrieval. Our key findings are as follows:
\begin{itemize}[leftmargin=*, itemsep=-0.3em, topsep=0.1em]
    \item \textbf{Superiority of Visual Retrievers}: Visual retrievers consistently outperform text retrievers across various domains and retrieval metrics, highlighting the advantage of using screenshots to retain multimodal cues often lost in text conversion.
    \item \textbf{Effectiveness of \dname}: 
    The visual retrievers trained on the \dname training set demonstrate superior performance, demonstrating the value of high-quality training data.
    \item \textbf{Effectiveness of VLM-Text}: Although VLM-text approaches underperform visual retrievers, they achieve much better performance than the OCR-text methods. This indicates benefits of using GPT-4o to preserve visual cues in text.
    \item \textbf{Effect of Token-level Embeddings}: Compared to dense-level retrievers (\eg BGE, DSE, DPR-Phi3$_{\mathrm{ours}}$), token-level retrievers (\eg ColBERT, ColPali, Col-Phi3$_{\mathrm{ours}}$) achieve more advantageous results in Recall@1 and have marginal performance improvement in Recall@3/5. However, token-level embedding can incur storage costs of 10 times more than a single embedding (DSE requires 0.24GB for indexing \dname while ColPali requires 10.0GB).
    \item \textbf{Top 5 Coverage}: Retrieving top 5 pages provides substantial coverage of relevant information.
\end{itemize}

\subsection{Main Results for Layout-level Retrieval}
Table~\ref{tab:main_layout_recall} shows the main results for layout-level retrieval. Our key findings are as follows:
\begin{itemize}[leftmargin=*, itemsep=-0.3em, topsep=0.1em]
    \item \textbf{Effectiveness of VLM-Text}: Interestingly, VLM-text approaches perform comparably to visual retrievers, demonstrating the promising image description capabilities of state-of-the-art VLM. This greatly benefits textual retrievers in multimodal understanding.
    \item \textbf{Comparison of Hybrid vs. Pure Image Sequences}: Visual retrievers relying on hybrid image-text sequences generally perform less effectively than those utilizing pure image sequences. This suggests that current VLMs may have stronger capabilities in modeling images than text within the multimodal framework. 
    \item \textbf{Effect of Token-level Embeddings}: For layout retrieval tasks, token-level retrievers marginally outperform dense-level retrievers, demonstrating its importance of such task.
    \item \textbf{Top 10 Coverage}: For layout retrieval tasks, retrieving top 10 layouts does not guarantee comprehensive coverage of the ground truth layout labels, emphasizing the complexity of the tasks.
\end{itemize}

\begin{table}[t] 
\small
\setlength{\tabcolsep}{3.5pt}
\renewcommand{\arraystretch}{0.85}
    \centering
    \resizebox{\linewidth}{!}{%
  \begin{tabu}{ll|ccc|ccc}
    \toprule

    \multicolumn{2}{c|}{\multirow{2}{*}{Method}}
     &  \multicolumn{3}{c|}{Page recall} &  \multicolumn{3}{c}{Layout recall} \\
     & & OCR & VLM & $\Delta$ & OCR & VLM & $\Delta$ \\
     
    \midrule
   
    \parbox[t]{3.0mm}{\multirow{6}{*}{\rotatebox[origin=c]{90}{$k=1$}}} &
    DPR            & 22.3 & 27.2 & +4.9 & 12.6 & 19.3 & +6.7 \\
    & ColBERT      & 40.3 & 45.8 & +5.5 & 19.8 & 31.3 & +11.5 \\
    & BGE          & 35.7 & 40.6 & +4.9 & 19.0 & 28.3 & +9.3 \\
    & E5           & 35.0 & 40.8 & +5.8 & 18.4 & 26.7 & +8.3 \\
    & Contriever   & 35.3 & 40.9 & +5.6 & 18.8 & 28.3 & +9.5 \\
    & GTE          & 35.4 & 38.9 & +3.5 & 18.2 & 26.1 & +7.9 \\
    \midrule
    
    \parbox[t]{3.0mm}{\multirow{6}{*}{\rotatebox[origin=c]{90}{$k=3 \text{ or } 5$}}} &
    DPR            & 39.4 & 46.3 & +6.9 & 23.7 & 39.8 & +16.1 \\
    & ColBERT      & 58.8 & 64.9 & +6.1 & 33.2 & 54.4 & +21.2 \\
    & BGE          & 55.4 & 59.7 & +4.3 & 32.7 & 51.8 & +19.1 \\
    & E5           & 54.8 & 60.3 & +5.5 & 33.3 & 51.1 & +17.8 \\
    & Contriever   & 54.9 & 60.6 & +5.7 & 31.7 & 51.7 & +20.0 \\
    & GTE          & 54.9 & 58.7 & +3.8 & 33.5 & 51.3 & +17.8 \\
    \midrule
    
    \parbox[t]{3.0mm}{\multirow{6}{*}{\rotatebox[origin=c]{90}{$k=5 \text{ or } 10$}}} &
    DPR            & 49.0 & 57.8 & +8.8 & 29.9 & 49.5 & +19.6 \\
    & ColBERT      & 66.0 & 72.3 & +6.3 & 37.6 & 61.9 & +24.3 \\
    & BGE          & 62.7 & 68.4 & +5.7 & 37.8 & 60.3 & +22.5 \\
    & E5           & 64.1 & 69.1 & +5.0 & 39.0 & 60.4 & +21.4 \\
    & Contriever   & 63.1 & 69.2 & +6.1 & 37.3 & 60.4 & +23.1 \\
    & GTE          & 63.2 & 67.6 & +4.4 & 40.9 & 60.7 & +19.8 \\

  \bottomrule
  \end{tabu}
  }
\vspace{-1em}
\caption{Results of text retrievers using OCR/VLM-text.}
\vspace{-1em}
\label{tab:ocr_llm_compare}
\end{table}

\label{ssec:main_layout}

\subsection{Text Retrieval: OCR-text vs VLM-Text}
Text retrievers leveraging VLM-text significantly outperform those using OCR-text in both tasks. Based on results, OCR-text is insufficient for multimodal retrieval, while VLM-text retains richer multimodal information.
Although VLM-text offers much more comprehensive text information than OCR-text, it also introduces higher computational overhead and longer inference time.

Most text retrievers based on on BERT~\cite{devlin-etal-2019-bert}, truncate input that exceed 512 tokens (approximately 380 english words). As shown in Figure \ref{fig:subimg3}, there are many pages containing more than 380 words (62.9\% for OCR-page and 61.1\% for VLM-page). Those pages suffer from critical information loss during page retrieval if the ground truth evidence is in the truncated part. In contrast, only a small fraction of layouts contain more than 380 tokens (3.9\% for OCR-text, 4.8\% for VLM-text, 0.5\% for natural-text). Hence, as reflected in Table \ref{tab:main_page_recall} and \ref{tab:main_layout_recall}, text retriever demonstrates stronger performance on layout-level retrieval than on page-level retrieval.

\begin{table}[t] 
\small
\setlength{\tabcolsep}{5.5pt}
\renewcommand{\arraystretch}{0.85}
    \centering
  \begin{tabu}{ll|ccc}
    \toprule

    \multicolumn{2}{c|}{\multirow{2}{*}{Method}}
     &  \multicolumn{3}{c}{Layout recall} \\
     &  & Hybrid & Image & $\Delta$ \\
     
    \midrule
   
    \parbox[t]{1.0mm}{\multirow{5}{*}{\rotatebox[origin=c]{90}{$k=1$}}} &
    DSE$_{\mathrm{wiki-ss}}$       & 24.6 & 28.2 & +3.6 \\
    & DSE$_{\mathrm{docmatix}}$    & 27.5 & 27.9 & +0.4 \\
    & ColPali                      & 28.5 & 32.7 & +4.2 \\
    & DPR-Phi3$_{\mathrm{ours}}$   & 28.9 & 29.5 & +0.6 \\
    & Col-Phi3$_{\mathrm{ours}}$   & 29.8 & 31.1 & +1.3 \\
    \midrule
    
    \parbox[t]{1.0mm}{\multirow{5}{*}{\rotatebox[origin=c]{90}{$k=5$}}} &
    DSE$_{\mathrm{wiki-ss}}$       & 46.7 & 50.2 & +3.5 \\
    & DSE$_{\mathrm{docmatix}}$    & 48.2 & 49.8 & +1.6 \\
    & ColPali                      & 52.2 & 54.0 & +1.8 \\
    & DPR-Phi3$_{\mathrm{ours}}$   & 50.1 & 51.6 & +1.5 \\
    & Col-Phi3$_{\mathrm{ours}}$   & 50.0 & 52.3 & +2.3 \\
    \midrule
    
    \parbox[t]{1.0mm}{\multirow{5}{*}{\rotatebox[origin=c]{90}{$k=10$}}} &
    DSE$_{\mathrm{wiki-ss}}$       & 55.8 & 58.5 & +2.7 \\
    & DSE$_{\mathrm{docmatix}}$    & 57.4 & 57.5 & +0.1 \\
    & ColPali                      & 60.0 & 62.0 & +2.0 \\
    & DPR-Phi3$_{\mathrm{ours}}$   & 55.5 & 60.2 & +4.7 \\
    & Col-Phi3$_{\mathrm{ours}}$   & 58.7 & 61.1 & +2.4 \\

  \bottomrule
  \end{tabu}
\vspace{-1em}
\caption{Results of visual retrievers: image vs hybrid.}
\vspace{-1em}
\label{tab:hyrid_compare}
\end{table}

\subsection{Visual Retrieval: Image vs Hybrid input}
\label{ssec:hybrid_analysis}

Visual retrievers tend to perform better when encoding text as images via visual encoders, rather than processing native textual input with LLM backbones. This advantage largely stems from their training setup: visual retrievers are typically optimized using text queries paired with image-based passages or documents, but are not fine-tuned directly on purely textual passages.
However, encoding text as images incurs substantial computational overhead. Representing text as image tokens requires significantly more resources than native text encoding.
To address this inefficiency and promote balanced retrieval capabilities \cite{dumitru2025copyspec, liang-etal-2025-adaptive}, we advocate for future visual retrievers to be jointly trained on both text and visual retrieval tasks using SFT or RL \cite{duong2025improving}. Such hybrid training would enable models to efficiently process text when appropriate, without compromising performance on visual inputs.

\subsection{Cascade Retrieval}
\label{ssec:cascade_retrieval}

As shown in Table~\ref{tab:main_layout_recall}, directly performing layout retrieval can be challenging. Hence, we propose alternative methods, by perform page-retrieval first, subsequently followed by layout-retrieval within the retrieved page. We term such retrieval to be cascade retrieval. Note that the page retrieval is not perfect, such error can propagate to layout retrieval and affect the final results.

\begin{table}[h]
\centering
\small
    \begin{tabular}{ll|c|c|c}
    \toprule
    \textbf{page(1st)} & \textbf{layout(2nd)} & \textbf{Top1} & \textbf{Top5} & \textbf{Top10} \\
    \midrule
    \multicolumn{2}{l|}{BGE: direct layout} & 29.0 & 53.2 & 62.4 \\
    BGE & BGE & 24.3 & 49.0 & 58.6 \\
    \midrule
    \multicolumn{2}{l|}{E5: direct layout} & 26.4 & 51.8 & 61.2 \\
    E5 & E5 & 22.2 & 47.8 & 58.2 \\
    \midrule
    \multicolumn{2}{l|}{ColBERT: direct layout} & 31.4 & 56.0 & 63.7 \\
    ColBERT & ColBERT & 28.5 & 53.0 & 61.3 \\
    \midrule
    \multicolumn{2}{l|}{ColPali: direct layout} & 32.5 & 54.3 & 63.2 \\
    ColPali & ColPali & 32.7 & 54.5 & 63.2 \\
    ColPali & ColBERT & 31.8 & 57.0 & 64.2 \\
    \midrule
    \multicolumn{2}{l|}{DSE: direct layout} & 29.1 & 51.9 & 59.9 \\
    DSE & DSE & 29.6 & 54.0 & 62.0 \\
    DSE & ColBERT & 29.0 & 56.4 & 64.4 \\
    \midrule
    \multicolumn{2}{l|}{DPR-Phi3: direct layout} & 30.2 & 53.9 & 62.8 \\
    DPR-Phi3 & DPR-Phi3 & 31.1 & 54.2 & 61.7 \\
    DPR-Phi3 & ColBERT & 30.6 & 56.6 & 64.5 \\
    \midrule
    \multicolumn{2}{l}{Col-Phi3: direct layout} & 31.6 & 54.5 & 63.3 \\
    Col-Phi3 & Col-Phi3 & 33.3 & 58.6 & 63.7 \\
    Col-Phi3 & ColBERT & \textbf{35.3} & \textbf{58.8} & \textbf{65.4} \\
    \bottomrule
\end{tabular}
\caption{Comparison of 1-stage vs 2-stage approaches across different models}
\label{tab:model_cascade}
\end{table}

In this setting, we retrieve top-$k$ pages first, then rerank all layouts belonging to retrieved $k$ pages, and get top-$k$ layouts. The cascade retrieval results are shown in Table~\ref{tab:model_cascade}. We can observe that method with high page retrieval recall can significantly improve layout retrieval in the reranking paradigm.

\begin{table}
\small
    \centering
    \begin{tabu}{l|ll|ccc}
        \toprule
         & & Model & Store  & Index  & Search \\
         & &  & (GB) & (MM:SS) &  (MM:SS) \\
        \midrule
        \parbox[t]{2.5mm}{\multirow{7}{*}{\rotatebox[origin=c]{90}{Page Processing}}} &

        \parbox[t]{2.5mm}{\multirow{4}{*}{\rotatebox[origin=c]{90}{Text}}} &
        DPR           & \textbf{0.06} & \textbf{6:53} & \textbf{00:02} \\
        &&ColBERT     & 3.45& 14:12 & 00:04 \\
        &&BGE         & \underline{0.08} & \underline{7:31} & \underline{00:03} \\
        &&E5          & \underline{0.08} & 8:26 & \underline{00:03} \\
 
        \tabucline[.2pt on 1pt off 1.5pt]{2-6} \\[-7.5pt]
        
        & \parbox[t]{2.5mm}{\multirow{3}{*}{\rotatebox[origin=c]{90}{Image}}} &
        DPR-Phi3       & 0.24  & 101:20 & 00:04   \\
        && ColPali        & 10.00   & 47:14 & 00:05   \\
        && Col-Phi3     & 24.56   & 106:23 & 00:07   \\
        
        \midrule
        
        \parbox[t]{2.5mm}{\multirow{10}{*}{\rotatebox[origin=c]{90}{Layout Processing}}} &

        \parbox[t]{2.5mm}{\multirow{4}{*}{\rotatebox[origin=c]{90}{Text}}} &
        DPR           & \textbf{0.51} & \textbf{41:33} &\textbf{ 00:15} \\
        &&ColBERT     & 26.72& 94:56 & 02:25 \\
        &&BGE         & \underline{0.66} & \underline{55:05} & \underline{00:18} \\
        &&E5          & \underline{0.66} & 60:21 & \underline{00:18} \\
 
        \tabucline[.2pt on 1pt off 1.5pt]{2-6} \\[-7.5pt]
        
        & \parbox[t]{2.5mm}{\multirow{3}{*}{\rotatebox[origin=c]{90}{Image}}} &
        DPR-Phi3       & 1.99   & 735:51 & 00:44   \\
        && ColPali        & 83.50   & 262:29 & 09:06   \\
        && Col-Phi3       & 204.32   & 784:07 & 10:56   \\
        \tabucline[.2pt on 1pt off 1.5pt]{2-6} \\[-7.5pt]
        
         & \parbox[t]{2.5mm}{\multirow{3}{*}{\rotatebox[origin=c]{90}{Hybrid}}} &
        DPR-Phi3       & 1.84   & 130:38 & 01:09  \\
        && ColPali     & 12.06   & 73:50 & 04:24   \\
        && Col-Phi3    & 22.72  & 140:44 & 02:41  \\
        \bottomrule
    \end{tabu}
\vspace{-1em}
\caption{Efficiency analysis different retrievers.}
\vspace{-1em}
\label{tab:efficiency_stats}
\end{table}

\subsection{Efficiency Analysis}
\label{ssec:efficiency_analysis}

We evaluate the inference efficiency by measuring three key metrics :
storage consumption, indexing time and, retrieval latency, as shown in Table ~\ref{tab:efficiency_stats}. Experiments are conducted with batch size of 4 for image and 256 for text. DPR-styled retrievers which generate single vector embeddings, demonstrates higher efficiency and lower computation across all metrics, compared to ColBERT-styled retrievers that produce token-level embeddings. Although DPR-styled retrievers slightly underperform in retrieval accuracy, their smaller embeddings size provide a significant advantage in the inference stage when storage space and inference time are concerned.

Another key finding is that textual inputs are significantly more efficient than the visual inputs across all metrics. Meanwhile, hybrid retrieval system, which processes text in the image through LLM rather than visual encoders, further reduces memory and time consumption. Hence, future works on training hybrid retrieval system are encouraged as it offers a strong balance between computational efficiency and retrieval performance.

\section{Related Work}
\label{sec:related}

\textbf{DocCVQA}~\cite{tito2021doccvqa} proposes extracting information from a document image collection. However, it is limited by its small question set (20 questions).
While \textbf{PDF-MVQA}~\cite{ding2024mmvqa} is tailored for multimodal retrieval in biomedical articles, it is annotated by GPT-3.5-turbo rather than experts.
\textbf{SciMMIR}~\cite{wu-etal-2024-scimmir} also investigates multimodal retrieval but only provides image-caption pairs, lacking user queries paired with the corresponding document pages.
\citet{ma2024dse} introduces two relevant datasets, namely Wiki-SS and DocMatix-IR.
\textbf{Wiki-SS} is derived from natural questions~\cite{kwiatkowski-etal-2019-natural} , wherein evidence passages are screenshots of Wikipedia pages. However, natural questions are primarily designed for text retrieval, and the provided screenshots may not consistently capture the ground-truth evidence, as only the front page is considered.
\textbf{DocMatix-IR} is constructed from the large-scale DocMatix~\cite{laurençon2024building} dataset using filtering and hard negative mining. However, the questions are generated by \texttt{Phi-3-small}~\cite{abdin2024phi3} rather than human experts, and are not de-contextualized for retrieval task.
\textbf{MMDocRAG}~\cite{dong2025mmdocrag} is constructed upon \dname to support multimodal generation.
\textbf{ViDoRe}~\cite{faysse2024colpali} is the most relevant benchmark to \dname.
It integrates multiple DocVQA datasets and provides new industrial documents.
Upon a thorough examination of the 2,400 questions, we find that over 80\% questions exhibit notable limitations in terms of their complexity, contextual clarity, and the absence of complete document corpora. Refer to Appendix \ref{appendix:vidore} for detailed quantitative and qualitative analysis of ViDoRe.

\section{Conclusion}
\label{sec:conclusion}
In conclusion, multimodal document retrieval presents a complex challenge that requires both understanding and integrating diverse data modalities beyond plain text.
To more effectively evaluate these capabilities, we introduce the \dname benchmark, which features the innovative dual‐task retrieval capabilities targeting page-level and layout-level document granularity.
The \dname includes a rich dataset featuring expertly annotated labels for 1,685 questions and bootstrapped labels for 73,843 questions, serving as a valuable resource for both training and evaluation of multimodal document retrieval.
Our comprehensive empirical studies show that visual-driven retrievers significantly outperform text-driven ones, underscoring the importance of visual information in improving retrieval performance. Future work can expand upon these findings by optimizing retrieval algorithms to enhance both accuracy and efficiency of multimodal document retrieval systems, as well as the multilingual capability \cite{liang-etal-2020-monolingual}.

\section*{Limitations}
The limitations of \dname are summarized as follows:
\begin{itemize}[leftmargin=*, itemsep=0.4em, topsep=0.1em]
    \item \textbf{Incomplete layout label annotations for training set}: For 3 out of 7 training subsets, our semi-automated pipelines could not extract layout labels. These pipelines are optimized for datasets with single text or image layouts and cannot handle complex or cross-modal layouts. Future work should explore leveraging advanced vision-language models (VLMs) to facilitate annotation of layout labels for these subsets.

    \item \textbf{Lack of joint text and visual training}: As demonstrated in Section~\ref{ssec:hybrid_analysis}, all visual retrievers are suboptimal at modeling text passages, compared to modeling text as image screenshots. Our current visual retrievers do not explicitly utilize text query-document pairs to address this limitation. Future research should consider integrating both text and visual passages for joint training or finetuning to improve performance on both retrieval tasks.
\end{itemize}

\bibliography{custom}


\appendix
\section*{Appendix Table of Contents}

Due to page limits, we provide meaningful descriptions, results, implementation details, case studies, and other supplementary materials in Appendix. Our 21-page appendix is organized as follows:
\begin{itemize}[leftmargin=*, itemsep=0.0em]
    \item \textbf{Appendix~\ref{appendix:exp_results}}: Supplementary experimental results for page and layout retrieval.
    
    \item \textbf{Appendix~\ref{appendix:dataset}}: process and implementation for \dname curation.
    \begin{itemize}[leftmargin=1em, itemsep=0.0em]
        \item Related DocVQA benchmarks~(\S\ref{appendix:docvqa_datasets})
        \item Multimodal corpora selection~(\S\ref{appendix:eval_doc_corpora})
        \item Question filtering for IR context~(\S\ref{appendix:dataset_filter})
        \item Collection of training multimodal documents~(\S\ref{appendix:train_doc})
        \item Curation of ground-truth training labels~(\S\ref{appendix:labels})
        \item Selection of hard negatives~(\S\ref{appendix:hard_neg})
        \item Modality distribution~(\S\ref{appendix:dataset_modality_distribution})
        \item Resource and Artifact URLs~(\S\ref{appendix:artifacts_url})
    \end{itemize}
    
    \item \textbf{Appendix~\ref{appendix:train}}: Training visual retrievers with \dname training set.
    \begin{itemize}[leftmargin=1em, itemsep=0.0em]
        \item Encoding text queries and images~(\S\ref{appendix:encode})
        \item Similarity calculation~(\S\ref{appendix:similarity})
        \item Training objectives~(\S\ref{appendix:loss})
        \item Implementation details~(\S\ref{appendix:train_imp})
    \end{itemize}
    
    \item \textbf{Appendix~\ref{appendix:retrievers}}: Text and visual retriever details.
    \begin{itemize}[leftmargin=1em, itemsep=0.0em]
        \item Text-centric retrieval~(\S\ref{appendix:text_retriever})
        \item Vision-driven retrieval~(\S\ref{appendix:visual_retriever})
        \item Retriever implementation~(\S\ref{appendix:retriever_impl})
    \end{itemize}
    
    \item \textbf{Appendix~\ref{appendix:data_demo}}: Examples from \dname.
    \begin{itemize}[leftmargin=1em, itemsep=0.0em]
        \item Document pages (10 domains)~(\S\ref{appendix:data_demo_pages})
        \item Document layouts~(\S\ref{appendix:data_demo_layout})
        \item Page and layout annotations~(\S\ref{appendix:data_demo_annotation})
    \end{itemize}
    
    \item \textbf{Appendix~\ref{appendix:vidore}}: Analyses on ViDoRe benchmark.
    \begin{itemize}[leftmargin=1em, itemsep=0.0em]
        \item Search query and new annotation~(\S\ref{appendix:vidore_query})
        \item Corpora analysis~(\S\ref{appendix:vidore_analysis})
    \end{itemize}

    \item \textbf{Appendix~\ref{appendix:license}}: License agreements.

    \item \textbf{Appendix~\ref{appendix:ethical}}: Ethical considerations.
    
\end{itemize}

\section{Supplementary Experimental Results}
\label{appendix:exp_results}
In this section, we provide the full results of page-level and layout-level retrievals, which supplement our main results discussion in  Section~\ref{ssec:main_page} and Section~\ref{ssec:main_layout} respectively.

Specifically, Table~\ref{tab:appendix_page_recall} extends the main page-level results shown in Table~\ref{tab:main_page_recall} with the results of text retrievers using OCR-text.
Table~\ref{tab:appendix_layout_recall} extends the main layout-level results shown in Table~\ref{tab:main_layout_recall} with the results of (i) text retrievers using OCR-text and (ii) visual retrievers using hybrid inputs.

\begin{table*}[t] 
\small
\setlength{\tabcolsep}{2.5pt}
    \centering
    \resizebox{\linewidth}{!}{%
  \begin{tabu}{lll|cccccccccc|cc}

    \toprule

    \multicolumn{3}{c}{\multirow{2}{*}{\diagbox{Method}{Domain}}} 
     & Resear. & Admin & Tutori.\& & Acade. & \xspace Broch- & Finance & Guide- & Govern- & \multirow{2}{*}{\xspace Laws \xspace} & \multirow{2}{*}{\xspace News\xspace}  & \multicolumn{2}{c}{Average} \\
     & & & Report & \&Indu. & Worksh. & Paper & ure & Report & book & ment & & & \xspace Macro & Micro \xspace \\

    \midrule
   
    \parbox[t]{2.5mm}{\multirow{17}{*}{\rotatebox[origin=c]{90}{Recall@$k=1$}}} &
    \parbox[t]{2.0mm}{\multirow{6}{*}{\rotatebox[origin=c]{90}{OCR-text}}}
     & DPR          & 21.2 & 22.1 & 27.7 & 23.3 & 24.4 & 16.7 & 21.1 & 20.7 & 31.0 & 15.1 & 22.3 & 21.7 \\
    && ColBERT      & 43.8 & 39.8 & 42.4 & 39.3 & 39.2 & 38.7 & 46.3 & 50.6 & 46.1 & 17.1 & 40.3 & 40.0 \\
    && BGE          & 45.5 & 29.0 & 41.5 & 33.6 & 40.8 & 32.7 & 40.0 & 42.8 & 36.4 & 15.1 & 35.7 & 35.2 \\
    && E5           & 44.2 & 30.8 & 39.9 & 33.2 & 33.0 & 32.3 & 40.4 & 41.7 & 38.9 & 15.8 & 35.0 & 34.7 \\
    && Contriever   & 39.1 & 33.3 & 44.0 & 34.2 & 43.9 & 26.4 & 40.6 & 39.4 & 37.0 & 15.1 & 35.3 & 33.6 \\
    && GTE          & 44.6 & 32.6 & 45.0 & 33.2 & 37.2 & 31.8 & 40.0 & 39.9 & 35.2 & 14.5 & 35.4 & 34.6 \\
    \tabucline[.2pt on 1pt off 1.5pt]{2-15} \\[-7.5pt]  
    
    & \parbox[t]{2.0mm}{\multirow{6}{*}{\rotatebox[origin=c]{90}{VLM-text}}}
     & DPR          & 32.3 & 25.5 & 27.0 & 31.0 & 28.4 & 18.8 & 23.5 & 31.2 & 38.3 & 16.1 & 27.2 & 26.9 \\
    && ColBERT      & 48.6 & 42.8 & 51.1 & 46.2 & 36.0 & 36.8 & 49.6 & \underline{60.9} & 59.5 & 26.3 & 45.8 & 44.9 \\
    && BGE          & 48.8 & 30.9 & 47.1 & 40.8 & 37.6 & 28.4 & 43.4 & 51.9 & 48.9 & 28.5 & 40.6 & 39.6 \\
    && E5           & 48.1 & 30.0 & 50.4 & 39.4 & 41.1 & 29.7 & 40.9 & 52.8 & 51.1 & 24.1 & 40.8 & 39.5 \\
    && Contriever   & 45.5 & 31.2 & 49.8 & 41.5 & 39.4 & 29.4 & 45.2 & 55.3 & 51.1 & 20.4 & 40.9 & 39.7 \\
    && GTE          & 46.5 & 26.3 & 48.7 & 38.9 & 35.9 & 27.0 & 46.2 & 50.1 & 45.8 & 24.1 & 38.9 & 37.9 \\
    \tabucline[.2pt on 1pt off 1.5pt]{2-15} \\[-7.5pt]  
    
    & \parbox[t]{1.5mm}{\multirow{5}{*}{\rotatebox[origin=c]{90}{Image}}}
    & DSE$_{\mathrm{wiki-ss}}$  & 53.0 & 50.0 & 54.0 & 48.7 & 45.1 & 43.0 & 51.5 & 46.9 & 54.2 & 33.6 & 48.0 & 47.5 \\
    && DSE$_{\mathrm{docmatix}}$  & 52.3 & 40.4 & 56.1 & 51.7 & 45.8 & 43.5 & 53.8 & 53.7 & 58.3 & \underline{46.7} & 50.2 & 50.1 \\
    && ColPali  & 56.0 & \textbf{51.8} & \textbf{58.6} & 55.9 & 52.0 & \underline{47.2} & 57.9 & 53.9 & \underline{64.0} & 32.8 & 53.0 & 52.7 \\
    && DPR-Phi3$_{\mathrm{ours}}$  & \textbf{58.9} & \underline{50.4} & \underline{57.4} & \underline{59.0} & \textbf{57.3} & 44.6 & \textbf{63.8} & 50.5 & \textbf{64.4} & 35.0 & \underline{54.1} & \underline{53.7} \\
    && Col-Phi3$_{\mathrm{ours}}$ & \underline{56.7} & \underline{50.4} & 56.9 & \textbf{61.3} &\underline{54.8} & \textbf{50.7} & \underline{60.8} & \textbf{61.3} & 63.6 & \textbf{54.0} & \textbf{57.0} & \textbf{57.1} \\
    \midrule

    \parbox[t]{1.2mm}{\multirow{17}{*}{\rotatebox[origin=c]{90}{Recall@$k=3$}}} &
    \parbox[t]{1.5mm}{\multirow{6}{*}{\rotatebox[origin=c]{90}{OCR-text}}}
     & DPR          & 46.1 & 40.6 & 38.9 & 46.7 & 43.9 & 32.4 & 38.4 & 37.0 & 50.0 & 20.4 & 39.4 & 39.8 \\
    && ColBERT      & 72.6 & 59.7 & 57.8 & 66.7 & 60.0 & 53.7 & 63.8 & 68.5 & 61.4 & 23.7 & 58.8 & 59.5 \\
    && BGE          & 69.8 & 57.7 & 56.3 & 58.6 & 60.7 & 48.5 & 57.9 & 60.9 & 62.7 & 20.4 & 55.4 & 55.0 \\
    && E5           & 66.6 & 48.7 & 59.0 & 58.0 & 60.9 & 48.8 & 63.7 & 61.4 & 60.8 & 20.4 & 54.8 & 54.6 \\
    && Contriever   & 70.2 & 55.8 & 60.4 & 56.6 & 62.1 & 43.0 & 60.0 & 56.8 & 61.4 & 22.4 & 54.9 & 53.6 \\
    && GTE          & 69.2 & 47.0 & 58.7 & 59.5 & 61.8 & 46.6 & 65.5 & 59.1 & 61.4 & 19.7 & 54.9 & 54.7 \\
    \tabucline[.2pt on 1pt off 1.5pt]{2-15} \\[-7.5pt]  
    
    & \parbox[t]{2.0mm}{\multirow{6}{*}{\rotatebox[origin=c]{90}{VLM-text}}}
     & DPR          & 52.2 & 44.2 & 43.5 & 54.6 & 52.0 & 35.1 & 44.4 & 53.9 & 57.2 & 25.5 & 46.3 & 46.2 \\
    && ColBERT      & 70.1 & 64.4 & 70.3 & 72.3 & 59.1 & 55.3 & 71.1 & \textbf{81.3} & 70.8 & 34.3 & 64.9 & 64.8 \\
    && BGE          & 71.5 & 48.2 & 68.8 & 65.7 & 56.2 & 46.5 & 66.1 & 69.9 & 72.0 & 32.1 & 59.7 & 59.6 \\
    && E5           & 68.4 & 45.7 & 68.1 & 63.7 & 60.1 & 44.0 & 69.3 & 72.3 & 78.8 & 32.8 & 60.3 & 59.3 \\
    && Contriever   & 69.4 & 55.3 & 68.3 & 64.9 & 56.9 & 46.2 & 69.9 & 71.1 & 72.0 & 32.1 & 60.6 & 59.7 \\
    && GTE          & 71.1 & 44.5 & 67.2 & 64.4 & 54.3 & 43.0 & 70.6 & 71.9 & 68.2 & 31.4 & 58.7 & 58.3 \\
    \tabucline[.2pt on 1pt off 1.5pt]{2-15} \\[-7.5pt] 
    
    & \parbox[t]{1.5mm}{\multirow{5}{*}{\rotatebox[origin=c]{90}{Image}}}
    & DSE$_{\mathrm{wiki-ss}}$  & 75.4 & 65.0 & 73.9 & 79.8 & 69.5 & 63.5 & 75.4 & 71.5 & 81.4 & 50.4 & 70.6 & 71.4 \\
    && DSE$_{\mathrm{docmatix}}$  & 75.4 & 67.5 & 73.3 & 80.0 & 66.3 & 61.6 & 72.8 & 76.4 & 82.6 & \underline{57.7} & 71.4 & 71.8 \\
    && ColPali  & 77.6 & \underline{71.8} & \textbf{79.4} & 83.4 & \textbf{72.6} & \underline{66.1} & \textbf{80.0} & \underline{80.4} & \textbf{86.4} & 49.6 & \underline{74.7} & \underline{75.0} \\
    && DPR-Phi3$_{\mathrm{ours}}$  & \textbf{80.3} & 66.5 & \underline{77.6} & \underline{83.9} & \underline{71.9} & 63.8 & \underline{79.8} & 71.4 & \underline{84.5} & 55.5 & 73.5 & 74.3 \\
    && Col-Phi3$_{\mathrm{ours}}$ & \underline{80.2} & \textbf{74.1} & 77.4 & \textbf{84.8} & 69.1 & \textbf{67.7} & 78.7 & 79.5 & 81.8 & \textbf{69.3} & \textbf{76.3} & \textbf{76.8} \\
    \midrule

    \parbox[t]{1.2mm}{\multirow{17}{*}{\rotatebox[origin=c]{90}{Recall@$k=5$}}} &
    \parbox[t]{1.5mm}{\multirow{6}{*}{\rotatebox[origin=c]{90}{OCR-text}}}
     & DPR          & 59.5 & 55.8 & 43.4 & 59.1 & 56.2 & 41.2 & 50.7 & 45.5 & 56.0 & 23.0 & 49.0 & 49.4 \\
    && ColBERT      & 78.4 & 71.1 & 63.3 & 75.2 & 68.8 & 60.5 & 72.0 & 72.7 & 67.5 & 30.3 & 66.0 & 66.5 \\
    && BGE          & 79.3 & 65.9 & 62.1 & 69.7 & 69.8 & 56.5 & 68.0 & 62.8 & 66.3 & 26.3 & 62.7 & 62.9 \\
    && E5           & 79.3 & 62.4 & 67.0 & 70.3 & 71.8 & 57.6 & 72.5 & 67.1 & 67.5 & 25.7 & 64.1 & 64.2 \\
    && Contriever   & 79.9 & 62.7 & 64.7 & 71.7 & 71.1 & 48.8 & 72.4 & 65.8 & 67.5 & 26.3 & 63.1 & 62.5 \\
    && GTE          & 78.3 & 61.9 & 67.3 & 72.3 & 68.7 & 55.2 & 72.6 & 64.0 & 67.5 & 24.3 & 63.2 & 63.5 \\
    \tabucline[.2pt on 1pt off 1.5pt]{2-15} \\[-7.5pt]  

    & \parbox[t]{2.0mm}{\multirow{6}{*}{\rotatebox[origin=c]{90}{VLM-text}}}
     & DPR          & 66.5 & 60.1 & 56.0 & 68.9 & 58.8 & 43.8 & 57.1 & 68.6 & 64.8 & 33.6 & 57.8 & 57.8 \\
    && ColBERT      & 78.8 & 74.0 & 78.7 & 82.3 & 66.1 & 60.8 & 77.0 & \textbf{88.5} & 78.0 & 38.7 & 72.3 & 72.3 \\
    && BGE          & 79.5 & 65.8 & 71.3 & 76.8 & 62.4 & 56.0 & 77.2 & 77.4 & 79.5 & 38.0 & 68.4 & 68.5 \\
    && E5           & 76.9 & 64.2 & 75.3 & 74.4 & 67.4 & 52.0 & 78.5 & 78.6 & 82.6 & 40.9 & 69.1 & 67.9 \\
    && Contriever   & 77.2 & 67.1 & 76.7 & 75.2 & 65.1 & 53.7 & 75.4 & 79.2 & 83.3 & 39.4 & 69.2 & 68.3 \\
    && GTE          & 77.4 & 62.6 & 74.7 & 75.8 & 62.0 & 51.8 & 77.8 & 80.0 & 75.0 & 39.4 & 67.6 & 67.2 \\
    \tabucline[.2pt on 1pt off 1.5pt]{2-15} \\[-7.5pt] 
    
    & \parbox[t]{1.5mm}{\multirow{5}{*}{\rotatebox[origin=c]{90}{Image}}}
    & DSE$_{\mathrm{wiki-ss}}$  & 84.0 & \textbf{80.2} & 78.7 & 87.0 & 75.7 & \underline{73.0} & 82.0 & 77.3 & 88.3 & 58.4 & 78.5 & 79.2 \\
    && DSE$_{\mathrm{docmatix}}$  & 82.1 & 77.2 & 79.6 & 87.8 & 73.9 & 72.4 & 81.7 & 83.1 & 89.4 & \underline{67.9} & 79.5 & 80.1 \\
    && ColPali  & 84.6 & \underline{79.3} & \underline{82.3} & 89.0 & \underline{79.8} & 72.1 & \underline{86.7} & \underline{84.9} & \textbf{92.4} & 56.9 & 80.8 & 81.0 \\
    && DPR-Phi3$_{\mathrm{ours}}$  & \textbf{86.9} & 76.2 & \textbf{85.3} & \underline{91.9} & \textbf{80.0} & 71.2 & \textbf{87.1} & 79.5 & \underline{92.0} & 61.3 & \underline{81.1} & \underline{81.8} \\
    && Col-Phi3$_{\mathrm{ours}}$ & \underline{86.3} & 78.8 & 81.2 & \textbf{92.4} & 79.0 & \textbf{73.8} & 85.3 & \textbf{85.1} & 87.1 & \textbf{73.0} & \textbf{82.2} & \textbf{83.0} \\

  \bottomrule
  \end{tabu}
  }
\vspace{-1em}
\caption{Main results for page-level retrieval. ``\textit{OCR-text}'' and ``\textit{VLM-text}'' refer to converting multi-modal content using OCR and VLM respectively. ``\textit{Image}'' refers to processing document page as screenshot image.}
\vspace{-1em}
\label{tab:appendix_page_recall}
\end{table*}

\begin{table*}[t] 
\renewcommand{\arraystretch}{0.95 }
\setlength{\tabcolsep}{2.5pt}
\small
    \centering
    \resizebox{\linewidth}{!}{%
  \begin{tabu}{llll|cccccccccc|cc}
    \toprule

    \multicolumn{4}{c}{\multirow{2}{*}{\diagbox{Method}{Domain}}} 
     & Resear. & Admin & Tutori.\& & Acade. & \xspace Broch- & Finance & Guide- & Govern- & \multirow{2}{*}{\xspace Laws \xspace} & \multirow{2}{*}{\xspace News\xspace}  & \multicolumn{2}{c}{Average} \\
     & & & & Report & \&Indu. & Worksh. & Paper & ure & Report & book & ment & & & \xspace Macro & Micro \xspace \\
     
    \midrule
   
    \parbox[t]{2.5mm}{\multirow{22}{*}{\rotatebox[origin=c]{90}{Recall@$k=1$}}} &

    \parbox[t]{2.5mm}{\multirow{12}{*}{\rotatebox[origin=c]{90}{Textual Retrieval}}} &
    
    \parbox[t]{2.0mm}{\multirow{6}{*}{\rotatebox[origin=c]{90}{OCR-text}}}
     & DPR          & 3.4 & 7.2 & 1.2 & 11.3 & 3.0 & 9.8 & 8.2 & 24.7 & 30.9 & 26.3 & 12.6 & 12.4 \\
    &&& ColBERT      & 5.0 & 8.8 & 4.7 & 16.4 & 2.0 & 13.2 & 4.6 & 50.8 & 47.7 & \textbf{45.3} & 19.8 & 19.1 \\
    &&& BGE          & 7.0 & 10.9 & 3.7 & 14.3 & 2.3 & 16.3 & 8.4 & 46.1 & 45.5 & 35.8 & 19.0 & 18.5 \\
    &&& E5           & 6.3 & 6.0 & 2.6 & 14.0 & 3.5 & 14.4 & 6.1 & 44.7 & 45.4 & 40.5 & 18.4 & 17.9 \\
    &&& Contriever   & 6.7 & 7.0 & 3.8 & 14.3 & 4.1 & 13.3 & 8.3 & 43.6 & 43.9 & 42.9 & 18.8 & 18.1 \\
    &&& GTE          & 5.7 & 7.0 & 3.9 & 17.2 & 2.8 & 15.8 & 9.4 & 40.7 & 41.6 & 38.3 & 18.2 & 18.4 \\
    \tabucline[.2pt on 1pt off 1.5pt]{3-16} \\[-7.5pt] 

    & & \parbox[t]{2.0mm}{\multirow{6}{*}{\rotatebox[origin=c]{90}{VLM-text}}}
    & DPR          & 11.6 & 9.5 & 19.2 & 19.2 & 14.9 & 15.9 & 15.8 & 25.6 & 34.7 & 27.0 & 19.3 & 19.2 \\
    &&& ColBERT      & 22.0 & 14.9 & 28.0 & 28.3 & 17.9 & 29.7 & 21.1 & \textbf{52.6} & \textbf{54.5} & \underline{44.5} & \underline{31.3} & 31.4 \\
    &&& BGE          & 19.2 & 15.2 & 24.6 & 28.7 & 12.8 & 27.6 & 19.7 & 47.0 & 52.3 & 35.8 & 28.3 & 29.0 \\
    &&& E5           & 15.9 & 8.8 & 27.7 & 24.3 & 14.6 & 21.8 & 14.7 & 45.6 & \underline{53.0} & 40.5 & 26.7 & 26.4 \\
    &&& Contriever   & 23.4 & 7.5 & 28.2 & 26.8 & 17.1 & 25.7 & 16.1 & 43.6 & 51.5 & 42.9 & 28.3 & 28.9 \\
    &&& GTE          & 17.5 & 10.5 & 23.0 & 27.2 & 14.5 & 26.3 & 14.4 & 39.8 & 49.2 & 38.3 & 26.1 & 27.1 \\
    \tabucline[.2pt on 1pt off 1.5pt]{2-16} \\[-7.5pt]  

    & \parbox[t]{2.5mm}{\multirow{10}{*}{\rotatebox[origin=c]{90}{Visual Retrieval}}}

    & \parbox[t]{1.5mm}{\multirow{5}{*}{\rotatebox [origin=c]{90}{Pure-Image}}}
    & DSE$_{\mathrm{wiki-ss}}$  & 20.6 & 15.1 & 31.0 & 31.1 & 20.1 & 29.2 & 22.0 & 39.3 & 37.5 & 35.8 & 28.2 & 29.2 \\
    &&& DSE$_{\mathrm{docmatix}}$  & 19.9 & 11.4 & 31.5 & 30.1 & 17.8 & 30.0 & 20.8 & 46.5 & 39.4 & 31.4 & 27.9 & 29.1 \\
    &&& ColPali   & 22.5 & 21.3 & 36.6 & 30.9 & 26.8 & \textbf{32.1} & 19.3 & \underline{52.5} & 51.8 & 33.6 & \textbf{32.7} & \textbf{32.5} \\
    &&& DPR-Phi3$_{\mathrm{ours}}$  & 21.1 & \textbf{22.1} & \underline{36.8} & \textbf{35.2} & 25.6 & 28.7 & 24.1 & 38.3 & 35.4 & 27.4 & 29.5 & 30.2  \\
    &&& Col-Phi3$_{\mathrm{ours}}$ & 22.6 & \underline{22.0} & \textbf{37.5} & \underline{34.9} & 28.9 & \underline{30.3} & 22.7 & 50.2 & 45.1 & 26.3 & 31.1 & \underline{31.6}\\
    \tabucline[.2pt on 1pt off 1.5pt]{3-16} \\[-7.5pt]  

    && \parbox[t]{1.5mm}{\multirow{5}{*}{\rotatebox[origin=c]{90}{Hybrid}}}
    & DSE$_{\mathrm{wiki-ss}}$  & 14.0 & 10.4 & 29.8 & 18.0 & 13.7 & 20.4 & 13.5 & 46.0 & 45.1 & 34.7 & 24.6 & 23.4 \\
    &&& DSE$_{\mathrm{docmatix}}$  & 18.2 & 11.6 & 32.7 & 24.0 & 17.7 & 27.2 & 16.7 & 48.1 & 45.5 & 33.0 & 27.5 & 27.4 \\
    &&& ColPali   &  17.7 & 12.3 & 30.0 & 18.4 & 19.0 & 25.5 & 20.6 & 49.7 & 51.2 & 40.9 & 28.5 & 27.1 \\
    &&& DPR-Phi3$_{\mathrm{ours}}$  & \textbf{28.3} & 11.1 & 35.5 & 18.8 & \underline{29.3} & 24.0 & \textbf{27.4} & 38.0 & 41.9 & 34.5 & 28.9 & 27.3\\
    &&& Col-Phi3$_{\mathrm{ours}}$ & \underline{26.4} & 12.6 & 33.7 & 27.3 & \textbf{30.1} & 27.9 & \underline{24.6} & 46.2 & 47.4 & 21.9 & 29.8 & 29.6 \\
    \midrule

    \parbox[t]{2.5mm}{\multirow{22}{*}{\rotatebox[origin=c]{90}{Recall@$k=5$}}} &

    \parbox[t]{2.5mm}{\multirow{12}{*}{\rotatebox[origin=c]{90}{Textual Retrieval}}} &
    
    \parbox[t]{2.0mm}{\multirow{6}{*}{\rotatebox[origin=c]{90}{OCR-text}}}
     & DPR          & 7.3 & 12.5 & 5.6 & 24.0 & 8.9 & 16.9 & 13.6 & 47.2 & 50.2 & 51.1 & 23.7 & 23.5 \\
    &&& ColBERT      & 10.9 & 23.8 & 10.2 & 32.2 & 6.8 & 25.5 & 17.0 & \underline{78.7} & 63.4 & 63.2 & 33.2 & 32.6 \\
    &&& BGE          & 11.9 & 20.3 & 13.6 & 30.0 & 11.7 & 27.7 & 18.8 & 68.5 & 65.4 & 59.1 & 32.7 & 32.2 \\
    &&& E5           & 12.8 & 16.2 & 8.9 & 31.9 & 10.9 & 23.6 & 19.9 & 76.1 & 68.8 & \underline{63.9} & 33.3 & 32.7 \\
    &&& Contriever   & 11.9 & 17.9 & 11.9 & 28.8 & 9.3 & 24.6 & 18.1 & 64.3 & 68.0 & 62.7 & 31.7 & 31.2 \\
    &&& GTE          & 10.0 & 18.2 & 12.8 & 32.9 & 15.2 & 29.4 & 21.0 & 67.7 & 65.3 & 62.4 & 33.5 & 33.5 \\
    \tabucline[.2pt on 1pt off 1.5pt]{3-16} \\[-7.5pt] 

    & & \parbox[t]{2.0mm}{\multirow{6}{*}{\rotatebox[origin=c]{90}{VLM-text}}}
    & DPR          & 31.0 & 25.7 & 36.7 & 44.9 & 33.0 & 34.1 & 34.9 & 49.9 & 56.3 & 51.1 & 39.8 & 40.4 \\
    &&& ColBERT      & 41.8 & 37.7 & 53.7 & \underline{61.8} & 35.1 & 52.4 & 46.1 & 83.2 & 70.1 & 62.5 & \textbf{54.4} & \textbf{56.0} \\
    &&& BGE          & 41.0 & 28.1 & 52.7 & 59.2 & 36.7 & 46.0 & \textbf{50.7} & 72.0 & 71.5 & 59.9 & 51.8 & 53.2 \\
    &&& E5           & 35.4 & 28.1 & 51.7 & 58.5 & 33.2 & 41.2 & 40.2 & \textbf{79.7} & \textbf{77.9} & \textbf{64.6} & 51.1 & 51.8 \\
    &&& Contriever   & 40.2 & 29.2 & 54.1 & 57.9 & 36.8 & 47.1 & 44.6 & 68.8 & 76.2 & 61.9 & 51.7 & 53.0 \\
    &&& GTE          & 36.7 & 25.6 & 51.2 & 56.6 & 39.7 & 46.6 & 46.4 & 72.2 & 74.3 & 63.2 & 51.3 & 52.3 \\
    \tabucline[.2pt on 1pt off 1.5pt]{2-16} \\[-7.5pt]  

    & \parbox[t]{2.5mm}{\multirow{10}{*}{\rotatebox[origin=c]{90}{Visual Retrieval}}}

    & \parbox[t]{1.5mm}{\multirow{5}{*}{\rotatebox [origin=c]{90}{Pure-Image}}}
    & DSE$_{\mathrm{wiki-ss}}$  & 42.4 & 32.9 & \underline{56.3} & 58.5 & 39.8 & 50.6 & 41.6 & 68.6 & 60.9 & 50.0 & 50.2 & 52.1 \\
    &&& DSE$_{\mathrm{docmatix}}$  & 39.6 & 36.2 & 53.9 & 57.5 & 33.7 & \underline{52.5} & 42.8 & 69.4 & 63.1 & 48.9 & 49.8 & 51.9 \\
    &&& ColPali   & 40.7 & \textbf{45.9} & 54.9 & 58.5 & 42.6 & 51.2 & 45.7 & 76.8 & \underline{74.5} & 48.9 & \underline{54.0} & 54.3 \\
    &&& DPR-Phi3$_{\mathrm{ours}}$  & 45.5 & 37.7 & \textbf{57.0} & \textbf{62.9} & 41.4 & 51.1 & 45.5 & 65.1 & 60.8 & 49.3 & 51.6 & 53.9 \\
    &&& Col-Phi3$_{\mathrm{ours}}$ & 46.4 & 38.2 & 53.1 & \underline{61.8} & \textbf{45.0} & \textbf{54.6} & 45.7 & 68.8 & 65.7 & 43.8 & 52.3 & \underline{54.5} \\
    \tabucline[.2pt on 1pt off 1.5pt]{3-16} \\[-7.5pt]  

    && \parbox[t]{1.5mm}{\multirow{5}{*}{\rotatebox[origin=c]{90}{Hybrid}}}
    & DSE$_{\mathrm{wiki-ss}}$  & 31.8 & 29.5 & 51.1 & 43.0 & 34.5 & 39.3 & 38.3 & 71.3 & 71.3 & 57.1 & 46.7 & 45.6 \\
    &&& DSE$_{\mathrm{docmatix}}$  & 37.3 & 26.7 & 48.2 & 49.7 & 34.5 & 48.6 & 41.0 & 72.2 & 69.4 & 54.2 & 48.2 & 49.3 \\
    &&& ColPali   & 40.1 & \underline{38.3} & 55.2 & 49.2 & \underline{42.7} & 47.6 & 40.8 & 78.6 & 68.7 & 61.0 & 52.2 & 51.4 \\
    &&& DPR-Phi3$_{\mathrm{ours}}$  & \textbf{54.2} & 27.4 & 53.8 & 39.9 & 36.6 & 45.4 & \underline{49.6} & 67.2 & 66.8 & 59.8 & 50.1 & 49.3 \\
    &&& Col-Phi3$_{\mathrm{ours}}$ & \underline{50.9} & 25.5 & 49.1 & 58.3 & 41.9 & 48.1 & 49.2 & 62.3 & 60.6 & 48.5 & 50.0 & 51.8 \\
    \midrule

    \parbox[t]{2.5mm}{\multirow{22}{*}{\rotatebox[origin=c]{90}{Recall@$k=10$}}} &

    \parbox[t]{2.5mm}{\multirow{12}{*}{\rotatebox[origin=c]{90}{Textual Retrieval}}} &
    
    \parbox[t]{2.0mm}{\multirow{6}{*}{\rotatebox[origin=c]{90}{OCR-text}}}
     & DPR          & 10.5 & 21.1 & 8.8 & 32.0 & 14.9 & 19.6 & 17.0 & 59.7 & 56.4 & 58.9 & 29.9 & 29.3 \\
    &&& ColBERT      & 14.1 & 31.1 & 13.1 & 38.4 & 9.1 & 31.0 & 22.4 & 83.9 & 67.9 & 65.4 & 37.6 & 37.4 \\
    &&& BGE          & 15.7 & 24.3 & 15.9 & 35.9 & 17.9 & 31.6 & 25.3 & 76.3 & 73.4 & 62.1 & 37.8 & 37.3 \\
    &&& E5           & 16.9 & 24.5 & 13.8 & 40.1 & 15.8 & 26.7 & 24.7 & 83.5 & 77.9 & 66.1 & 39.0 & 38.3 \\
    &&& Contriever   & 15.1 & 25.7 & 14.2 & 36.8 & 15.9 & 27.9 & 24.2 & 76.8 & 71.8 & 64.9 & 37.3 & 36.6 \\
    &&& GTE          & 19.1 & 29.0 & 21.4 & 39.0 & 19.2 & 32.9 & 29.2 & 78.8 & 74.5 & 66.2 & 40.9 & 40.1 \\
    \tabucline[.2pt on 1pt off 1.5pt]{3-16} \\[-7.5pt] 

    & & \parbox[t]{2.0mm}{\multirow{6}{*}{\rotatebox[origin=c]{90}{VLM-text}}}
     & DPR          & 42.2 & 33.1 & 52.1 & 56.2 & 39.9 & 43.5 & 44.0 & 62.8 & 61.7 & 59.7 & 49.5 & 50.5 \\
    &&& ColBERT      & 51.0 & 48.7 & 60.6 & 69.8 & 43.9 & \textbf{61.6} & 53.7 & \textbf{88.4} & 74.8 & \underline{66.4} & \underline{61.9} & \textbf{63.7} \\
    &&& BGE          & 51.1 & 38.7 & 62.1 & 71.5 & 41.9 & 55.6 & \textbf{58.7} & 80.8 & 78.7 & 63.5 & 60.3 & 62.4 \\
    &&& E5           & 45.3 & 38.6 & 62.0 & 70.5 & 45.6 & 50.0 & 55.3 & \underline{87.1} & \underline{82.4} & \textbf{66.8} & 60.4 & 61.2 \\
    &&& Contriever   & 49.9 & 41.3 & 62.0 & 70.5 & 44.8 & 56.5 & 54.5 & 81.3 & 78.0 & 64.9 & 60.4 & 62.2 \\
    &&& GTE          & 48.6 & 41.1 & 61.5 & 68.8 & 44.3 & 56.9 & 58.0 & 83.0 & 77.5 & 66.9 & 60.7 & 62.2 \\
    \tabucline[.2pt on 1pt off 1.5pt]{2-16} \\[-7.5pt]  

    & \parbox[t]{2.5mm}{\multirow{10}{*}{\rotatebox[origin=c]{90}{Visual Retrieval}}}

    & \parbox[t]{1.5mm}{\multirow{5}{*}{\rotatebox [origin=c]{90}{Pure-Image}}}
    & DSE$_{\mathrm{wiki-ss}}$  & 55.9 & 41.3 & 61.5 & 68.1 & 47.8 & \underline{60.7} & 54.2 & 72.9 & 68.3 & 54.4 & 58.5 & 61.1 \\
    &&& DSE$_{\mathrm{docmatix}}$  & 53.7 & 43.3 & 59.6 & 66.5 & 44.7 & 59.1 & 50.3 & 75.4 & 69.2 & 53.7 & 57.5 & 59.9  \\
    &&& ColPali   & 53.6 & \textbf{54.1} & 64.4 & 69.5 & 48.8 & \underline{60.7} & 54.0 & 81.9 & \textbf{82.5} & 50.4 & \textbf{62.0} & 63.2 \\
    &&& DPR-Phi3$_{\mathrm{ours}}$  &  58.1 & 49.1 & \textbf{67.0} & \underline{74.7} & 48.4 & 57.9 & \underline{57.8} & 68.7 & 66.2 & 54.4 & 60.2 & 62.8 \\
    &&& Col-Phi3$_{\mathrm{ours}}$ & 57.7 & \underline{50.5} & \underline{66.6} & 72.3 & \textbf{50.7} & 59.3 & 53.6 & 68.5 & 74.8 & 57.5 & 61.1 & \underline{63.3} \\
    \tabucline[.2pt on 1pt off 1.5pt]{3-16} \\[-7.5pt]  

    && \parbox[t]{1.5mm}{\multirow{5}{*}{\rotatebox[origin=c]{90}{Hybrid}}}
    & DSE$_{\mathrm{wiki-ss}}$  & 44.1 & 34.3 & 57.6 & 56.3 & 42.6 & 50.7 & 48.6 & 81.1 & 79.1 & 63.7 & 55.8 & 56.0 \\
    &&& DSE$_{\mathrm{docmatix}}$  & 49.9 & 37.3 & 57.3 & 61.3 & 45.9 & 57.9 & 50.1 & 77.9 & 74.9 & 61.5 & 57.4 & 58.9 \\
    &&& ColPali   & 52.1 & 46.4 & 65.0 & 64.4 & \textbf{50.7} & 53.8 & 51.0 & 82.7 & 71.4 & 62.5 & 60.0 & 60.2 \\
    &&& DPR-Phi3$_{\mathrm{ours}}$  & \textbf{65.2} & 33.7 & 60.3 & 51.3 & 42.4 & 52.4 & 52.9 & 79.1 & 72.5 & 65.6 & 55.5 & 53.4 \\
    &&& Col-Phi3$_{\mathrm{ours}}$ & \underline{59.1} & 38.7 & 57.0 & \textbf{77.7} & 43.9 & 57.7 & 51.7 & 72.4 & 68.0 & 60.7 & 58.7 & 62.8 \\

  \bottomrule
  \end{tabu}
  }
  \vspace{-1em}
\caption{Main results for layout-level retrieval. ``\textit{Pure-Image}'' and ``\textit{Hybrid}'' refer to reading textual layouts in image and text format respectively.}
\vspace{-1em}
\label{tab:appendix_layout_recall}
\end{table*}

\clearpage

\section{Dataset Construction} \label{appendix:dataset}

\subsection{Related DocVQA Benchmarks}
\label{appendix:docvqa_datasets}
Early DocVQA benchmarks primarily address single-page visual question answering (VQA), exemplified by datasets such as DocVQA~\cite{2020docvqa}, InfoVQA~\cite{mathew22infographicsvqa}, and TAT-DQA~\cite{zhu2022tatdqa}. To overcome the limitations of single-page inputs, subsequent datasets like DUDE~\cite{landeghem2023dude}, MP-DocVQA~\cite{tito2023mp-docvqa}, and SlideVQA~\cite{tanaka2023slidevqa} have extended the context length to an average of 5.7, 8.3, and 20 pages, respectively. More recent benchmarks, including MMLongBench-Doc~\cite{ma2024mmlongbenchdoc} and DocBench~\cite{zou2024docbench}, treat DocVQA as a long-context task, accommodating entire documents that average between 50 to 70 pages in length. As document lengths increase, retrieval becomes essential. Relevant pages must first be identified, followed by answer generation based on the retrieved multimodal evidence.

\subsection{Document Corpora Collection Criteria} \label{appendix:eval_doc_corpora}

Early document related benchmarks \cite{dong-etal-2021-docoie} are mostly textual only, which are not considered in \dname.
To facilitate the development of \dname, we leverage visually-rich documents from recent DocVQA benchmarks described in Appendix \ref{appendix:docvqa_datasets}.
Despite not being curated for IR, they offer valuable document corpora and questions that can be adapted for IR tasks.
We select relevant DocVQA datasets based on the following criteria:
\begin{itemize}[leftmargin=*, itemsep=-0.5em, topsep=0.0em]
    \item \textbf{Document Source}: The dataset must include accessible original documents or sources for these documents. We need to access and enrich them to support more complex retrieval tasks.
    \item \textbf{Diverse Domain/Modality}: The document collections must (1) encompass diverse domains suitable for generalized evaluation, 
    and (2) contain multiple modalities, such as text, figures, tables, charts, and layouts. 
    \item \textbf{Long Document}: We choose documents with extensive texts as longer texts pose more significant challenges. This criterion can evaluate models in handling complex and lengthy documents.
    \item \textbf{Question Diversity and Comprehensiveness:} The questions included in the dataset should be diverse and challenging. For example, cross-modal questions require reasoning across both text and visual tables/figures; multi-hop questions require reasoning over multiple steps; multi-page questions require combining information from multiple pages.
\end{itemize}

Considering these criteria, we utilize document corpora and questions from datasets as follows:
\begin{itemize}[leftmargin=*, itemsep=0.0em, topsep=0.0em]

    \item \textbf{Evaluation}: MMLongBench-Doc \cite{ma2024mmlongbenchdoc} and DocBench \cite{zou2024docbench}. 
    
    \item \textbf{Training}: MP-DocVQA \cite{tito2023mp-docvqa}, SlideVQA \cite{tanaka2023slidevqa}, TAT-DQA \cite{zhu2022tatdqa}, SciQAG \cite{wan2024sciqag}, DUDE \cite{landeghem2023dude}, and CUAD \cite{2021HendrycksCUAD}.
\end{itemize}

\subsection{Question Filtering Guidelines} \label{appendix:dataset_filter}
We filter questions based on the following criteria:
\begin{itemize}[leftmargin=*, itemsep=0.0em, topsep=0.0em]
\item Summarization Questions: Questions such as ``\textit{What does this book mainly illustrate?}''``\textit{What does this story mainly tell?}'' require understanding of large sections or even the entire document. The broad scope makes it hard to pinpoint specific content and contradicts the IR nature of our task.

\item Overwhelm Statistical Questions: Questions that demand extensive data computation or collation, such as ``\textit{How many words are there in total in the paper?}''``\textit{How many pictures are there in total in the document?}'' are also excluded from our scope. 

\item Online Search Questions: Questions like ``\textit{What is the Google Scholar citation count of the author?}'' rely on information from external online resources. 
We focus only on retrieving information within the documents, and therefore exclude these questions.

\item Unanswerable Questions: These are designed to test if models generate answers based on non-existent information (model hallucinations). Since they do not facilitate the retrieval of factual document-based information, these questions are excluded.
\end{itemize}

\subsection{Training Document Collection} \label{appendix:train_doc}

We collect the training datasets as follows:
\begin{itemize}[leftmargin=*, itemsep=0.0em, topsep=0.0em]

\item \textbf{MP-DocVQA}~\cite{tito2023mp-docvqa} contains 47,952 images collected from Industry Documents Library (IDL) \footnote{\url{https://www.industrydocuments.ucsf.edu/}}.
IDL is a crucial resource for public health research, containing millions of documents produced by industries such as tobacco, drug, chemical, and food, which have had significant impacts on public health. 
We group the 47,952 document images into separate document files, and obtain 875 long documents (46.8 pages on average) with 15,266 QA pairs.

\item \textbf{SlideVQA}~\cite{tanaka2023slidevqa} contains 2,619 slide documents collected from slideshare \footnote{\url{https://www.slideshare.net/}} and covering 39 topics. 
SlideVQA hosts a wide variety of slide presentations across various categories such as business, mobile, social media, marketing, technology, arts, career, design, education, and government \& nonprofit, among others, which can enrich the diversity of our corpus.
Note that SlideVQA contains only the first 20 pages for each slide deck. In our research, we manually collect the remaining missing pages, and obtain 2,011 long documents (averaging 49.3 pages) with 11,066 QA pairs.
SlideVQA requires complex reasoning, including single-hop, multi-hop, and numerical reasoning, and also provides annotated arithmetic expressions of numerical answers for enhancing the ability of numerical reasoning.

\item \textbf{TAT-DQA}~\cite{zhu2022tatdqa} consists of 3,067 document pages from financial reports~\footnote{https://www.annualreports.com/}, dated between 2018 and 2020. 
AnnualReports.com provides access to a comprehensive collection of corporate annual reports from over 10,320 companies worldwide. 
Note that neither original documents nor links are provided. We use OCR to extract text in the pages, and use search engine to find relevant documents. After careful tracing and recognition, we identify 163 original documents (averaging 147.3 pages) with 15,814 QA pairs.

\item \textbf{arXivQA}~\cite{li2024arxiv} comprises 32k figures cropped from academic pages~\footnote{https://arxiv.org/}. 
The papers on arXiv cover a wide range of disciplines including physics, mathematics, computer science, quantitative biology, quantitative finance, statistics, engineering, and systems science, and economics, etc. 
We use the arXiv DOIs provided to collect the academic papers.
Due to the missing of paper versions, extra efforts are made to identify paper versions.
After careful tracing, recognition, and document length filtering, we identify 1,579 documents averaging 18.4 pages.

\item \textbf{SciQAG}~\cite{wan2024sciqag} consists of 22,728 papers and 188,042 QA pairs 
in 24 scientific disciplines, collected from Web of Science (WoS) Core Collection database.
WoS provides comprehensive scientific literature in natural sciences, social sciences, arts, and humanities. 
We sample 50 documents from each discipline, and manually collect 1,197 papers using the DOIs provided.

\item \textbf{DUDE}~\cite{landeghem2023dude} provides 5,019 documents from aggregator websites\footnote{1: \url{archive.org}, 2: \url{http://commons.wikimedia.org/}, 3: \url{http://documentcloud.org/}}. 
It covers a broad range of domains, including medical, legal, technical, and financial, among others, to evaluate models’ ability to handle diverse topics and the specific knowledge each requires. 
We filter out short documents and obtain 779 relatively long documents (averaging 15.6 pages) with 3,173 QA pairs.

\item \textbf{CUAD}~\cite{2021HendrycksCUAD} provides 510 commercial legal contracts, collected from Electronic Data Gathering, Analysis, and Retrieval (EDGAR)\footnote{https://www.sec.gov/search-filings}.
EDGAR contracts are usually more complex and heavily negotiated than the general population of all legal contracts. 
We filter out short documents in CUAD and obtain 274 long documents (29.6 pages on average) with 11,234 QA pairs. 

\end{itemize}

\begin{figure*}
    \centering
    \begin{subfigure}[b]{1\linewidth}
         \caption{313 documents in \dname evaluation set in Section~\ref{sec:eval_dataset}}.
         \includegraphics[width=1\linewidth]{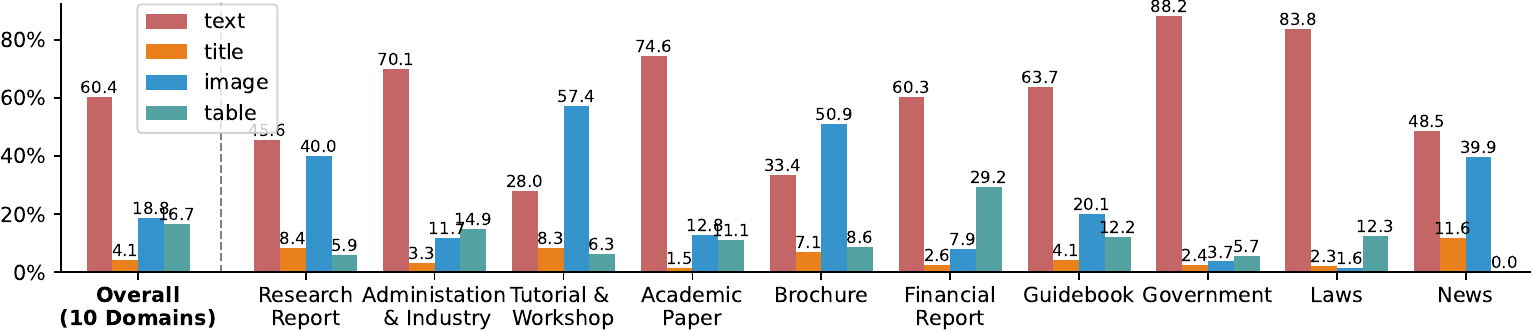}
         \vspace{-1em}
         \label{subfig:mmdocir_eval}
    \end{subfigure}
    \begin{subfigure}[b]{0.83\linewidth}
         \caption{313 documents in \dname training set in Section~\ref{sec:train_dataset}}.
         \includegraphics[width=1.0\linewidth]{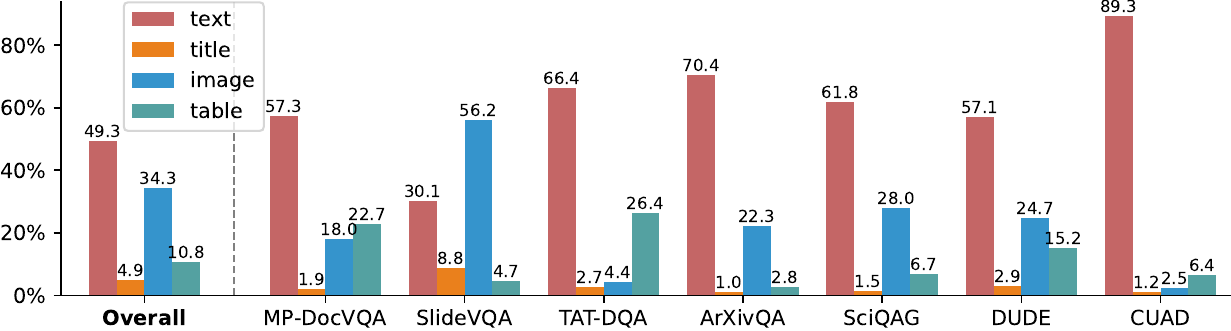}
         \vspace{-1.5em}
         \label{subfig:mmdocir_train}
    \end{subfigure}

    \caption{Area ratio of different modalities (1) in overall and (2) by domains/datasets in \dname evaluation and training set. Note that white spaces, headers, and footers are excluded from the area calculations.}
    \label{fig:mmdocir_layout}
\end{figure*}

\subsection{Training Dataset Label Construction} \label{appendix:labels}

The page labels can be directly obtained in the MP-DocVQA, SlideVQA, and DUDE datasets. Among these, only DUDE provides layout labels.

\textbf{SciQAG} provides only question and answer in texts. We use these information to infer the page-level and layout-level labels. Specifically, we first use MinerU to obtain layout-level passage chunks. For each QA pair, we deploy E5 and BGE retrievers to obtain question-passage and answer-passage similarity scores against all extracted passage chunks. If both scores rank within top 3 for a specific passage chunk, we assign this layout as the layout-level labels for the given QA pair. 

Similarly, \textbf{arXivQA} provides only cropped images, without document page/layout labels.
We first use MinerU to obtain layout-level images. For each cropped image, we calculate its similarity against all extracted images using brute-force matcher\footnote{https://opencv.org/}, and select the most similar one. Subsequently, we manually examine if the selected image matches the cropped image. In this way, we filter around 20\% unmatched images, resulting 1,579 questions with page and layout level labels.

For \textbf{TAT-DQA}, layout-level labels are provided for each sampled page. To localize the page index of the sampled pages, we first utilize PDF mapping tool\footnote{\url{https://github.com/pymupdf/PyMuPDF}} to retrieve best matched page in the document. Then, we manually verify whether the retrieved page matches the given page, and correct the labels if there were any errors.


\subsection{Hard Negative Sampling} \label{appendix:hard_neg}
In addition to annotating ground truth (positive) page labels, we enhance our training data with negative labels \cite{li-etal-2025-coir}. In the context of retrieval, \emph{hard negatives} are particularly informative non-relevant documents that closely resemble true positives according to the model's current scoring function. Unlike randomly selected negatives, hard negatives are challenging for the model to distinguish from relevant documents, thus providing stronger supervision.

In our framework, hard negatives are crucial for improving retrieval performance. By training the model on these challenging examples, we encourage it to learn more discriminative representations, ultimately enhancing its robustness and reducing false positives during retrieval.

As described in Appendix \ref{appendix:loss}, training is conducted using a contrastive loss, where the model aims to separate relevant documents from irrelevant ones. Specifically, we obtain hard negatives using the ColPali retriever \cite{faysse2024colpali}, which scores all document pages for a given query. The irrelevant pages with the highest top-$k$ scores (\ie those most likely to be confused with positives) are selected as hard negatives for training.
In the future, we consider to incorporate content planning~\cite{bao-etal-2022-aeg} and synthetic methods \cite{bao-etal-2023-synthetic, bao-etal-2022-generative} for hard negative generation.

\begin{table*}
\small
    \centering
    \resizebox{\linewidth}{!}{%
    \begin{tabular}{l|c|c|l}
        \toprule
        Artifacts & Purpose & Referred Section & Resource URL\\
        \midrule
        MMLong'-Doc       & Eval-set curation  & \multirow{2}{*}{Section \ref{sec:eval_dataset}} &  \url{https://github.com/mayubo2333/MMLongBench-Doc} \\
        DocBench       & Eval-set curation & & \url{https://github.com/Anni-Zou/DocBench} \\

        \cmidrule(lr){1-3}
        
        MP-DocVQA	& Train-set curation & 
        \multirow{7}{*}{
        \begin{tabular}[c]{@{}c@{}}
            Section \ref{sec:train_dataset} \\
            Appendix \ref{appendix:train_doc}
        \end{tabular}
        } & \url{https://rrc.cvc.uab.es/?ch=17&com=tasks} \\
        SlideVQA	& Train-set curation & & \url{https://github.com/nttmdlab-nlp/SlideVQA} \\
        TAT-DQA	& Train-set curation & & \url{https://github.com/NExTplusplus/TAT-DQA} \\
        arXivQA	& Train-set curation & & \url{https://huggingface.co/datasets/taesiri/arxiv_qa} \\
        SciQAG	& Train-set curation & & \url{https://github.com/MasterAI-EAM/SciQAG} \\
        DUDE	& Train-set curation & & \url{https://github.com/duchallenge-team/dude} \\
        CUAD	& Train-set curation & & \url{https://www.atticusprojectai.org/cuad} \\

        \cmidrule(lr){1-3}

        MinerU		& Doc Parsing & Section \ref{ssec:dataset_label} & \url{https://github.com/opendatalab/MinerU} \\
        \cmidrule(lr){1-3}
        Tesseract OCR & OCR-text & \multirow{3}{*}{Section \ref{ssec:mm-text}} & \url{https://github.com/tesseract-ocr/tesseract} \\
        GPT-4o		& VLM-text & & \url{https://openai.com/index/hello-gpt-4o/} \\
        QwenVL2.5	& VLM-text & & \url{https://github.com/QwenLM/Qwen2.5-VL} \\
        \cmidrule(lr){1-3}
        PyMuPDF		& Page matching & \multirow{3}{*}{Appendix \ref{appendix:labels}} & \url{https://github.com/pymupdf/PyMuPDF} \\
        OpenCV		& Image matching & & \url{https://opencv.org/} \\
        Text Retriever  & Layout location &  & BGE and E5 (see Table \ref{tab:retriever-implementation-details}) \\
        \cmidrule(lr){1-3}
        Visual Retriever & Hard negatives & Appendix \ref{appendix:hard_neg} & Colpali (see Table \ref{tab:retriever-implementation-details}) \\
        \bottomrule
    \end{tabular}
    }
\caption{Artifacts used to facilitate construction of \dname evaluation \& train set.}
\label{tab:artifacts}
\end{table*}

\subsection{Fine-grained Modality Distribution} \label{appendix:dataset_modality_distribution}
\dname evaluation set includes 313 long documents with an average length of 65.1 pages, categorized into ten main domains: research reports, administration\&industry, tutorials\&workshops, academic papers, brochures, financial reports, guidebooks, government documents, laws, and news articles. Overall, the modality distribution is: Text (60.4\%), Image (18.8\%), Table (16.7\%), and other modalities (4.1\%), as shown in Figure~\ref{subfig:mmdocir_eval}
Different domains exhibit different distributions of multimodal information. For instance, research reports, tutorials, workshops, and brochures predominantly contain images, whereas financial and industry documents are table-rich. In contrast, government and legal documents primarily comprise text.

\dname training set includes 6,878 long documents with an average length of 32.6 pages, categorized into seven Document VQA or QA datasets: MP-DocVQA~\cite{tito2023mp-docvqa}, SlideVQA~\cite{tanaka2023slidevqa}, TAT-DQA~\cite{zhu2022tatdqa}, arXivQA~\cite{li2024arxiv}, SciQAG~\cite{wan2024sciqag}, and DUDE~\cite{landeghem2023dude}.
Overall, the modality distribution is: Text (49.3\%), Image (34.3\%), Table (10.8\%), and other modalities (4.9\%), as shown in Figure~\ref{subfig:mmdocir_train}. 
Each dataset features unique distributions of multimodal content.
The legal documents and academic papers are text-intensive. The slides consist mostly of visual features. Industrial documents and financial reports are table-intensive.

\subsection{Resource URL of Artifacts} \label{appendix:artifacts_url}
In this section, we summarize the artifacts used to facilitate the construction of \dname's evaluation and train set, as shown in Table \ref{tab:artifacts}.
These artifacts mainly includes: datasets used for curating \dname evaluation and training sets, tools for parsing documents, packages for locating evidence, and etc. 

\section{Model Training: DPR-Phi3\&Col-Phi3} 
\label{appendix:train}

To evaluate the effectiveness of the \dname training set, we train two visual retrievers based on \texttt{Phi3-Vision}~\cite{abdin2024phi3}.
\texttt{Phi3-Vision} ($\mathbf{M_{phi3v}}$) reuses the image tokenizer from \texttt{clip-vit-large}\footnote{ViT-Large: \url{https://huggingface.co/openai/clip-vit-large-patch14-336}} ($\mathbf{M_{vit}}$). It can deal with high-resolution images by cropping them into sub-images, where each sub-image has $336\times336$ pixels.

\subsection{Document/Query Encoding}
\label{appendix:encode}

\texttt{DPR-Phi3} and \texttt{Col-Phi3} represent document page or query using a single dense embedding (following DPR~\cite{karpukhin-etal-2020-dense}) and a list of token-level embeddings (following ColBERT~\cite{khattab-etal-2020-colbert}), respectively.
Specifically, we follow \citet{ma2024dse} to concatenate document image with a text prompt: ``\textit{<s><d> What is shown in this image?</s>}''. Here, the <d> token is a special placeholder token and is replaced by the sequence of patch latent embeddings from the vision encoder. We consider only text queries and use text prompt: ``\textit{<s> query: \text{<q>} </s>}''. Similarly, the placeholder <q> token is replaced by input query.
We encode query $q$ and document $d$ in two ways:
\begin{equation}
    \begin{split}
    \mathrm{E_{d}^{dpr}} & = \mathbf{M_{phi3v}} \big( \mathbf{M_{vit}}(d), \text{prompt} \big) [-1], \in \mathbb{R}^{D_1} \\ 
    \mathrm{E_{q}^{dpr}} & = \mathbf{M_{phi3v}} \big( q, \text{prompt} \big) [-1], \in \mathbb{R}^{D_1}
    \end{split}
\end{equation}
where the end-of-sequence token </s> from the last hidden state ($D_1 = 3072$) of $\mathbf{M_{phi3v}}$ is used to represent $\mathrm{E_{d}^{dpr}}$ and $\mathrm{E_{q}^{dpr}}$. 
\begin{equation}
    \begin{split}
    \mathrm{E_{d}^{col}} & = \mathbf{M_{proj}} \cdot \mathbf{M_{phi3v}} \big( \mathbf{M_{vit}}(d), \text{prompt} \big) \\
    \mathrm{E_{q}^{col}} & = \mathbf{M_{proj}} \cdot \mathbf{M_{phi3v}} \big( q, \text{prompt} \big)
    \end{split}
\end{equation}
where $\mathrm{E_{d}^{col}} \in \mathbb{R}^{N_d \times D_2}$ and $\mathrm{E_{q}^{col}} \in \mathbb{R}^{N_q \times D_2} $, and $\mathbf{M_{proj}}$ is projection layer to map the last hidden states of $\mathbf{M_{phi3v}}$ into reduced dimension $D_2=128$. $N_d \approx 2500$ for a typical high-resolution page and $N_q$ is the number of query tokens.

\subsection{Query-Doc Similarity}
\label{appendix:similarity}

The similarity between the query and the document is computed as follows:
\begin{equation}
    \text{Sim}(q, d)_{dpr} = \frac{\langle \mathrm{E_{q}^{dpr}} | \mathrm{E_{d}^{dpr}} \rangle}{\left \| \mathrm{E_{q}^{dpr}} \right \|\cdot \left \| \mathrm{E_{d}^{dpr}} \right \|}
\end{equation}
where $Sim(q, d)_{dpr}$ is computed as the cosine similarity between their embeddings. and $\langle \cdot | \cdot \rangle$ is the dot product.
\begin{equation}
    \text{Sim}(q, d)_{col} = \sum_{i \in [1, N_q]} \max_{j \in [1, N_d]} \langle \mathrm{E_{q}^{col}}^{(i)} | \mathrm{E_{d}^{col}}^{(j)} \rangle
\label{eq:colbert_maxsim_score}
\end{equation}
where $Sim(q, d)_{col}$ is the sum over all query vectors $\mathrm{E_{q}^{col}}^{(i)}$, of its maximum dot product $\langle \cdot | \cdot \rangle$ with each of the $N_d$ document embedding vectors $\mathrm{E_d^{col}}^{(j)}$.

\subsection{Contrastive Loss} 
\label{appendix:loss}

Given the query $q$, we have the positive document $d^+$ and a set of negative documents $d^-$ including hard negatives and in-batch negatives. The hard negatives are negative pages within the document with highest $Sim(q, d^-)$ scored by ColPali~\cite{faysse2024colpali} retriever, refer to Appendix \ref{appendix:hard_neg} for more details on hard negative selection.
We calculate the loss as:
\begin{equation}\label{eq:dpr_obj}
    \mathcal{L}^{dpr}_{(q, d^+, d^-)} = -\log \frac{\exp(\text{Sim}^{dpr}_{(q, d^+)} /\tau )}{\sum_{d_i \in d^+ \cup d^-} \exp(\text{Sim}^{dpr}_{(q, d_i)} /\tau )}
\end{equation}
where \texttt{DPR-Phi3} is trained on the InfoNCE loss, and the temperature parameter $\tau=0.02$ in our experiments.
\begin{align}\label{eq:col_obj}
    \mathcal{L}^{col}_{(q, d^+, d^-)} = \ & \log \Big( 1 + \exp \big( \max_{d_i \in d^-} (\text{Sim}^{col}_{(q, d_i)}) \notag \\
    & - \text{Sim}^{col}_{(q, d^+)} \big) \Big)
\end{align}
where \texttt{Col-Phi3} is trained via the softplus loss based on the positive scores w.r.t. to the maximal negative scores.

\subsection{Training Implementation Details}
\label{appendix:train_imp}

In summary, we train two visual retrievers based on \texttt{Phi3-Vision}~\cite{abdin2024phi3}.
\texttt{DPR-Phi3} and \texttt{Col-Phi3} represent document page or query using a single dense embedding (following DPR~\cite{karpukhin-etal-2020-dense}) and a list of token-level embeddings (following ColBERT~\cite{khattab-etal-2020-colbert}), respectively.
To train the model, we employ memory-efficient techniques such as PERF~\cite{zhang2024gradient, zhang2024proactive}, LoRA~\cite{hu2022lora}, FlashAttention~\cite{dao2024flashattn2}, and DeepSpeed~\cite{rasley2020deepspeed}. 

The model is trained with a batch size of 64 for one epoch on \dname training set. The model weights are shared between the language models for document screenshot and query encoding.
In both tasks, each training query is paired with one positive document and one hard negative document. The document screenshots are resized to $1,344 \times 1,344$ pixels and cropped into $4 \times 4$ sub-images.

\section{Retrievers: Introduction and Implementation Details}
\label{appendix:retrievers}

\subsection{Text-Centric Document Retrieval} 
\label{appendix:text_retriever}

For text retrieval, the first step is to convert multimodal document into text using techniques, \eg Document Parsing~\cite{DBLP:conf/das/ChaoF04, wang2024mineru}, Optical Character Recognition (OCR)~\cite{Chaudhuri2017,DBLP:journals/corr/Borovikov14, 10.5555/319799}, Layout Detection~\cite{DBLP:conf/wincom/SassiouiBOECO23,2020layoutlm,2021layoutlm2}, Information extraction~\cite{dong-etal-2022-syntactic,dong-etal-2023-speculation}, Chunking~\cite{chen2023densexretrievalretrieval,raina2024questionbased}, and Image Captioning~\cite{DBLP:conf/cvpr/YouJWFL16,DBLP:conf/cvpr/AnejaDS18}. 
These steps are time-consuming and can introduce errors that impact the overall retrieval performance \cite{wu-etal-2025-from, nie-etal-2023-cross, li-etal-2023-from}.
Current text retrieval are primarily \xd{categorized as} sparse \xd{or} dense retrieval on chunks \cite{dong-etal-2023-open}. For two widely-used sparse retrievers: TF-IDF~\cite{TF-IDF} calculates the relevance via word frequency with the inverse document frequency, and BM25~\cite{bm25} introduces nonlinear word frequency saturation and length normalization.
Dense retrievers encode content into vector representations. DPR~\cite{karpukhin-etal-2020-dense} is the pioneering work of dense vector representations for QA tasks. Similarly, ColBERT~\cite{khattab-etal-2020-colbert} introduces an efficient question-document interaction model with late fine-grained term matching. 
Contriever~\cite{izacard-etal-2022-contriever} leverages contrastive learning to improve content dense encoding.
E5~\cite{wang2022-e5} and BGE~\cite{xiao2023-bge} propose novel training and data preparation techniques to enhance retrieval performance.
Moreover, GTE~\cite{li2023-gte} integrates graph-based techniques to enhance dense embedding. 
\xd{However, most} text retrieval systems \xd{overlook} valuable visual information present in documents.

\begin{table*}[t]
\small
\renewcommand{\arraystretch}{1.0 }
    \centering
\begin{tabu}{l|l|c|c|l}
\toprule
& \textbf{Model} & \textbf{Dimension} & \textbf{Base Model} & \textbf{HuggingFace Checkpoint}                               \\ 
\midrule

\parbox[t]{2.5mm}{\multirow{4}{*}{\rotatebox[origin=c]{90}{Text}}} &

DPR & 768 & BERT-base &
  \begin{tabular}[c]{@{}l@{}}
  \href{https://huggingface.co/facebook/dpr-ctx_encoder-multiset-base}{facebook/dpr-ctx\_encoder-multiset-base} \\ \href{https://huggingface.co/facebook/dpr-question_encoder-multiset-base}{facebook/dpr-question\_encoder-multiset-base}
  \end{tabular} \\ 
\tabucline[.2pt on 1pt off 1.5pt]{2-5} \\[-7.5pt]

& ColBERT  & $N_{\mathrm{tok}} \times$768 & BERT-base & \href{https://huggingface.co/colbert-ir/colbertv2.0}{colbert-ir/colbertv2.0} \\ 
\tabucline[.2pt on 1pt off 1.5pt]{2-5} \\[-7.5pt]
& Contriever & 768 & BERT-base & \href{https://huggingface.co/facebook/contriever-msmarco}{facebook/contriever-msmarco} \\ 
\tabucline[.2pt on 1pt off 1.5pt]{2-5} \\[-7.5pt]
& E5   & 1,024   & BERT-large & \href{https://huggingface.co/intfloat/e5-large-v2}{intfloat/e5-large-v2}  \\ 
\tabucline[.2pt on 1pt off 1.5pt]{2-5} \\[-7.5pt]
& BGE & 1,024 & RetroMAE   & \href{https://huggingface.co/BAAI/bge-large-en-v1.5}{BAAI/bge-large-en-v1.5}  \\ 
\tabucline[.2pt on 1pt off 1.5pt]{2-5} \\[-7.5pt]
& GTE  & 1,024 & BERT-large & \href{https://huggingface.co/thenlper/gte-large}{thenlper/gte-large}  \\
\midrule

\parbox[t]{2.5mm}{\multirow{3}{*}{\rotatebox[origin=c]{90}{Visual}}} &
DSE$_{\mathrm{wiki-ss}}$  & 3,072 & Phi-3-Vision & \href{https://huggingface.co/Tevatron/dse-phi3-v1.0}{Tevatron/dse-phi3-v1.0} \\
\tabucline[.2pt on 1pt off 1.5pt]{2-5} \\[-7.5pt]
& DSE$_{\mathrm{docmatix}}$ & 3,072 & Phi-3-Vision & \href{https://huggingface.co/Tevatron/dse-phi3-docmatix-v2}{Tevatron/dse-phi3-docmatix-v2} \\
\tabucline[.2pt on 1pt off 1.5pt]{2-5} \\[-7.5pt]
& ColPali & $N_{\mathrm{tok}}\times$128  & PaliGemma & \href{https://huggingface.co/vidore/colpali}{vidore/colpali} \\

\bottomrule
\end{tabu}
\caption{Implementation details for Text and Vision Retrieval Models}
\label{tab:retriever-implementation-details}
\end{table*}

\subsection{Vision-Driven Document Retrieval} 
\label{appendix:visual_retriever}

Vision Language Models (VLMs)~\cite{abdin2024phi3,beyer2024paligemma,bai2023qwenvl,chen2024internvl} can understand and generate text based on combined text and visual inputs.
This advancement has led to the development of cutting-edge visual-driven retrievers, such as ColPali~\cite{faysse2024colpali} and DSE~\cite{ma2024dse}. These models specifically leverage PaliGemma~\cite{beyer2024paligemma} and Phi3-Vision~\cite{abdin2024phi3} to directly encode document page screenshots for multimodal document retrieval.
ColPali adopts a similar question-document interaction as ColBERT, and represents each document page in token-level embeddings.
By contrast, DSE is similar to DPR \xd{in} that it encodes each page with a single dense embedding.
Visual retrievers \xd{are capable of modeling} useful visual information, allowing direct utilization of multimodal content without first converting it into text \xd{first}.
Despite these advancements, visual retrievers face challenges, particularly in dealing with text details when document page resolutions are high. The high resolution of document pages substantially increases the computational cost and complexity of the embedding process, which may hinder the model's performance.

\subsection{Implementation Details}
\label{appendix:retriever_impl}

In our experiments (refer to Section~\ref{ssec:baseline}), we implement 9 off-the-shelf retrievers including 6 text retrievers and 3 visual retrievers.
The text retrieval models deployed are namely DPR, ColBERT, Contriever, E5, BGE and GTE. These models use the WordPiece tokenizer from BERT and also inherit the maximum input length of 512 tokens from BERT~\cite{devlin-etal-2019-bert}. 
Additionally, we make use of the sentence-transformer library\footnote{\url{https://www.sbert.net/}} when deploying E5, BGE and GTE.
The visual retrieval models deployed are namely DSE$_{\mathrm{wiki-ss}}$, DSE$_{\mathrm{docmatix}}$, and ColPali.
We use pre-trained checkpoints available on HuggingFace \footnote{\url{https://huggingface.co/}}; the specific checkpoint information can be found in Table~\ref{tab:retriever-implementation-details} alongside other configuration details.

\section{Dataset Demonstration} \label{appendix:data_demo}

\subsection{Document Pages by Domains} \label{appendix:data_demo_pages}

\begin{figure*}[t]
    \centering
    \begin{subfigure}{0.80\linewidth}
        \caption{Page screenshots in Research Report domain.} 
        \includegraphics[width=\linewidth, trim={0em 0.65em 0em 0em}, clip]{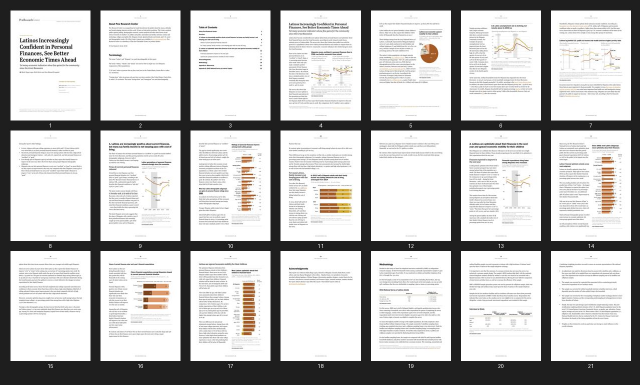}
        \label{fig:doc1}
    \end{subfigure}
    \begin{subfigure}{0.80\linewidth}
        \caption{Page screenshots in Administration \& Industry  domain.} 
        \includegraphics[width=\linewidth, trim={0em 0.65em 0em 0em}, clip]{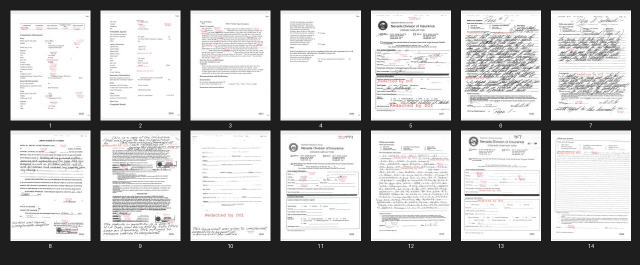}
         \label{fig:doc2}
    \end{subfigure}
    \begin{subfigure}{0.80\linewidth}
        \caption{Page screenshots in Tutorial \& Workshop domain.} 
        \includegraphics[width=\linewidth, trim={0em 0.65em 0em 0em}, clip]{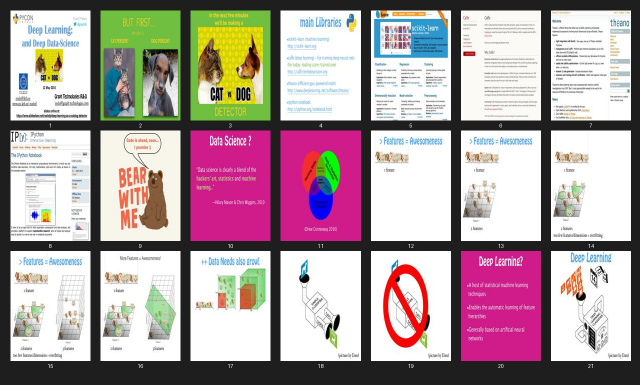}
        \label{fig:doc3}
    \end{subfigure} 

    \vspace{-0.5em}
    \caption{The screenshot examples of typical document pages for (a) Research Report, (b) Administration \& Industry, and (c) Tutorial \& Workshop domain.}
    \vspace{-1.0em}
    \label{fig:doc_examples1}
\end{figure*}

\begin{figure*}[t]
    \centering
    \begin{subfigure}{0.80\linewidth}
        \caption{Page screenshots in Academic Paper domain.} 
        \includegraphics[width=\linewidth, trim={0em 0.65em 0em 0em}, clip]{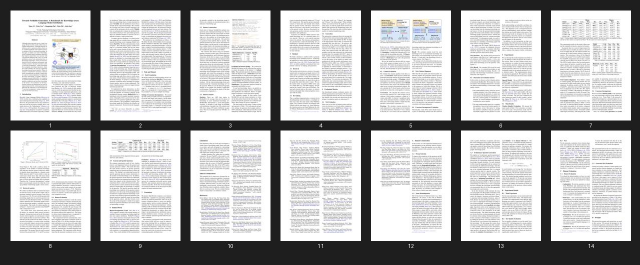}
        \label{fig:doc4}
    \end{subfigure}
    \begin{subfigure}{0.80\linewidth}
        \caption{Page screenshots in Brochure domain.} 
        \includegraphics[width=\linewidth, trim={0em 0.65em 0em 0em}, clip]{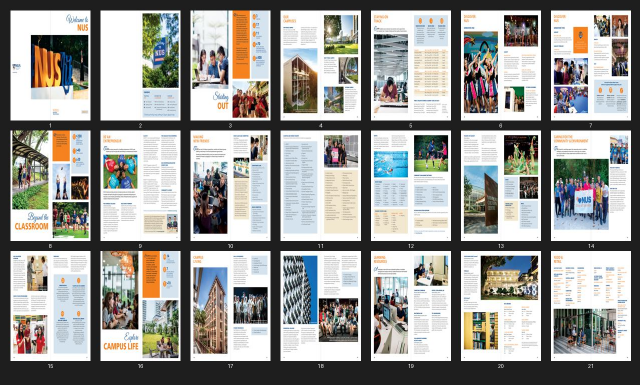}
        \label{fig:doc5}
    \end{subfigure}
    \begin{subfigure}{0.80\linewidth}
        \caption{Page screenshots in Financial Report domain.} 
        \includegraphics[width=\linewidth, trim={0em 0.65em 0em 0em}, clip]{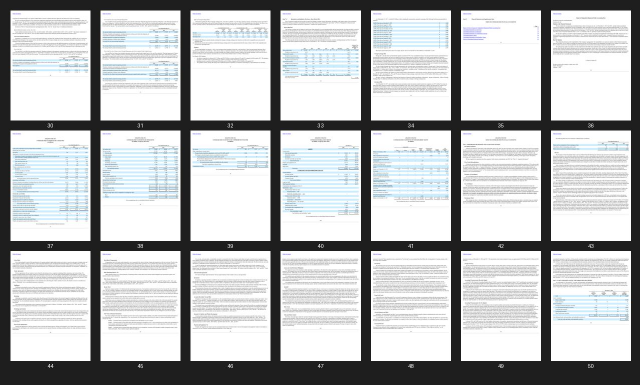}
        \label{fig:doc6}
    \end{subfigure}
    
    \vspace{-0.5em}
    \caption{The screenshot examples of typical document pages for (a) Academic Paper, (b) Brochure, and (c) Financial Report domain.}
    \vspace{-1.0em}
    \label{fig:doc_examples2}
\end{figure*}

\begin{figure*}[t]
    \centering
    \begin{subfigure}{0.80\linewidth}
        \caption{Page screenshots in Guidebook domain.} 
        \includegraphics[width=\linewidth, trim={0em 0.65em 0em 0em}, clip]{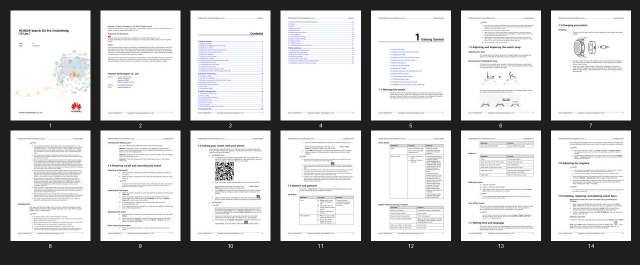}
         \label{fig:doc7}
    \end{subfigure}
    \begin{subfigure}{0.80\linewidth}
        \vspace{-1.0em}
        \caption{Page screenshots in Government domain.} 
        \includegraphics[width=\linewidth, trim={0em 0.65em 0em 0em}, clip]{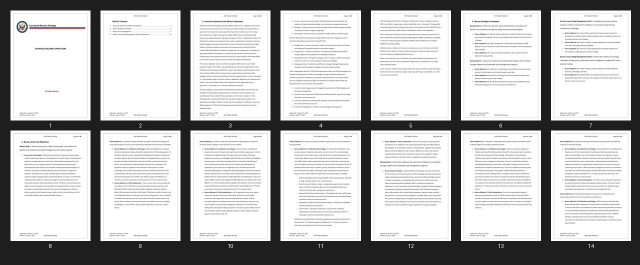}
        \label{fig:doc8}
    \end{subfigure} 
    \begin{subfigure}{0.80\linewidth}
        \vspace{-1.0em}
        \caption{Page screenshots in Laws domain.} 
        \includegraphics[width=\linewidth, trim={0em 0.65em 0em 0em}, clip]{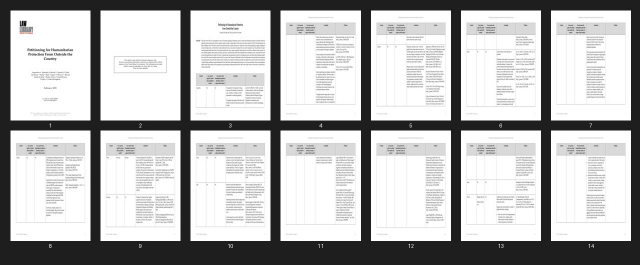}
        \label{fig:doc9}
    \end{subfigure}
    \begin{subfigure}{0.80\linewidth}
        \vspace{-1.0em}
        \caption{Page screenshots in News domain.} 
        \includegraphics[width=\linewidth, trim={0em 0.65em 0em 0em}, clip]{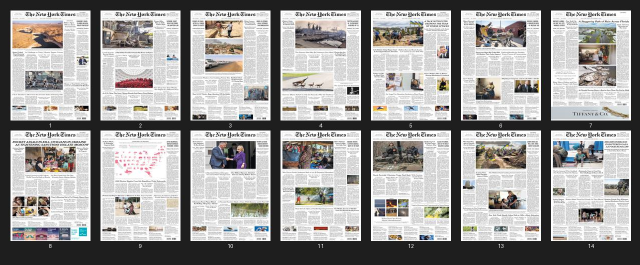}
        \label{fig:doc10}
    \end{subfigure}
    
    \vspace{-0.5em}
    \caption{The screenshot examples of typical document pages for (a) Guidebook, (b) Government, (c) Laws domain, and (d) News domain.}
    \vspace{-1.0em}
    \label{fig:doc_examples3}
\end{figure*}

The documents in MMDocIR can be categorized into 10 types. We \xd{provide} examples of each type as below.
\begin{itemize}[leftmargin=*, itemsep=0.0em, topsep=0.1em]

    \item \textbf{Admin \& Industry}: These documents primarily consist of instructional and overview content on industry, reflected by the dominance of text-based questions (78.0\%) and a smaller reliance on visual evidence (image questions only 20.3\%), which shows a text-heavy structure (70.1\%). \xd{Some} detailed \xd{examples are shown} in Figure~\ref{fig:doc2}.

    \item \textbf{Tut \& Workshop}: Documents in this category focus on slides or tutorials, which exhibit a balanced question modality: 61.7\% text, 24.5\% image, and 9.5\% table questions. Strong visual components \xd{are present}, with 57.4\% of its content being images—the highest among all categories. \xd{Some} detailed \xd{examples are shown} in Figure~\ref{fig:doc3}.

    \item \textbf{Academic Paper}: These documents are formal publications with structured layouts, citations, and academic pictures. The questions span multiple modalities: 28.8\% text, 25.7\% image, and 50.0\% table. Text modality dominates content distribution (74.6\%), with the presence of tables (11.1\%) and images (12.8\%) demonstrating rich multimodal alignment and explicit questions with answers. Some detailed examples are shown in Figure~\ref{fig:doc4}.

    \item \textbf{Brochure}: Designed for promotional purposes, the brochure category contains highly visual documents. Over 52.6\% of questions are image-based—the highest among all domains—while text-based questions account for only 60.5\%. Modality distribution is similarly diverse: 50.8\% image, showcasing their visually complex layout. Some detailed examples are shown in Figure~\ref{fig:doc5}.

    \item \textbf{Financial Report}: These documents involve massive numerical and quantitative data, reflected in a high proportion of table questions (54.5\%) and strong table content distribution (29.2\%). While text remains significant (60.3\%), the inclusion of tabular and numerical analysis is essential for understanding these documents. Some detailed examples are shown in Figure~\ref{fig:doc6}.

    \item \textbf{Guidebook}: Instruction manuals for electronics and tools, guidebooks exhibit the most balanced question modality: 51.8\% text, 54.4\% image, and 26.8\% table, indicating multimodal instructional designs. Some detailed examples are shown in Figure~\ref{fig:doc7}.

    \item \textbf{Government}: This category covers policy files and governmental reports. It is highly text-centric with 69.9\% text questions and 88.2\% text content. This reflects the formal and regulatory nature of such documents. Some detailed examples are shown in Figure~\ref{fig:doc8}.

    \item \textbf{Laws}: Legal documents exhibit strong textual dominance both in questions (62.1\%) and content (83.8\%), with very limited visual presence (image content only 1.6\%). They often maintain specific formats and focus on linguistic interpretation rather than visual layout. Some detailed examples are shown in Figure~\ref{fig:doc9}.

    \item \textbf{News}: Although based on only one document, the “News” domain shows notable multimodal richness. It includes a significant image portion (39.8\%), high text presence (48.5\%), and 11.6\% titles. This reflects the use of images and headlines typical of news articles. Some detailed examples are shown in Figure~\ref{fig:doc10}.
\end{itemize}

\subsection{Document Layouts}  \label{appendix:data_demo_layout}

\begin{figure*}[hbp]
    \centering

    \begin{minipage}{0.9\linewidth}
        \centering
        \caption*{(a) Example 1: original page vs. page highlighted with layout bounding boxes.} 
        \vspace{-1.0em}
        \begin{subfigure}{0.48\linewidth}
            \includegraphics[width=\linewidth, trim={0em 0em 0em 0em}, clip]{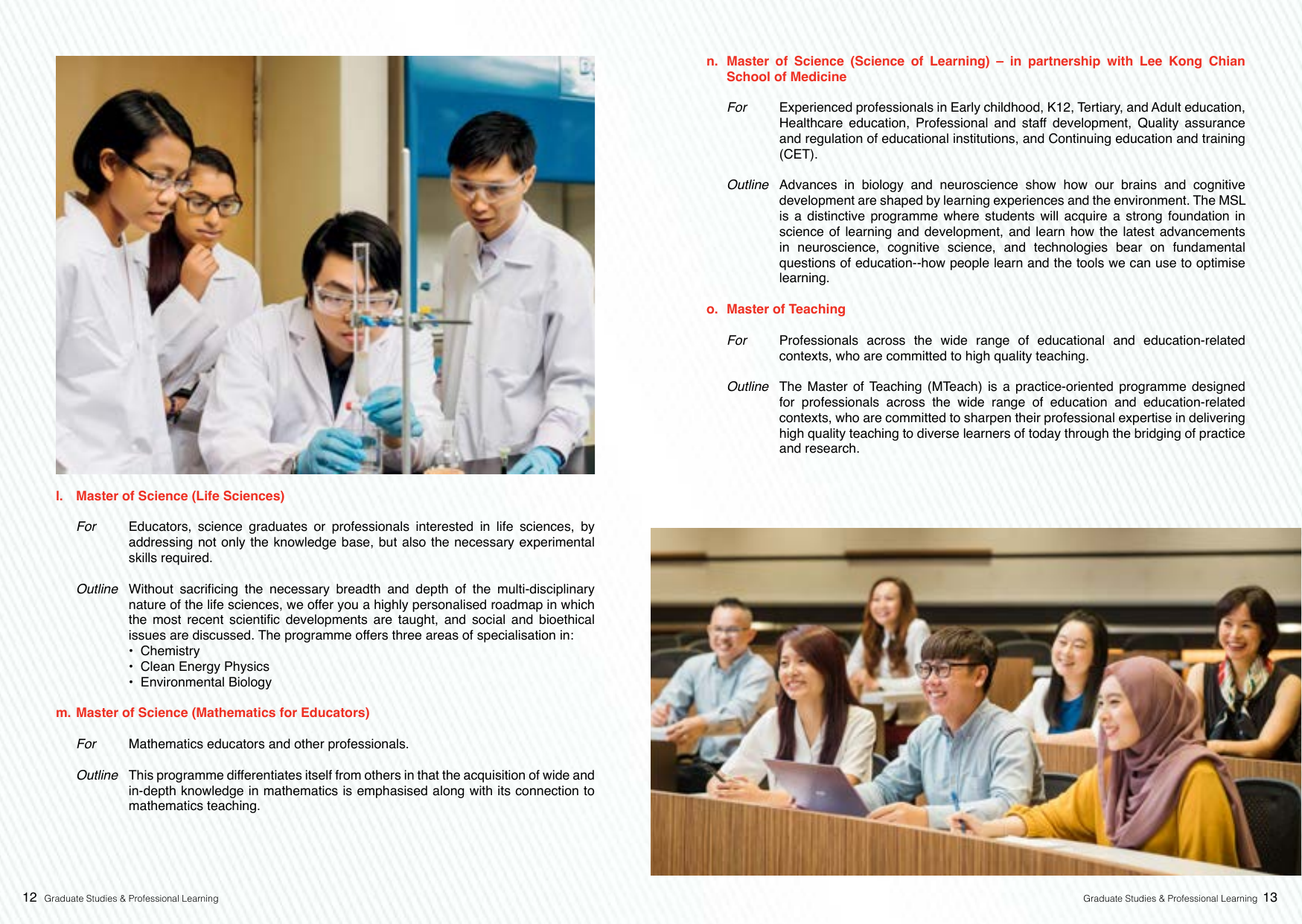}
        \end{subfigure}
        \hspace{0.02\textwidth}
        \begin{subfigure}{0.48\linewidth}
            \includegraphics[width=\linewidth, trim={0em 0em 0em 0em}, clip]{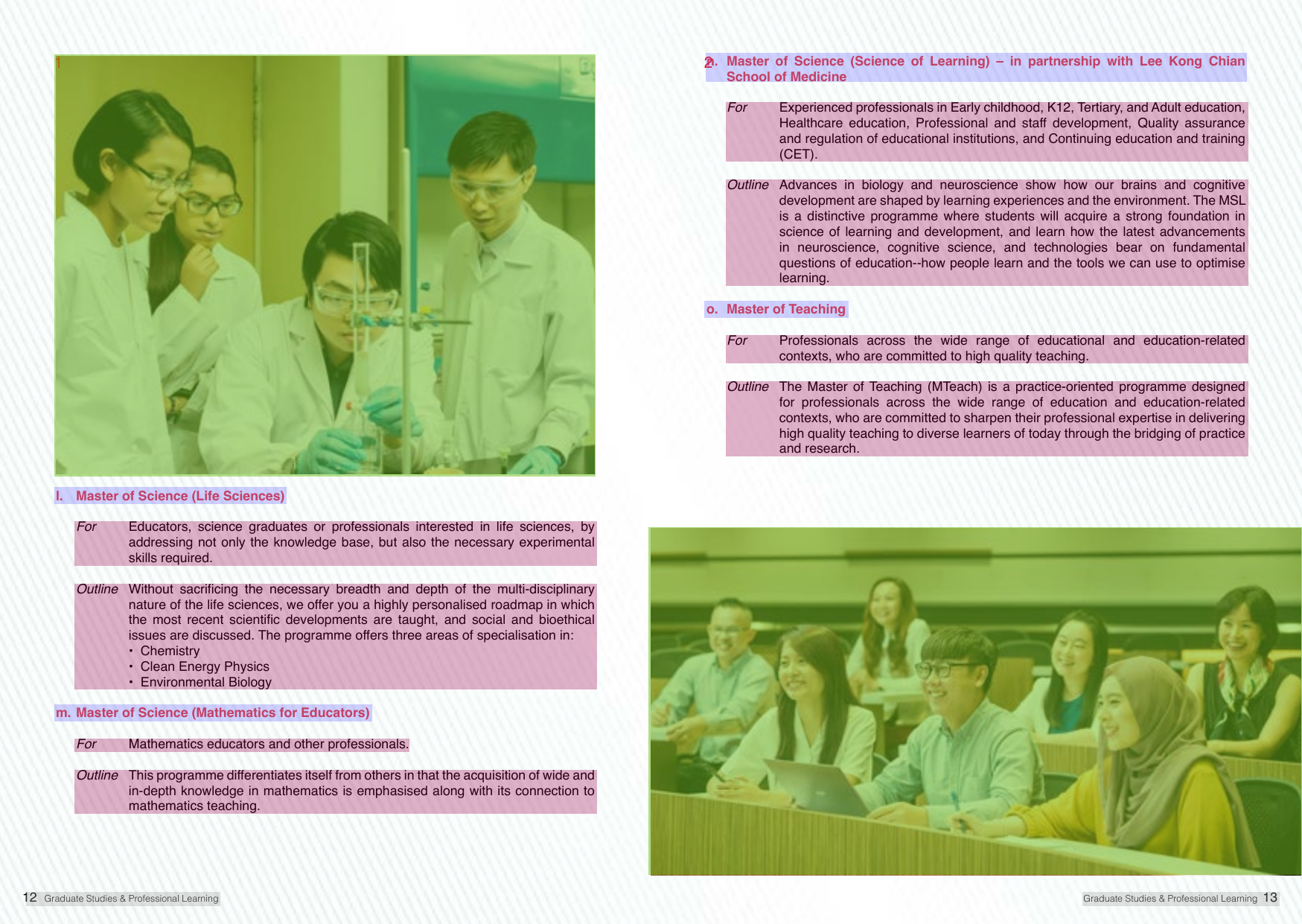}
        \end{subfigure}
        \label{fig:layout1}
    \end{minipage}

    \begin{minipage}{\linewidth}
        \centering
        \caption*{(b) Example 2: original page vs. page highlighted with layout bounding boxes.} 
        \begin{subfigure}{0.48\linewidth}
            \includegraphics[width=\linewidth, trim={0em 5em 0em 6.5em}, clip]{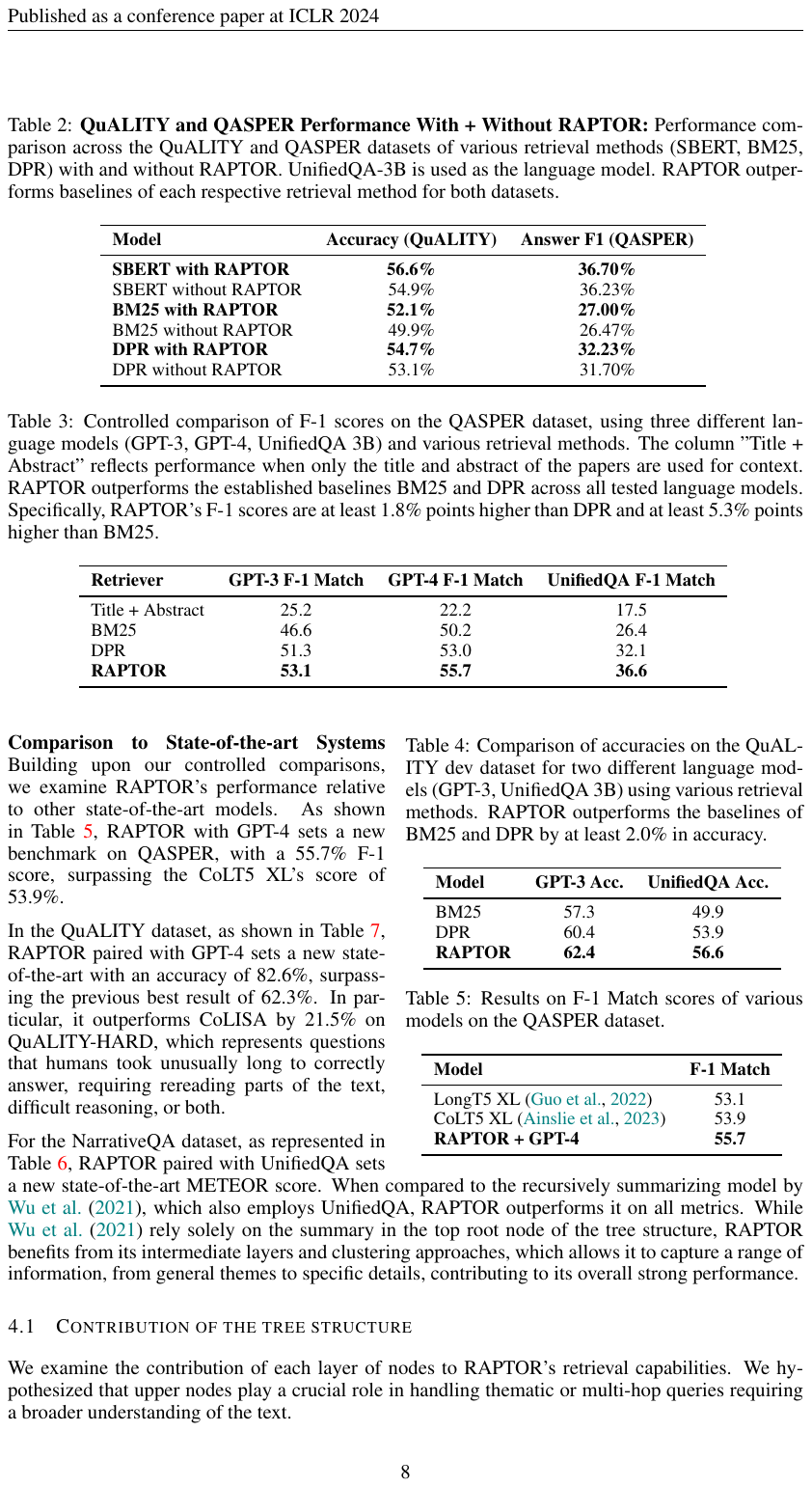}
        \end{subfigure}
        \hspace{0.02\textwidth}
        \begin{subfigure}{0.48\linewidth}
            \includegraphics[width=\linewidth, trim={0em 5em 0em 6.5em}, clip]{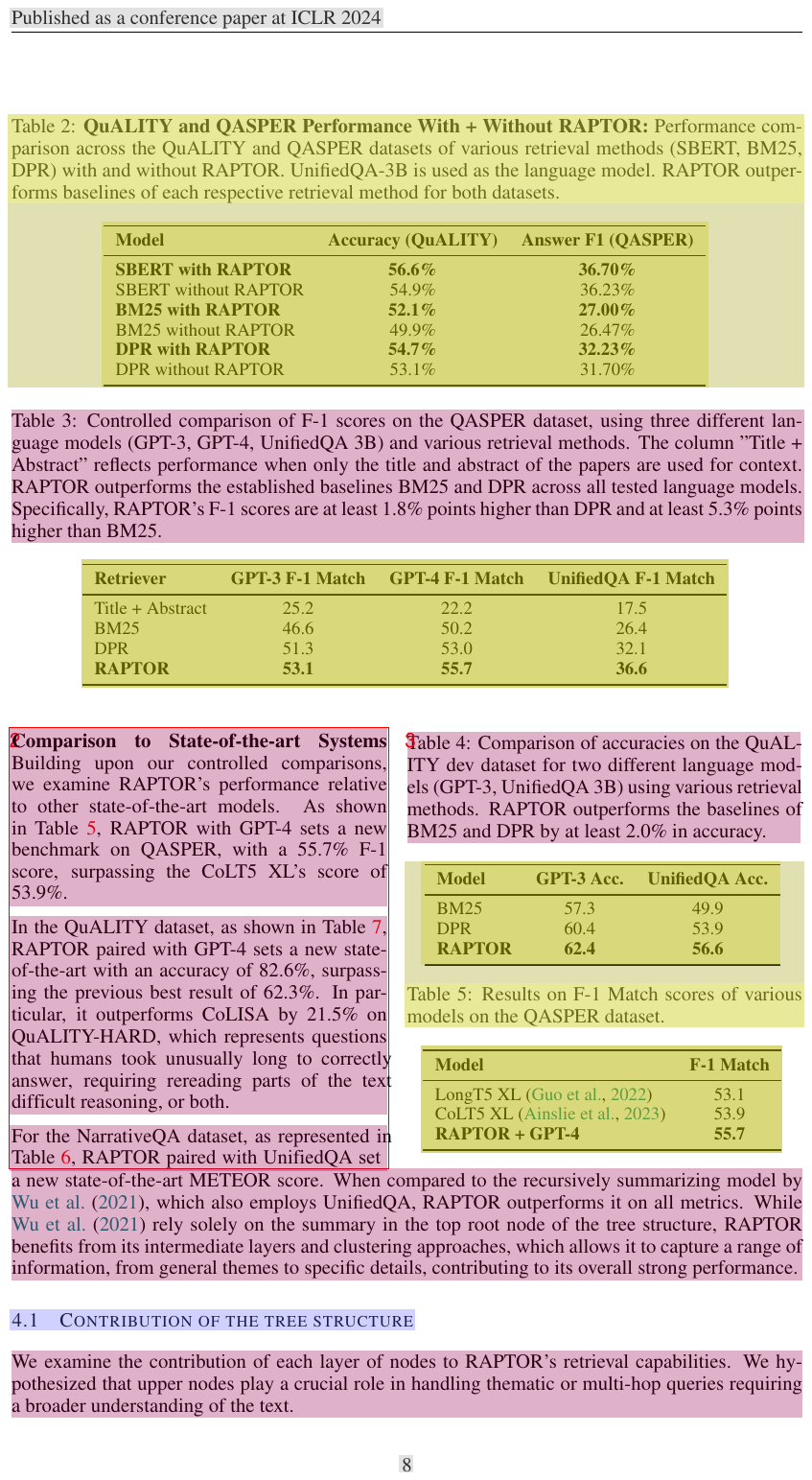}
        \end{subfigure}
        \label{fig:layout2}
    \end{minipage}

    \begin{minipage}{0.9\linewidth}
        \centering
        \caption*{(c) Example 3: original page vs. page highlighted with layout bounding boxes.} 
        \begin{subfigure}{0.48\linewidth}
            \includegraphics[width=\linewidth, trim={0em 8em 0em 3em}, clip]{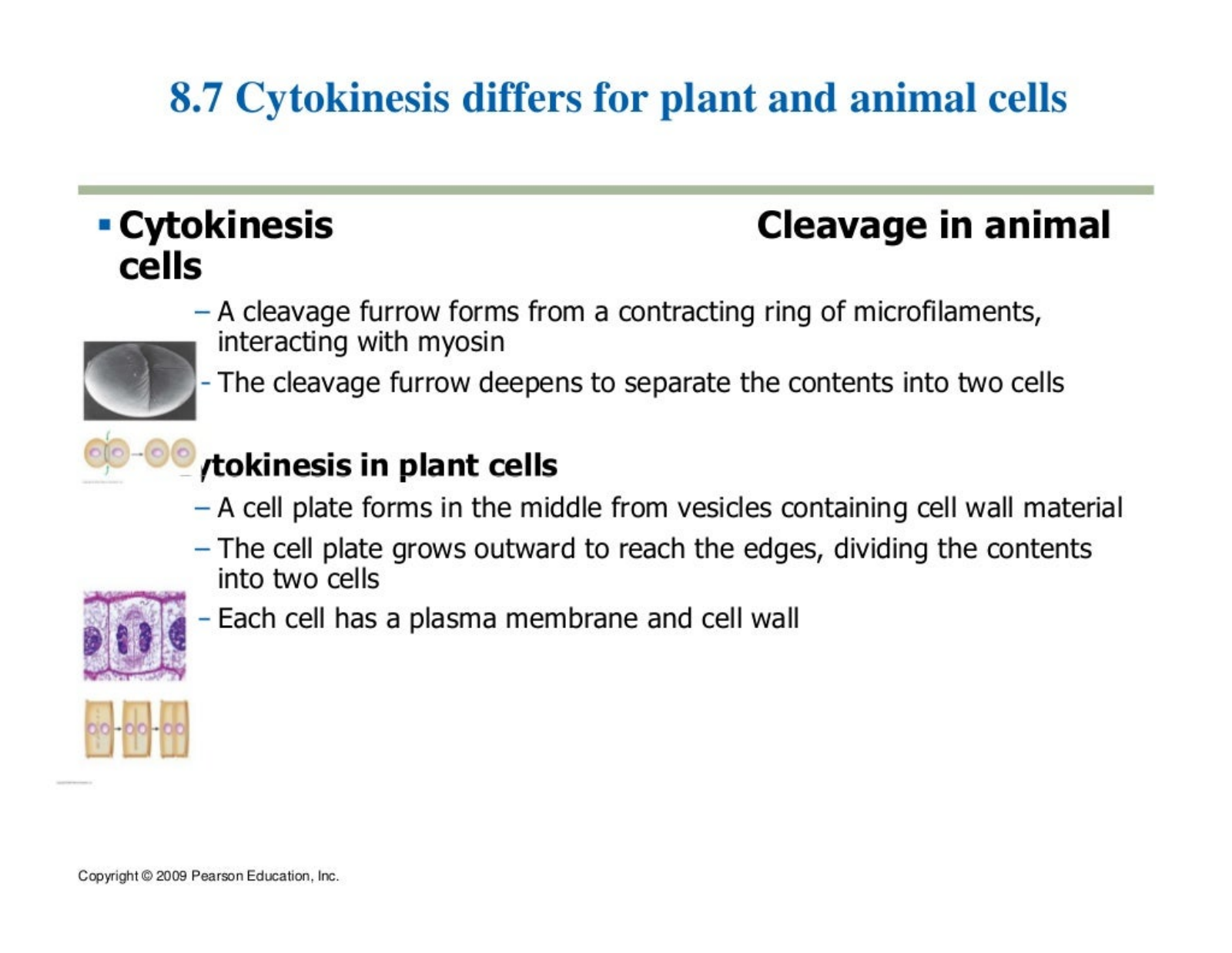}
        \end{subfigure}
        \hspace{0.02\textwidth}
        \begin{subfigure}{0.48\linewidth}
            \includegraphics[width=\linewidth, trim={0em 8em 0em 3em}, clip]{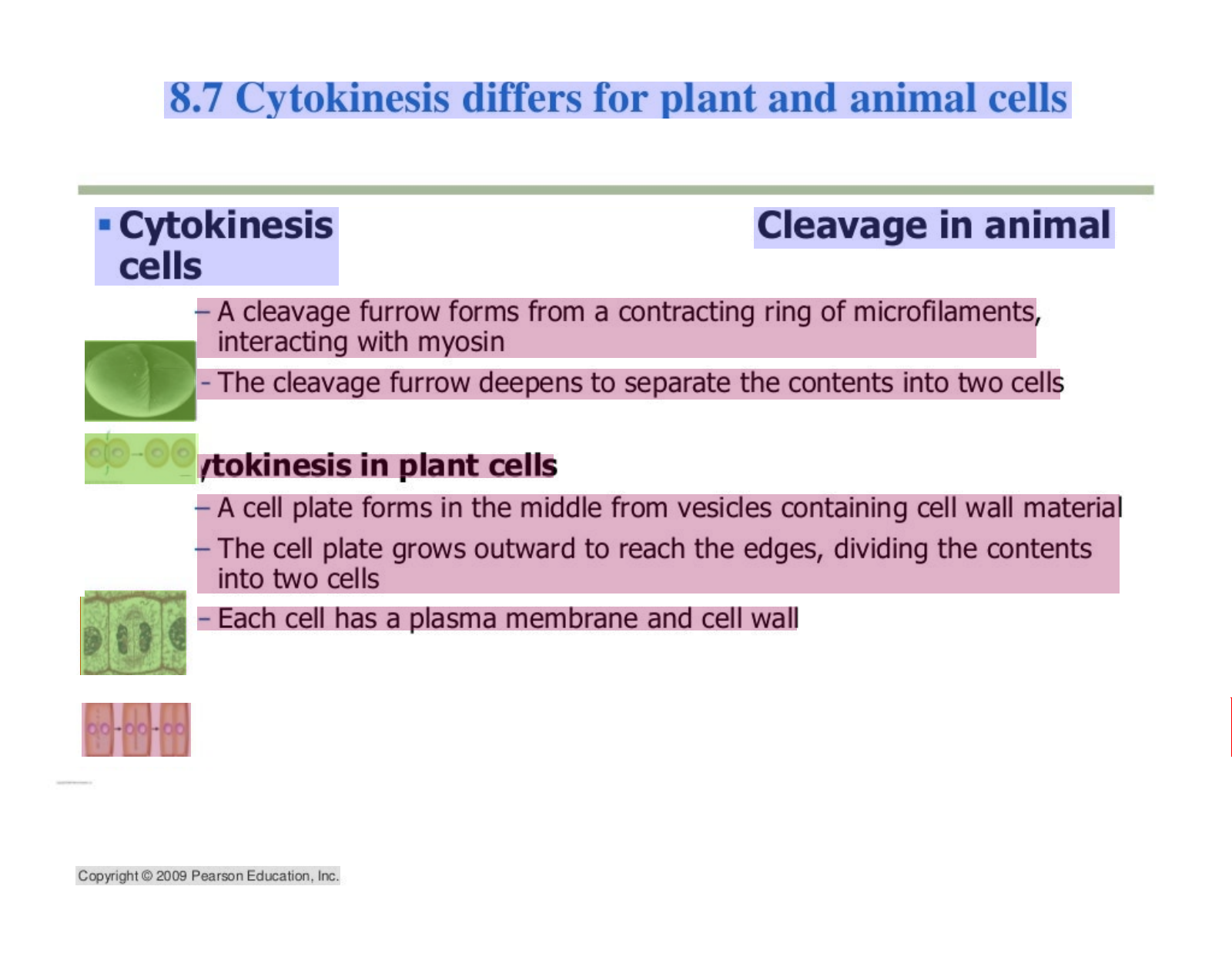}
        \end{subfigure}
        \label{fig:layout3}
    \end{minipage}

    \vspace{-0.5em}
    \caption{The 3 examples illustrate the function and effectiveness of layout detection on document pages.}
    \vspace{-1.0em}
    \label{fig:layout_examples1}
\end{figure*}

\begin{figure*}[hbp]
    \centering

    \begin{minipage}{0.95\linewidth}
        \centering
        \caption*{(a) Example 1: original page vs. page highlighted with layout bounding boxes.} 
        \vspace{-1.0em}
        \begin{subfigure}{0.48\linewidth}
            \includegraphics[width=\linewidth, trim={0em 19em 0em 1em}, clip]{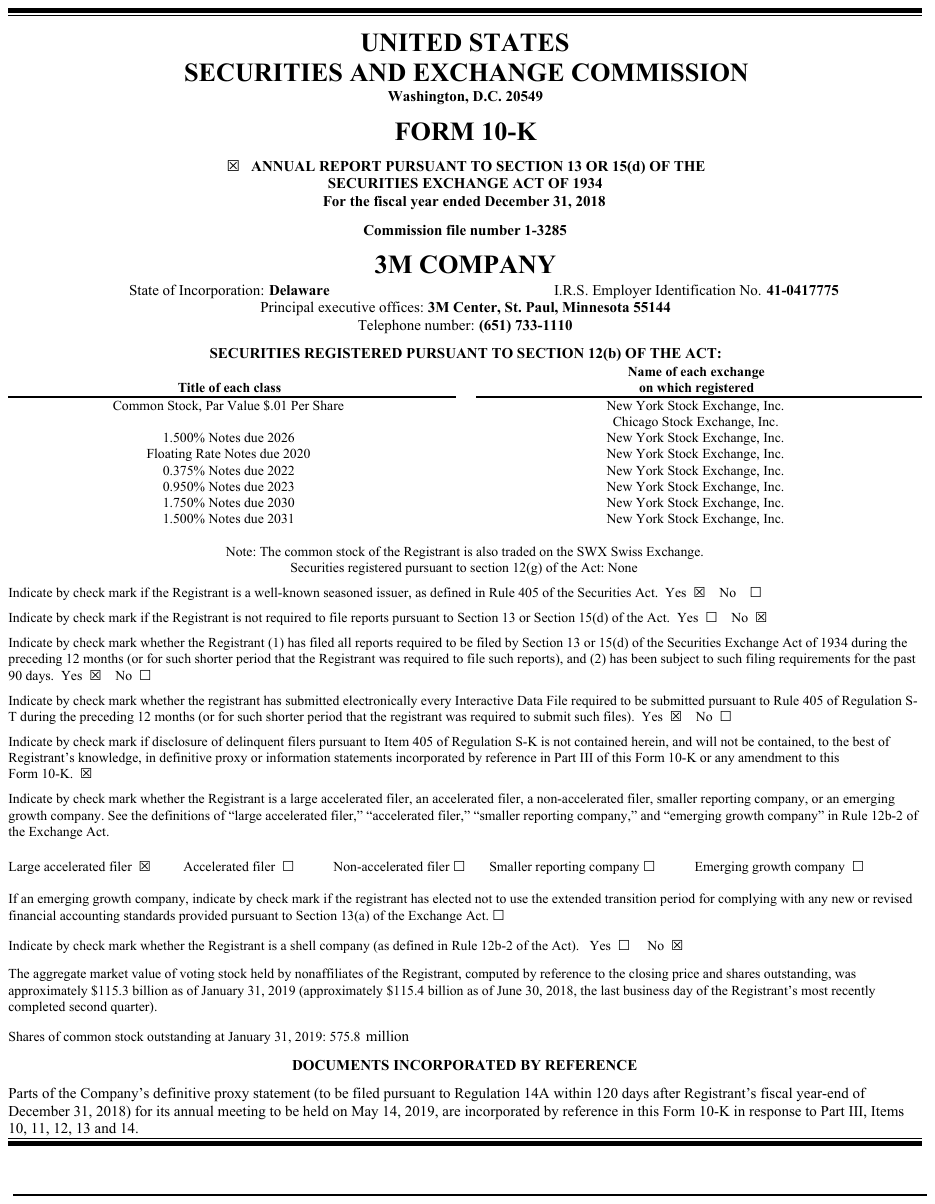}
        \end{subfigure}
        \hspace{0.02\textwidth}
        \begin{subfigure}{0.48\linewidth}
            \includegraphics[width=\linewidth, trim={0em 19em 0em 1em}, clip]{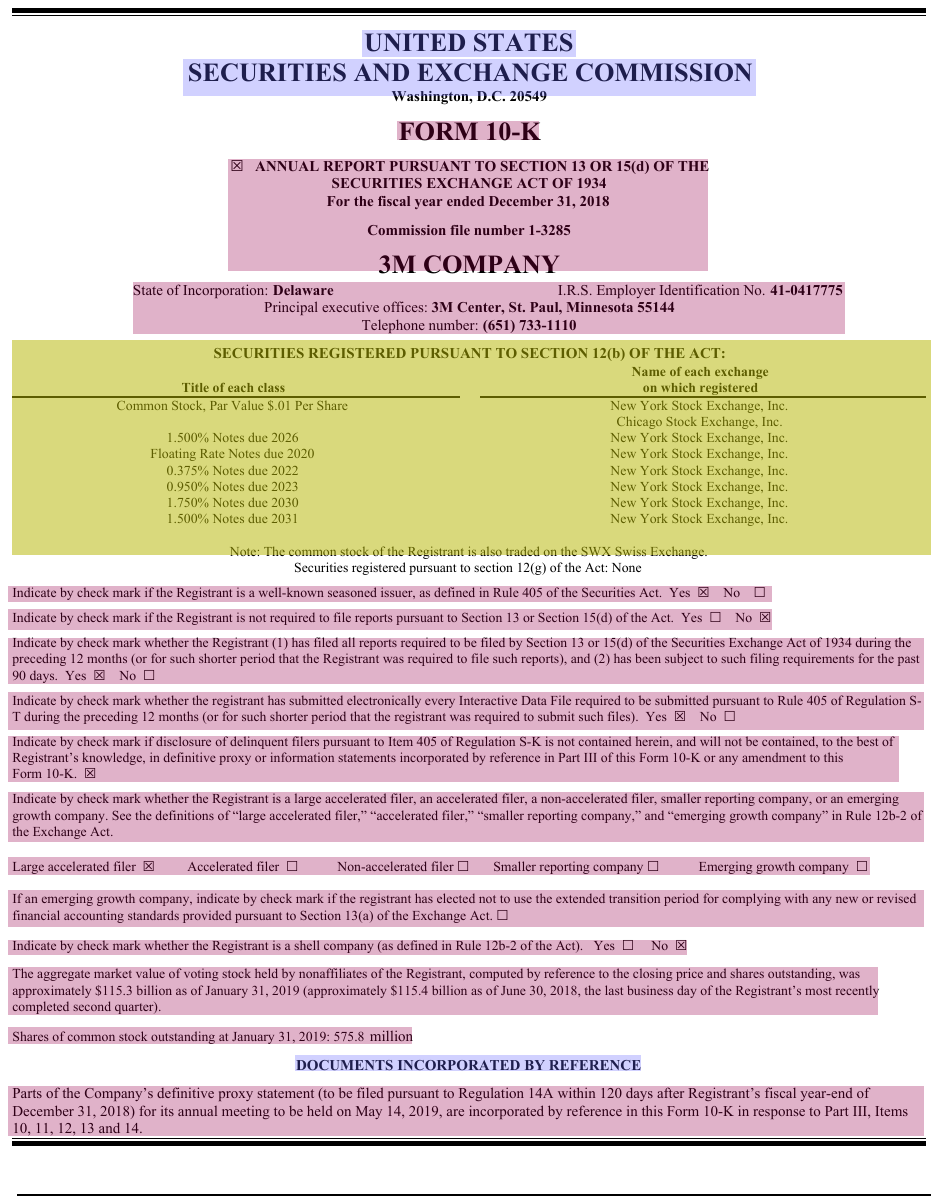}
        \end{subfigure}
        \label{fig:layout4}
    \end{minipage}

    \begin{minipage}{0.95\linewidth}
        \centering
        \caption*{(b) Example 2: original page vs. page highlighted with layout bounding boxes.} 
        \begin{subfigure}{0.48\linewidth}
            \includegraphics[width=\linewidth, trim={0em 24em 0em 5em}, clip]{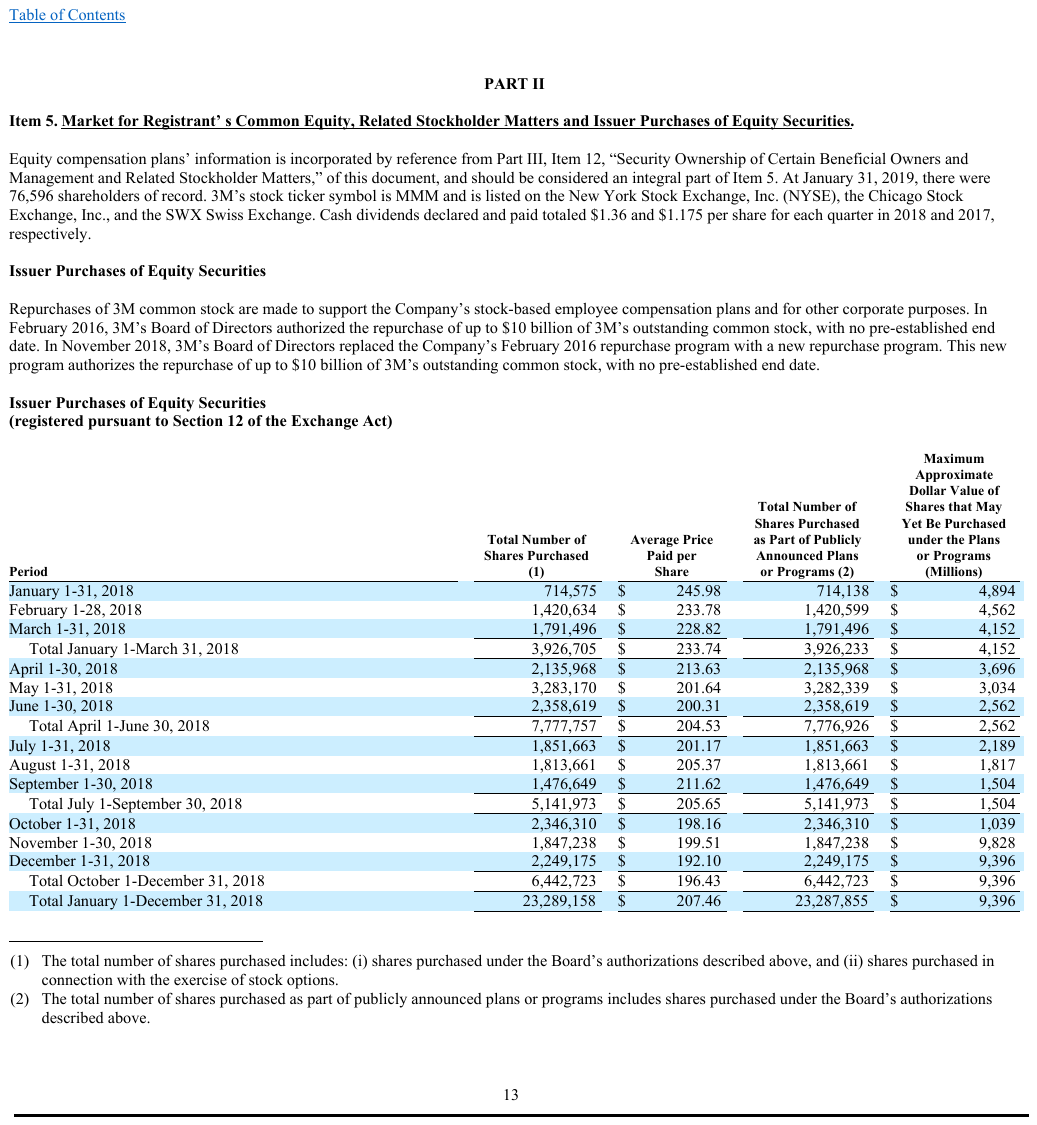}
        \end{subfigure}
        \hspace{0.02\textwidth}
        \begin{subfigure}{0.48\linewidth}
            \includegraphics[width=\linewidth, trim={0em 24em 0em 5em}, clip]{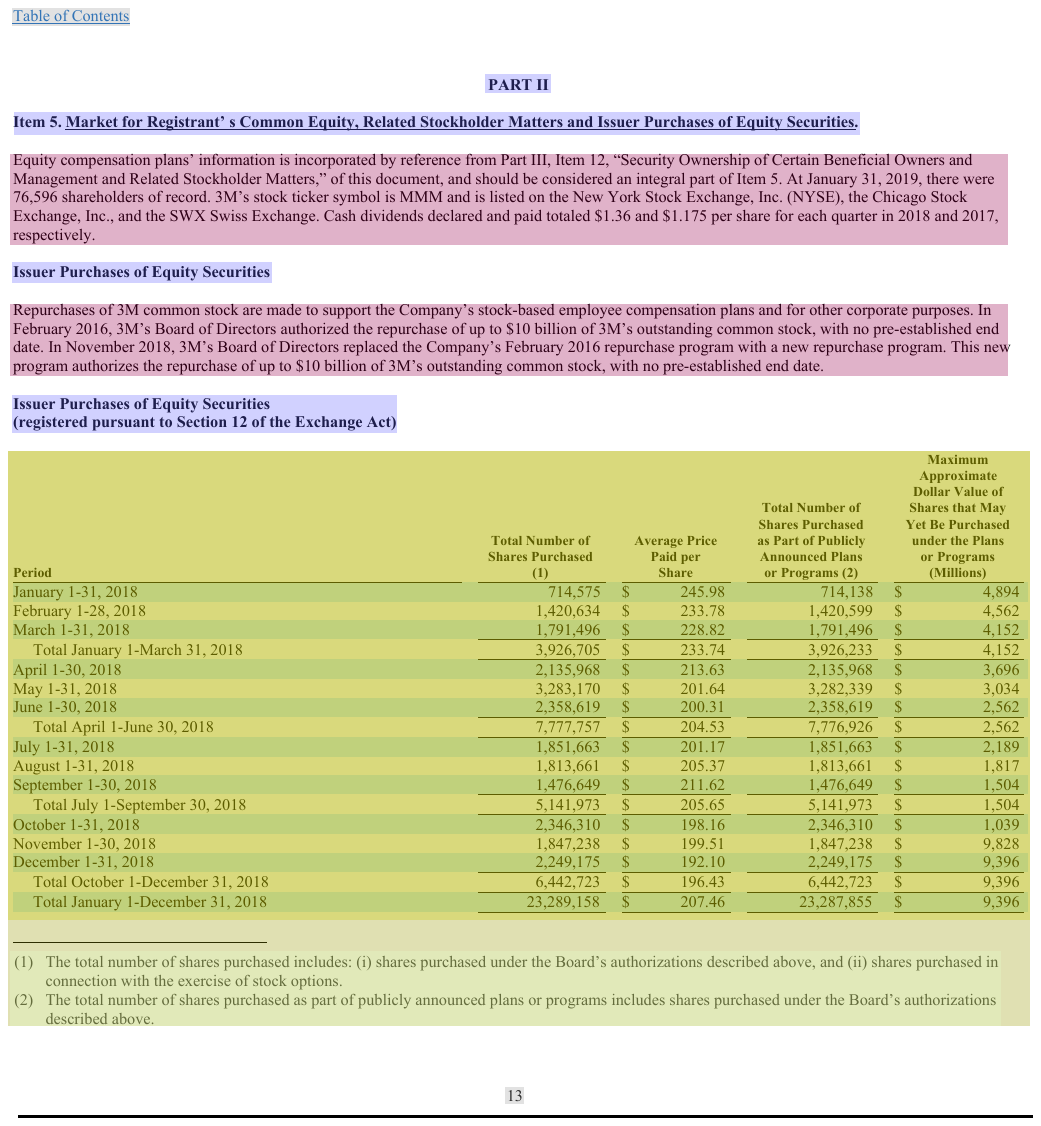}
        \end{subfigure}
        \label{fig:layout5}
    \end{minipage}

    \begin{minipage}{0.9\linewidth}
        \centering
        \caption*{(c) Example 3: original page vs. page highlighted with layout bounding boxes.} 
        \begin{subfigure}{0.48\linewidth}
            \includegraphics[width=\linewidth, trim={0em 8em 0em 3em}, clip]{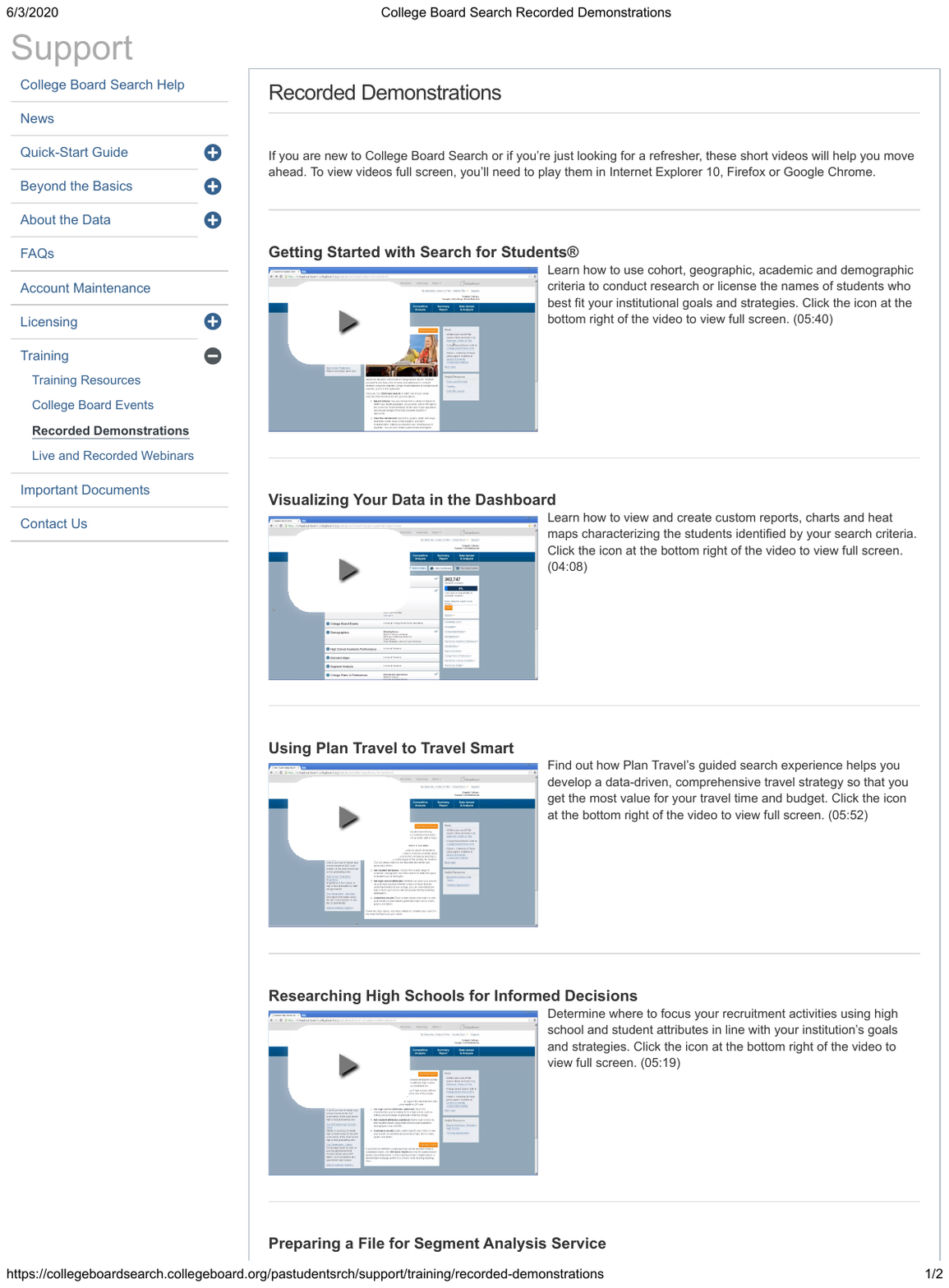}
        \end{subfigure}
        \hspace{0.02\textwidth}
        \begin{subfigure}{0.48\linewidth}
            \includegraphics[width=\linewidth, trim={0em 8em 0em 3em}, clip]{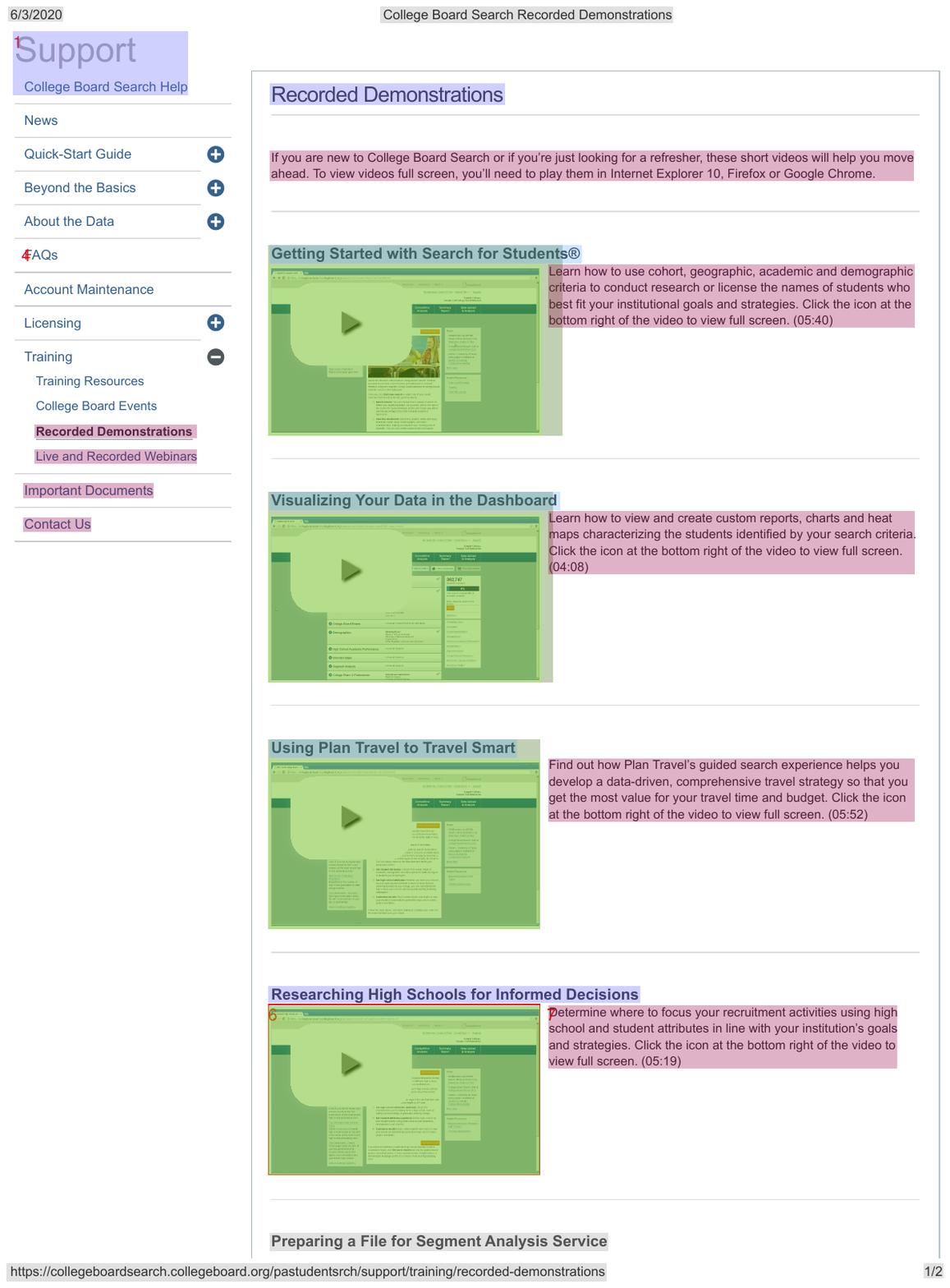}
        \end{subfigure}
        \label{fig:layout6}
    \end{minipage}

    \vspace{-0.5em}
    \caption{The 3 examples illustrate the function and effectiveness of layout detection on document pages.}
    \vspace{-1.0em}
    \label{fig:layout_examples2}
\end{figure*}

\begin{figure*}[hbp]
    \centering

    \begin{minipage}{0.8\linewidth}
        \centering
        \caption*{(a) Example 1: original page vs. page highlighted with layout bounding boxes.} 
        \vspace{-1.0em}
        \begin{subfigure}{0.48\linewidth}
            \includegraphics[width=\linewidth, trim={0em 4em 0em 0em}, clip]{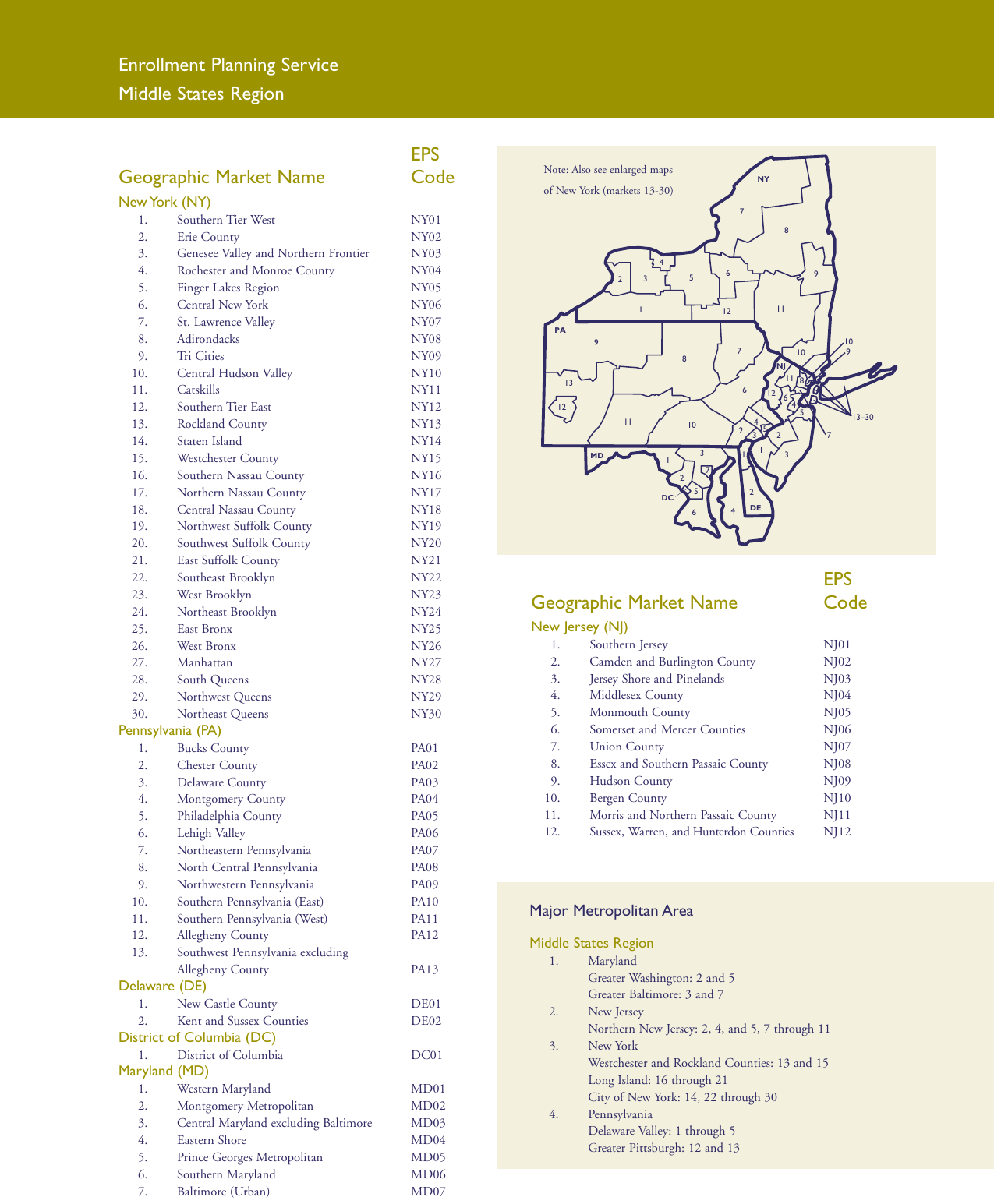}
        \end{subfigure}
        \hspace{0.02\textwidth}
        \begin{subfigure}{0.48\linewidth}
            \includegraphics[width=\linewidth, trim={0em 4em 0em 0em}, clip]{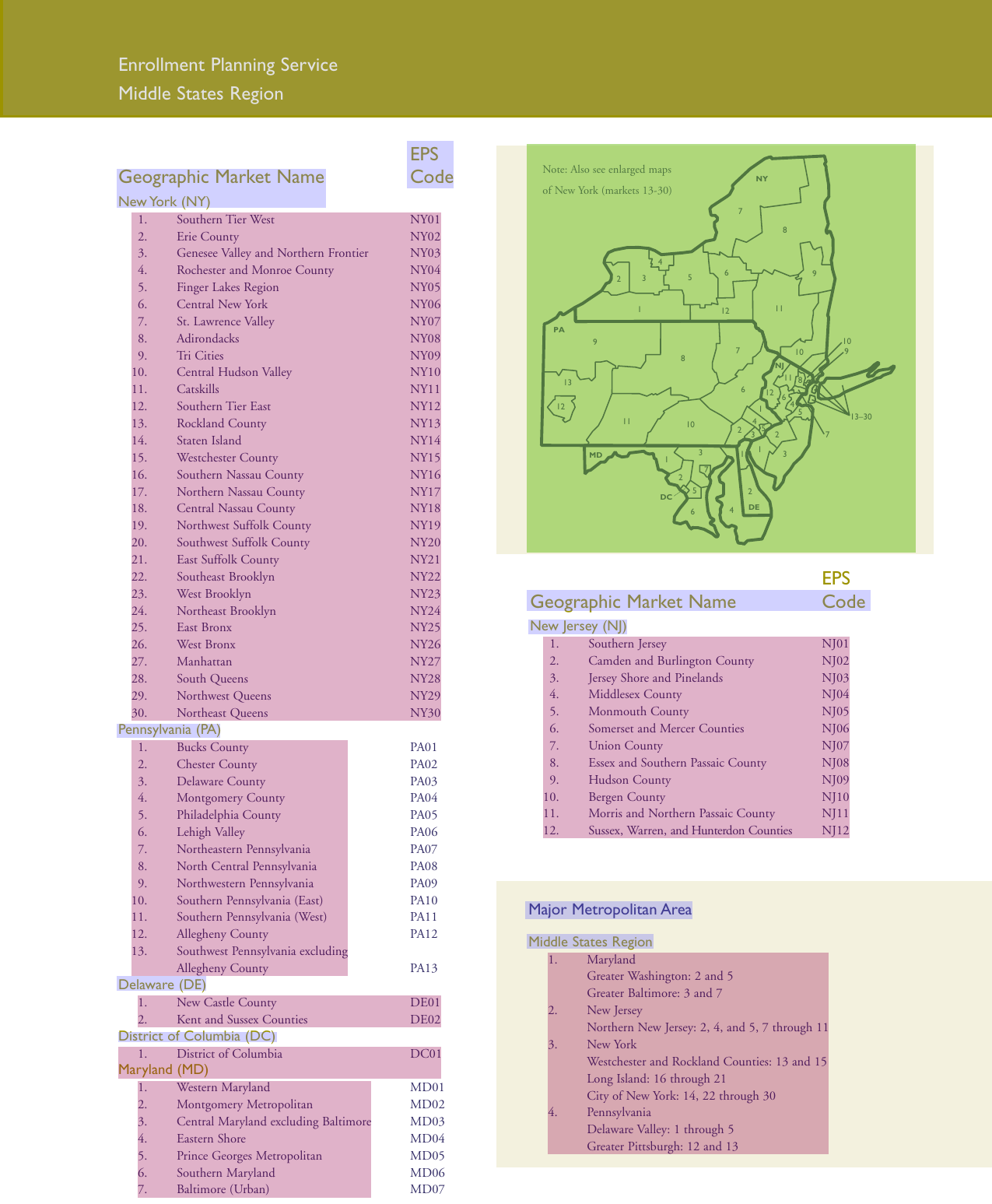}
        \end{subfigure}
        \label{fig:layout7}
    \end{minipage}
    
    \begin{minipage}{0.9\linewidth}
        \centering
        \vspace{-1.0em}
        \caption*{(b) Example 2: original page vs. page highlighted with layout bounding boxes.} 
        \vspace{-1.0em}
        \begin{subfigure}{0.48\linewidth}
            \includegraphics[width=\linewidth, trim={0em 5em 0em 1em}, clip]{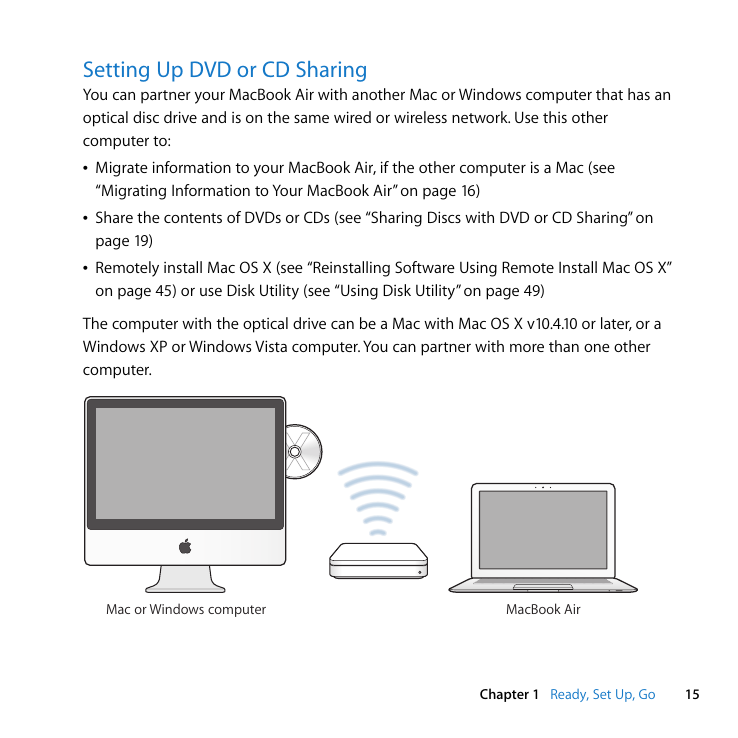}
        \end{subfigure}
        \hspace{0.02\textwidth}
        \begin{subfigure}{0.48\linewidth}
            \includegraphics[width=\linewidth, trim={0em 5em 0em 1em}, clip]{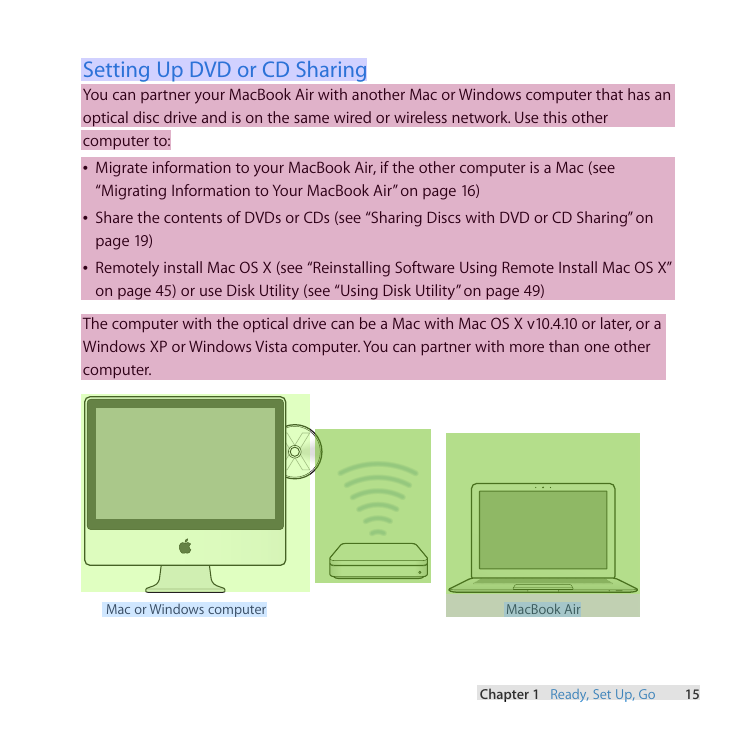}
        \end{subfigure}
        \label{fig:layout8}
    \end{minipage}

    \begin{minipage}{0.8\linewidth}
        \centering
        \caption*{(c) Example 3: original page vs. page highlighted with layout bounding boxes.} 
        \begin{subfigure}{0.48\linewidth}
            \includegraphics[width=\linewidth, trim={0em 8em 0em 3em}, clip]{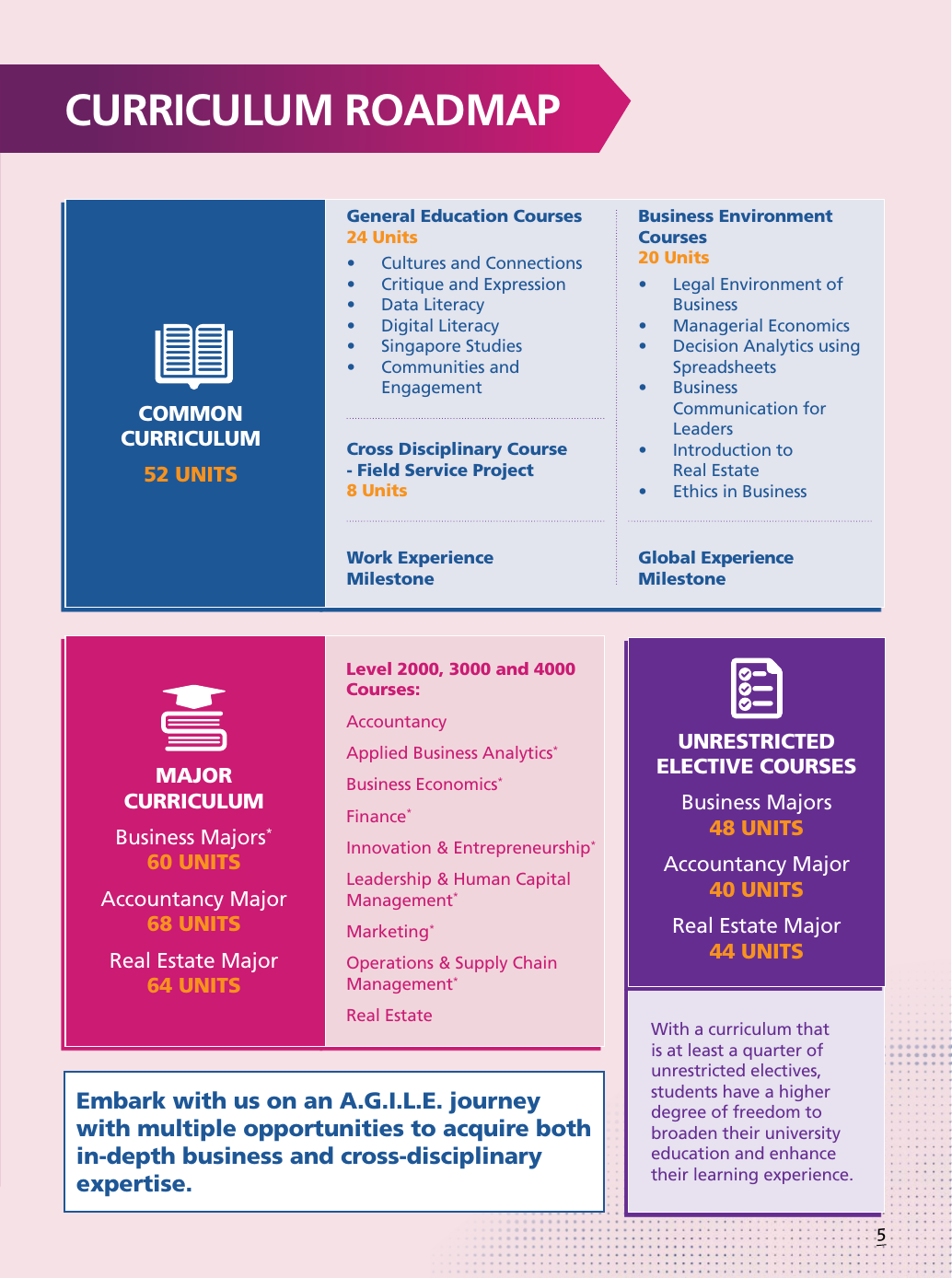}
        \end{subfigure}
        \hspace{0.02\textwidth}
        \begin{subfigure}{0.48\linewidth}
            \includegraphics[width=\linewidth, trim={0em 8em 0em 3em}, clip]{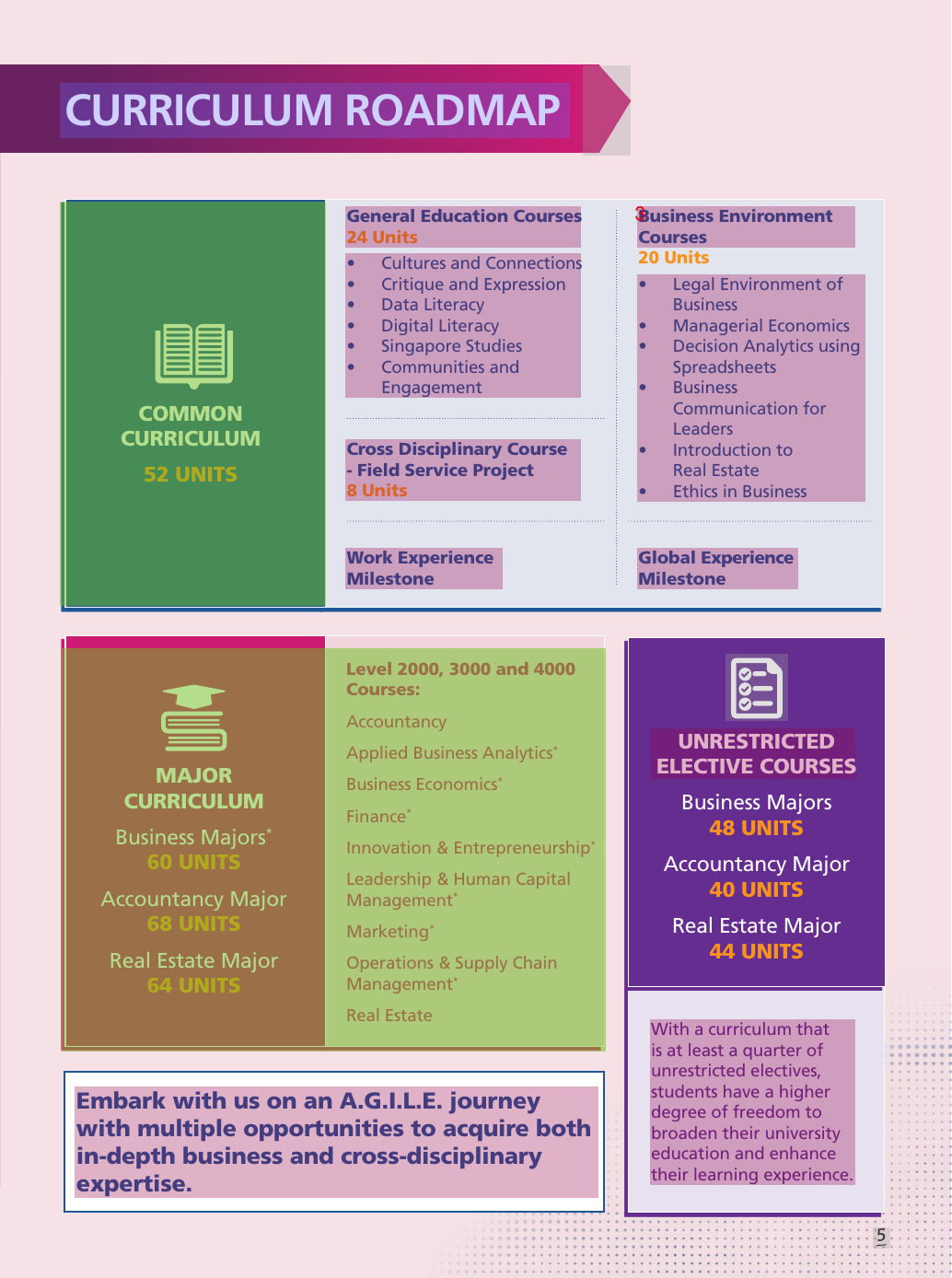}
        \end{subfigure}
        \label{fig:layout9}
    \end{minipage}

    \vspace{-0.5em}
    \caption{The 3 examples illustrate the function and effectiveness of layout detection on document pages.}
    \vspace{-1.0em}
    \label{fig:layout_examples3}
\end{figure*}

In this section, we present 9 pages along with their \xd{detected} layouts, \xd{which are} highlighted for better visualizations, as shown in Figure \ref{fig:layout_examples1}, \ref{fig:layout_examples2}, and \ref{fig:layout_examples3}.
Specifically, layout detection identifies the spatial location of different content types, such as images, tables, and text within a document. With the help of layout detection, we can precisely locate the specific position of an answer, whether it is an image, a text paragraph, or a table. This enables a more fine-grained layout-level \xd{evaluation of} multimodal retrieval capabilities.

\subsection{Annotation Examples}  \label{appendix:data_demo_annotation}

In this section, we present 4 annotation examples that illustrate typical multimodal retrieval and reasoning patterns, which help explain the construction and retrieval process. Each annotation includes the following primary components: question, answer, page-level labels, and layout-level labels. The page-level labels show the selected pages that contain ground truth evidence. Based on these selected pages, layout-level labels further display the specific layout box detection of evidence.  These examples frequently require reasoning across multiple pages and modalities. The evidence encompasses diverse formats such as figures, charts, tables, and texts, highlighting the complexity and richness of the multimodal retrieval tasks.


\begin{figure*}[htp]
    \centering
    \includegraphics[width=\linewidth]{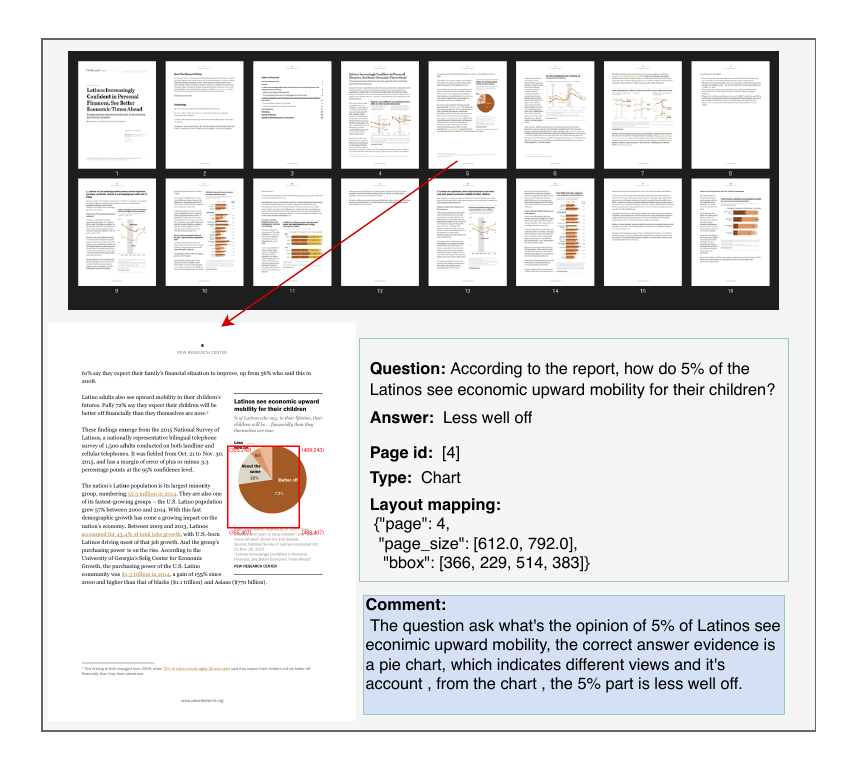}
    \caption{This example shows a typical image retrieval and reasoning task that requires synthesizing information from pie chart.}
    \label{fig:case1}
\end{figure*}

\begin{figure*}[htp]
    \centering
    \includegraphics[width=\linewidth]{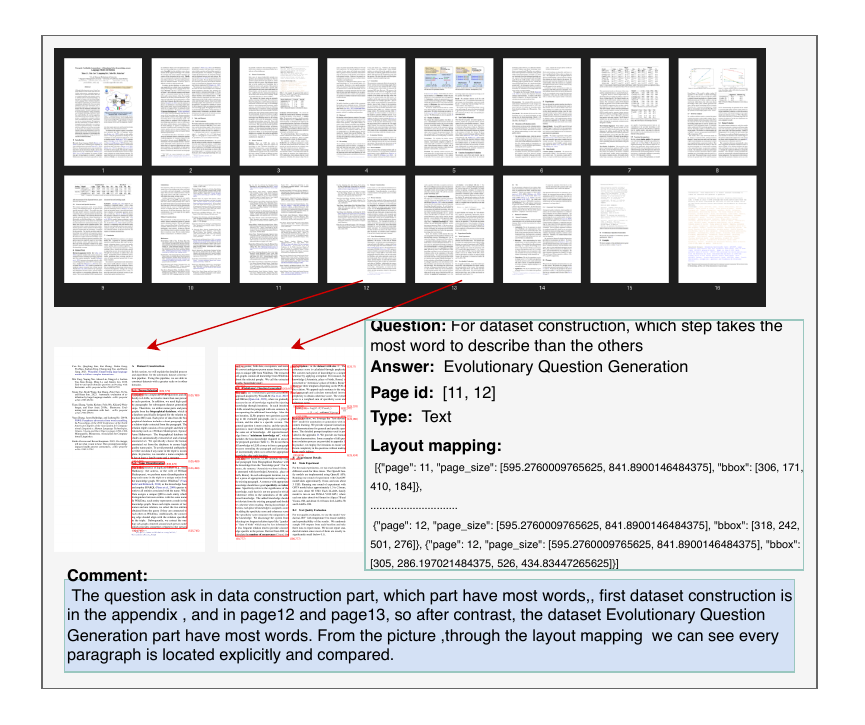}
    \caption{This example shows a typical multi-page retrieval task that requires synthesizing information from text passages across multiple pages.}
    \label{fig:case2}
\end{figure*}

\begin{figure*}[htp]
    \centering
    \includegraphics[width=\linewidth]{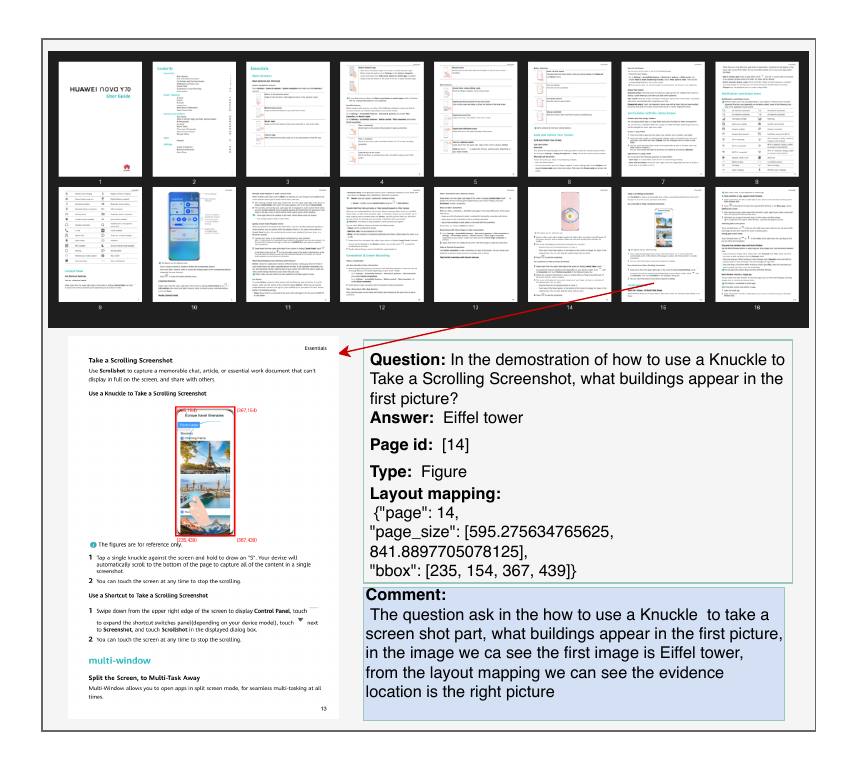}
    \caption{This example shows a typical image reasoning task that requires synthesizing information from specific image.}
    \label{fig:case3}
\end{figure*}

\begin{figure*}[htp]
    \centering
    \includegraphics[width=\linewidth]{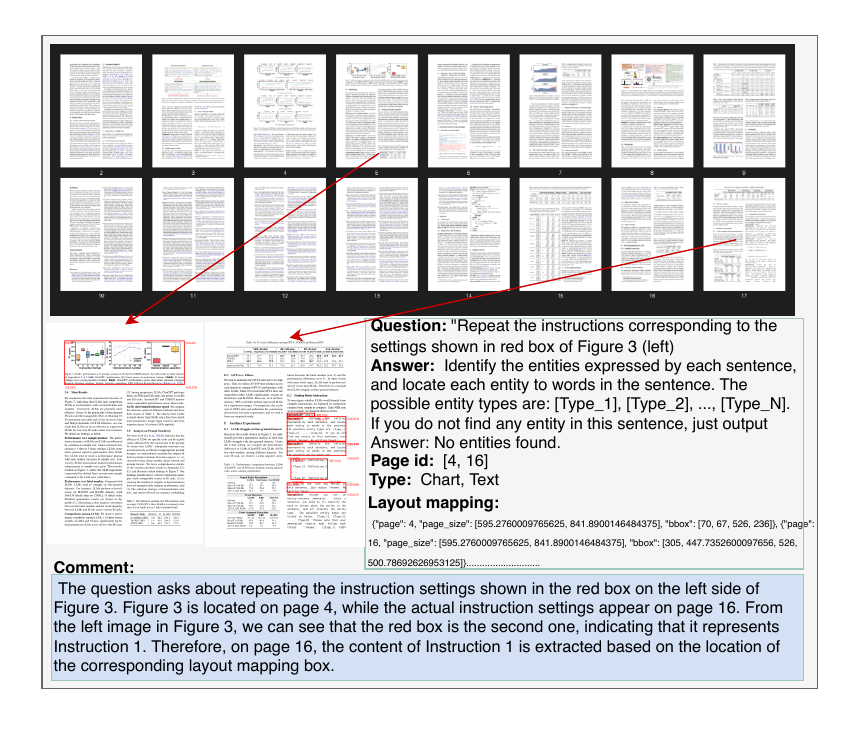}
    \caption{This example shows a typical multi-page image and text reasoning task that requires synthesizing cross-modal information from image and text.}
    \label{fig:case4}
\end{figure*}

\clearpage
\begin{table*}[t]
\small
    \centering
    \begin{tabular}{l|c@{\hskip 4pt}c@{\hskip 4pt}c|c|l}
        \toprule
        \multirow{2}{*}{Sub dataset} & \#Not  & \#Barely & \#Suit- & \multirow{2}{*}{\#Total} & \multirow{2}{*}{HuggingFace Resource}\\
        & suitable & suitable & able & \\
        \midrule
        arXiVQA       & 245   & 203 & 52  & 500 & 
        \href{https://huggingface.co/datasets/vidore/arxivqa_test_subsampled}{vidore/arxivqa\_test\_subsampled} \\
        DocVQA        & 345   & 130 & 25  & 500 & \href{https://huggingface.co/datasets/vidore/docvqa_test_subsampled}{vidore/docvqa\_test\_subsampled}\\
        InfoVQA       & 139   & 284 & 77  & 500 & \href{https://huggingface.co/datasets/vidore/infovqa_test_subsampled}{vidore/infovqa\_test\_subsampled}\\
        TAT-DQA       & 373   & 121 & 6   & 500 & \href{https://huggingface.co/datasets/vidore/tatdqa_test}{vidore/tatdqa\_test} \\
        
        \cmidrule(lr){2-6}
        
        Industrial  & 78    & 260 & 62  & 400 & 
        \begin{tabular}[c]{@{}l@{}}
            \href{https://huggingface.co/datasets/vidore/syntheticDocQA_energy_test}{vidore/syntheticDocQA\_energy\_test}\\ 
            \href{https://huggingface.co/datasets/vidore/syntheticDocQA_healthcare_industry_test}{vidore/syntheticDocQA\_healthcare\_industry\_test}\\
            \href{https://huggingface.co/datasets/vidore/syntheticDocQA_government_reports_test}{vidore/syntheticDocQA\_government\_reports\_test}\\
            \href{https://huggingface.co/datasets/vidore/syntheticDocQA_artificial_intelligence_test}{vidore/syntheticDocQA\_artificial\_intelligence\_test}
        \end{tabular} \\
        \midrule
        Sum           & 1180  & 998 & 222 & 2,400 & - \\
        Percentage    & 49.1\% & 41.5\% & 9.25\% & - & - \\
        \bottomrule
    \end{tabular}
\caption{Document statistics for ViDoRe Benchmark.}
\label{tab:vidorestats}
\end{table*}

\section{Detailed Analysis of ViDoRe Benchmark}
\label{appendix:vidore}

\subsection{Query and Annotation Analysis}
\label{appendix:vidore_query}

As mentioned in Section~\ref{sec:related}, \textbf{ViDoRe}~\cite{faysse2024colpali} is the most relevant benchmark to \dname.
It integrates several datasets such as DocVQA~\cite{2020docvqa}, InfoVQA~\cite{mathew22infographicsvqa}, TAT-DQA~\cite{zhu2022tatdqa}, arXiVQA~\cite{li2024arxiv}, and providing new documents in scientific, medical, administrative, and environment domains.
In this Appendix, we elaborate our analysis of the sampled 2,400 questions sampled from ViDoRe. The statistics are shown in Table~\ref{tab:vidorestats}.
ViDoRe test set contains questions in either English or French. In our work, we examine only the English questions.
For academic subsets, we examine all 1,500 questions: 500 questions from DocVQA, 500 questions from InfoVQA, and 500 questions from arXiVQA. 
In TAT-DQA, we sample and examine the first 500 questions.
For the industrial documents, we select 100 questions from each domain (\ie energy, healthcare, government, and artificial intelligence).
We examine sampled questions and summarize these questions into 3 categories:

\begin{itemize}[leftmargin=*]
    \item \textbf{Unsuitable Queries.} Queries that are not well-suited for IR systems can often burden these systems by generating numerous irrelevant results. For example, a query such as ``\textit{What’s the x-axis of the figure}'' is likely to prompt matches from multiple passages within document corpora that mention figures with an x-axis. This tends to happen because the query is overly broad and lacks contextual specificity. When such queries stem from Document Visual Question Answering (DocVQA) tasks targeting a single image, the challenge is exacerbated, as the reliance on precise context increases while the target remains too vague, undermining the fundamental principles of effective IR.

    \item \textbf{Barely Suitable Queries.} Queries that fall into this category provide some guidance towards locating useful passages, yet suffer from a lack of precise detail. These queries often fetch moderate number of passages, where both relevance and focus may not be as sharp. For example, the query ``\textit{What was the total assets from AMER in 2018?}'' is meant for Visual Question Answering (VQA) focused on a specific financial topic. Although this seems specific, the issue arises when multiple sections within an annual report discuss AMER's total assets for the year. This causes significant confusion since ViDoRe is set to acknowledge only a single passage as the verified answer. This lack of uniqueness in the ground truth makes it hard to evaluate the actual performance of IR system.

    \item \textbf{Suitable Queries.} 
    The most effective queries for IR systems are characterized by their specificity and ability to distinguish between different sections of texts. These queries often involve precise facts or detailed inquiries that facilitate pinpointing exact passages. For instance, the question ``\textit{What was the magnitude of the earthquake that occurred in Maule on 2/27/2010?}'' incorporates significant keywords and details that guide the retrieval system directly to the necessary data. Such queries align perfectly with the objectives of IR, leveraging specificity and detailed context to efficiently retrieve most relevant information.
\end{itemize}

The comprehensive analysis of our queries, as presented in Table~\ref{tab:vidorestats}, reveals a significant challenge in adapting questions from Visual Question Answering (VQA) datasets (such as DocVQA, InfoVQA, TAT-DQA, and arXiVQA) for Information Retrieval (IR) purposes. Only 8\% of these queries prove suitable for effective IR usage. In comparison, queries derived from industrial documents perform slightly better, with 15.5\% deemed suitable. A common issue identified is that these queries are either excessively simplistic or highly specific to a particular context. 
Our findings suggest that the primary difficulty stems from the inherent differences between DocIR and DocVQA. VQA queries are typically crafted to address content on a specific page or within a particular image, inherently limiting their scope and specificity. This specificity and simplism are functional within the confines of the intended VQA context but pose substantial limitations when such queries are repurposed for IR tasks.

\subsection{Document Corpora Analysis}
\label{appendix:vidore_analysis}

ViDoRe bootstrap document corpora directly from existing DocVQA benchmarks (\ie DocVQA, InfoVQA, TAT-DQA, arXiVQA) that perform single-page VQA. In the DocVQA setting, only selected pages are provided for VQA, rather than the entire document pages. For arXivQA, the retrieved passages are not document pages, but are cropped images (\eg figures, tables, and charts).
In our experiments, we need to rely on the entire documents pages to evaluate retrieval on long documents.
To bridge the gap of missing complete document corpora,  we put in considerable efforts to collect the original documents of existing DocVQA datasets, as mentioned in Section~\ref{ssec:doc_corpora}.

\section{License Agreements}
\label{appendix:license}

We ensure that the distribution of each dataset complies with the corresponding licenses, all of which are listed below:
\begin{itemize}[leftmargin=*, itemsep=0.0em, topsep=0.0em]
    \item MMLongBench-Doc: is under Apache-2.0 license agreement for academic research purposes.
    \item DocBench: we achieved the agreement of usage as academic research from the dataset's author. 
    \item MP-DocVQA: is under ``MIT License'' license agreement for academic research purposes.
    \item SlideVQA:  is under ``NTT License'' license agreement for academic research purposes.
    \item TAT-DQA: is under ``CC-BY-4.0'' license agreement for academic research purposes.
    \item ArXivQA: is under ``CC-BY-SA-4.0'' license agreement for academic research purposes.
    \item SciQAG: is under ``CC-BY-4.0'' license agreement for academic research purposes.
    \item DUDE: is under ``GPL-3.0'' license agreement for academic research purposes.
    \item CUAD: is under ``CC-BY-4.0'' license agreement for academic research purposes.
\end{itemize}

For the new annotations contributed in \dname, including but not limited to the questions, page and layout annotations, we make them available solely for research purposes. Users are permitted to use, modify, and share these annotations for academic and non-commercial research activities. Any other use, including commercial exploitation, is not permitted without explicit written permission from the authors.

\section{Ethical Considerations}
\label{appendix:ethical}

The introduction and broader adoption of \dname may have potential ethical impacts spanning both positive and negative dimensions. Below, we outline possible negative consequences and discuss potential mitigation strategies:

\textbf{Privacy Risks:}
\dname enables models to retrieve relevant information over lengthy, multimodal documents, which may include sensitive personal, financial, or health information. There is a risk that such technologies could be leveraged for large-scale surveillance, unauthorized extraction of personal data, or other privacy violations.

\textbf{Fairness and Bias:}
If benchmarked models are trained or evaluated on data that does not reflect diverse demographic, linguistic, and backgrounds, outputs may exhibit biases. This may lead to unfair decision-making or stereotypes.

\textbf{Mitigation Strategies:}
To mitigate these risks, we make sure that:
(i) Benchmark development uses only publicly available, carefully vetted datasets, with sensitive information anonymized or removed;
(ii) Retrieval outputs are monitored for bias and fairness.

We encourage researchers and practitioners employing \dname to be mindful of these factors and to actively work toward responsible development and deployment, including transparency about limitations and proactive safeguards where needed. We welcome community feedback and collaboration on best practices to further reduce risks as this technology evolves.

\end{document}